\documentclass[12pt]{article}
\usepackage[a4paper, width=170mm, top=22mm, bottom=22mm]{geometry}
\usepackage{suffix}     
\usepackage{ifthen}     

\usepackage{hyperref}
\usepackage{epigraph}   
\usepackage{csquotes}   
\usepackage{xcolor}     
\usepackage{enumitem}   
\setlist{listparindent=\parindent,parsep=0pt}
\usepackage{multicol}   

\usepackage[flushleft]{threeparttable}  

\usepackage{graphicx}
\graphicspath{ {images/} }
\usepackage{rotating}   
\usepackage{caption}
\usepackage{subcaption} 

\usepackage[backend=biber, bibencoding=utf8, style=authoryear, maxcitenames=2, uniquelist=false, sorting=nyt]{biblatex}   
\AtEveryBibitem{
    \clearfield{day}
    \clearfield{month}
}   

\usepackage{mathtools}
\usepackage{amsmath}
\usepackage{amssymb}
\usepackage{amsfonts}   
\usepackage{physics}    

\renewcommand{\vec}[1]{\ensuremath{\boldsymbol{\mathbf{#1}}}}

\newcommand{\unit}[1]{\ensuremath{\hat{\vec{#1}}}}
\newcommand{\diff}[2][]{%
    \ifthenelse{ \equal{#1}{} }
        {\ensuremath{\operatorname{d}\!{#2}}}
        {\ensuremath{\operatorname{d}^{#1}\!{#2}}}
}

\newcommand{\tensor}[1]{\ensuremath{\boldsymbol{\mathsf{#1}}}}
\newcommand{\contract}{\ensuremath{\mathbin{\boldsymbol{:}}}}
\newcommand*{\trans}{\ensuremath{^{\mathsf{T}}}}
\newcommand{\Dv}[2]{\ensuremath{\frac{\operatorname{D}\!{#1}}{\operatorname{D}\!{#2}}}}
\WithSuffix\newcommand\Dv*[2]{\ensuremath{\flatfrac{\operatorname{D}\!{#1}}{\operatorname{D}\!{#2}}}}
\newcommand{\DvMF}[3]{\ensuremath{\frac{\operatorname{D}_{#1}\!{#2}}{\operatorname{D}\!{#3}}}}
\WithSuffix\newcommand\DvMF*[3]{\ensuremath{\flatfrac{\operatorname{D}_{#1}\!{#2}}{\operatorname{D}\!{#3}}}}
\renewcommand{\laplacian}{\grad^2}

\newcommand{\laplacianhat}{\hat{\grad}^2}
\newcommand{\divtilde}[1]{\tilde{\grad}\mathbin{\vdot}{#1}}
\newcommand{\divhat}[1]{\hat{\grad}\mathbin{\vdot}{#1}}

\renewcommand{\Re}{\ensuremath{\operatorname{Re}}}  
\renewcommand{\Pr}{\ensuremath{\operatorname{Pr}}}  
\newcommand{\Ra}{\ensuremath{\operatorname{Ra}}}    
\newcommand{\Nu}{\ensuremath{\operatorname{Nu}}}    

\usepackage[title]{appendix}
\newif\ifverbose
\verbosefalse

\title{\vspace{-1.0cm}Two-fluid single-column modelling of Rayleigh-B\'{e}nard convection as a step towards multi-fluid modelling of atmospheric convection}
\author{Daniel Shipley, Hilary Weller, Peter Clark, William McIntyre}
\date{\today}

\begin{document}

\maketitle

\renewcommand{\abstractname}{Abstract}
\begin{abstract}
    \noindent Multi-fluid models have recently been proposed as an approach to improving the representation of convection in weather and climate models. 
    This is an attractive framework as it is fundamentally dynamical, removing some of the assumptions of mass-flux convection schemes which are invalid at current model resolutions. 
    However, it is still not understood how best to close the multi-fluid equations for atmospheric convection. 
    In this paper we develop a simple two-fluid, single-column model with one rising and one falling fluid. 
    No further modelling of sub-filter variability is included. 
    We then apply this model to Rayleigh-B\'{e}nard convection, showing that, with minimal closures, the correct scaling of the heat flux ($\Nu$) is predicted over six orders of magnitude of buoyancy forcing ($\Ra$). 
    This suggests that even a very simple two-fluid model can accurately capture the dominant coherent overturning structures of convection.
\end{abstract}

\setcounter{tocdepth}{2}
\tableofcontents

\section{Introduction}

Despite being an important part of the global circulation and local variability, atmospheric convection is one of the weakest aspects of Numerical Weather Prediction (NWP) and climate models \parencite{ar:StephensEtAl2010,ar:SherwoodEtAl2014,ar:SteinEtAl2015,ar:ClarkEtAl2016}. 
These difficulties are at least in part due to the ``grey zone'' problem: the resolution of current models is such that a typical grid spacing is neither much smaller, nor much larger, than a typical convective scale (say $\mathcal{O}(1\ \mathrm{km})$ for shallow convection), meaning that neither traditional parametrizations, nor so-called ``explicit convection'', adequately represent the flow \parencite{ar:Wyngaard2004,ar:HollowayEtAl2014,ar:ZhouEtAl2014,ar:ClarkEtAl2016}.
If a separation of scales exists between the clouds scale(s) and the grid spacing, simplifying assumptions may be made to aid in parametrization of sub-grid processes. 
In atmospheric convection, traditional closures have assumed some form of balance between the large-scale forcing and the convective response, and that the area of a ``grid box'' taken up by cloud is small \parencite{ar:ArakawaSchubert1974,ar:Plant2010}. 
Both of these assumptions break down in the grey zone of current model resolutions, leading to unrealistic behaviour of models with traditional parametrizations.

At the other end of the scale, for grid spacings much smaller than the convective length scale(s), techniques of large-eddy simulation (LES) become applicable \parencite[see][for a review]{ar:Mason1994}.
However, true LES requires very high resolution --- typically $\mathcal{O}(10\ \mathrm{m})$ for the dry convective boundary layer \parencite{ar:SullivanEtAl2011} --- which is beyond the computational capabilities of NWP and climate models for the foreseeable future. 
Many current operational forecasting models (e.g. the Met Office UKV configuration of the MetUM, DWD's ICON-D2 ) use ``explicit'' convection, where the convection scheme is turned off. 
Some form of 'turbulence' scheme is still required, often an LES-like eddy viscosity/diffusivity scheme.
While these perform better at grey zone resolutions, there are still undesirable effects, in particular the prediction of incorrect length scales (cloud size and inter-cloud spacing) typically larger than observed scales, even in cases where the model \textit{should} be able to resolve the smaller scale \parencite{ar:LeanEtAl2008}.
There is thus both a need for parametrization well into the future, and a need for new parametrization approaches in the grey zone.

Multi-fluid modelling has recently been proposed as an approach to representing convection in the grey zone \parencite{ar:Yano2014,ar:ThuburnEtAl2018,ar:TanEtAl2018}; 
similar equation sets are used for the modelling of multi-phase flows in engineering \parencite[e.g.][]{bk:StaedkeTwoPhase}.
In the convection context, this takes inspiration from traditional mass-flux parametrizations in splitting the fluid into multiple components, which may represent updrafts, environment, downdrafts etc..
The split is applied directly to the governing equations, which are then spatially filtered, allowing a fully 3D and time-dependent framework to be derived \parencite{ar:ThuburnEtAl2018,ar:ShipleyEtAl2021b_inPrep}.
Neither quasi-equilibrium nor small updraft fraction are assumed in the derivation.
Each ``fluid'' evolves according to its own prognostic equations, interacting with other fluids via the pressure gradient, and terms involving the exchange of mass, momentum, energy, and tracers. 
These exchange terms are the analogue of entrainment, detrainment, and cloud-base mass-flux in traditional models, and must be parametrized. 
Convection is inherently a part of the dynamics in this framework: there is no separate convection scheme which is called by the dynamical core.

The skewness of (joint) probability distribution functions of variables in convective flows is well known to be important \parencite{ar:LarsonEtAl2002,ar:ZhuZuidema2009} and is often poorly treated in first or second-order turbulence closures; one approach to modelling this variability is assuming bi-Gaussian joint probability distributions in PDF-based convective closures \parencite{ar:LarsonEtAl2012,ar:Fitch2019}.
Each Gaussian can be thought of as a different component of the fluid.
A potential advantage of the multi-fluid approach is that even the simplest possible multi-fluid model, a two fluid model, intrinsically captures information about odd-order moments. 
It is therefore possible that the multi-fluid method can provide a better low-order approximation for flows with bimodal distributions, or large skewness. 

In order to build a multi-fluid model of atmospheric convection, the multi-fluid equation set must be closed. 
The form of these closures directly depends on the definition of the fluid partitions \parencite{ar:deRooyEtAl2013,ar:ShipleyEtAl2021b_inPrep}.
For example, the single-column 2-fluid model of \textcite{ar:ThuburnEtAl2019} contains entrainment and detrainment closures designed to capture coherent structures in the convective boundary layer, whereas the closures in \textcite{ar:CohenEtAl2020} are designed to model a second fluid in the cloud layer only.
Perturbation pressure closures for the latter approach were suggested in \textcite{ar:HeEtAl2020_submitted}.
Entrainment and detrainment closures based on velocity divergence, and a bulk viscous parametrization for the perturbation pressure, were proposed and tested in \textcite{ar:WellerEtAl2020}, but the test cases used for comparison were non-turbulent, unlike the real atmosphere.
All of these multi-fluid models have been single-column, and used standard atmospheric test cases (e.g. dry rising bubble, dry convective boundary layer,  oceanic and continental shallow cumulus, diurnal deep convection) for verification. 
While prior work shows the considerable promise of the multi-fluid method, little work has been done testing the response of a specific multi-fluid scheme to a variety of forcings, or suggested how the closure constants should scale with that forcing.
Such investigation could lead to more consistent results compared to tuning a model to a handful of test cases.

To gain a better understanding of the multi-fluid equations, and how some of the new closure terms affect the solution, we present a single-column model of dry Rayleigh-B\'{e}nard convection (RBC) with one rising and one falling fluid. 
RBC is the simplest relevant convection problem: the equations and boundary conditions are as simple as possible while still allowing for a fully turbulent convective solution. 
RBC has been extensively studied, and a wealth of experimental, numerical, and theoretical results make it a well-constrained starting point \parencite{bk:ChandrasekharStability,ar:AhlersEtAl2009,ar:ChillaSchumacher2012}. 
In particular, the scaling of bulk buoyancy and momentum transport with the applied buoyancy forcing is well understood over at least ten orders of magnitude. 

It is important to understand the response of the model in a fully-parametrized equilibrium setting before moving to the grey zone.
This will help pin down the physics of a multi-fluid model of convection, free of the complexities --- especially microphysics and phase changes --- of the real atmosphere.

The paper begins with an overview of Rayleigh-B\'{e}nard convection in section~\ref{sec:RayleighBenard}, motivating its use as a reasonable testbed for developing insights into ``real-world'' convection. 
Results from 2D direct numerical simulations (DNS) of dry RBC are presented, and shown to agree with reference results. 
In section~\ref{sec:multiFluidDerivation} a multi-fluid Boussinesq equation set is presented, along with a discussion of how and why this equation set differs from previous papers on multi-fluid convection parametrization.
Closures for one rising and one falling fluid which attempt to capture the large-scale overturning circulation are presented in section~\ref{sec:closures}, and a scaling argument is presented for the magnitude of the pressure differences between the fluids.
The numerical method is then described in section~\ref{sec:numericalMethod}. 
In section~\ref{sec:results}, results of the two-fluid single-column model (section~\ref{sec:closures}) are compared with horizontally-averaged results from the DNS (section~\ref{sec:DNS}) over a range of buoyancy forcing spanning seven orders of magnitude ($10^{3} \leq \Ra \leq 10^{10}$), and the sensitivity of the model to its two dimensionless closure constants is investigated. 
The paper concludes with a summary of its results and their relevance to convection parametrizations, and a discussion of avenues for future research.

\section{Rayleigh-B\'{e}nard convection (RBC)}\label{sec:RayleighBenard}

The Rayleigh-B\'{e}nard problem is the simplest fluid dynamical model of convection. 
First studied experimentally by \textcite{ar:Benard1900}, the problem was given a theoretical treatment by \textcite{ar:Rayleigh1916} which has been the basis of over a century of investigation. 
\textcite{ar:Rayleigh1916} studied the motion of a Boussinesq fluid confined between two perfectly conducting horizontal plates of infinite extent, each held at a constant uniform temperature. 
For mathematical tractability he considered stress-free velocity boundary conditions at the plates; the no-slip case was tackled by \textcite{ar:Jeffreys1926,ar:Jeffreys1928}. 
RBC has long been of interest to the meteorological community, being the basis of the \textcite{ar:Lorenz1963} seminal discovery of deterministic chaos, and a key component of our understanding of convective systems \parencite[][ch.~3]{bk:EmanuelConvection}. 
Moist extensions of the model have been considered to gain insight into moist convection, though far less work has been performed on moist versions of the problem than on the dry case \parencite{ar:Bretherton1987,ar:Bretherton1988,ar:PauluisSchumacher2010,ar:WeidauerSchumacher2012,ar:VallisEtAl2019}. 
In this section, the classical results relevant to this paper are collected. 
The canonical text covering stability and the onset of convection is \textcite{bk:ChandrasekharStability}; recent reviews covering fully turbulent convection are \textcite{ar:AhlersEtAl2009,ar:ChillaSchumacher2012}.

The setup of the Rayleigh-B\'{e}nard problem is as follows. 
A Boussinesq fluid is confined between two smooth, flat, horizontal plates, a fixed distance $H$ apart. 
Each of these is held at a fixed buoyancy, $\pm \flatfrac{\Delta B}{2}$, with no-slip, no-normal flow velocity boundary conditions. 
For both analytical and numerical simplicity we choose the lateral boundaries to be periodic in all fields. 
The motion of the fluid is described by the following Boussinesq equations of motion:
\begin{align}
    \Dv{\vec{u}}{t}
    &=
    b \vec{k}
  - \grad{P}
  + \nu \laplacian{\vec{u}},
    \label{eq:SFboussinesq_momentum}
    \\
    \Dv{b}{t}
    &=
    \kappa \laplacian{b},
    \label{eq:SFboussinesq_buoyancy}
    \\
    \div{\vec{u}}
    &=
    0.
    \label{eq:SFboussinesq_mass}
\end{align}
Here $\vec{u}$ denotes the velocity field of the fluid; $b \coloneqq \flatfrac{g(\rho_\text{ref} - \rho)}{\rho_\text{ref}}$ its buoyancy\footnote{For our desired application to atmospheric convection, $\flatfrac{(\rho_\text{ref} - \rho)}{\rho_\text{ref}} \simeq \flatfrac{(\theta - \theta_\text{ref})}{\theta_\text{ref}}$, where $\theta$ is the potential temperature, though much previous work on Rayleigh-B\'{e}nard convection (RBC) is performed in terms of temperature, using the approximation $\flatfrac{(\rho_\text{ref} - \rho)}{\rho_\text{ref}} \simeq \flatfrac{(T - T_\text{ref})}{T_\text{ref}}$. The equation set retains the same form.}; $P \coloneqq \flatfrac{p}{\rho_\text{ref}}$ its pressure potential; $\nu$ its kinematic viscosity; $\kappa$ its buoyancy diffusivity; and $\vec{k}$ is a unit vector antiparallel to gravity, defining the vertical ($z$) direction. 
All variables are defined relative to a resting,  uniformly constant-density, hydrostatically-balanced pressure reference state.

A diffusive nondimensionalization of equations~\eqref{eq:SFboussinesq_momentum}-\eqref{eq:SFboussinesq_mass} \parencite[as][but choosing a diffusive rather than viscous time-scale]{bk:ChandrasekharStability,bk:EmanuelConvection} by the external parameters, $\unit{x} \coloneqq \flatfrac{\vec{x}}{H}, \hat{b} \coloneqq \flatfrac{b}{\Delta B}, \hat{t} \coloneqq \flatfrac{t \kappa}{H^2}, \unit{u} \coloneqq \flatfrac{\vec{u} H}{\kappa}, \hat{P} \coloneqq \flatfrac{H^2}{\kappa \nu}$, shows that two dimensionless parameters govern the flow (the boundary conditions are given for completeness):
\begin{align}
    \Dv{\hat{\vec{u}}}{\hat{t}}
    &=
    \Pr
    \left(
        \Ra \hat{b} \vec{k}
      - \hat{\grad}{\hat{P}}
      + \laplacianhat{\hat{\vec{u}}}
    \right),
    \label{eq:SFboussinesq_momentum_dimless}
    \\
    \Dv{\hat{b}}{\hat{t}}
    &=
    \laplacianhat{\hat{b}},
    \label{eq:SFboussinesq_buoyancy_dimless}
    \\
    \divhat{\hat{\vec{u}}}
    &=
    0,
    \label{eq:SFboussinesq_mass_dimless}
    \\
    \hat{b}(\hat{z}=0) = \frac{1}{2},
    &\qquad
    \hat{b}(\hat{z}=1) = - \frac{1}{2},
    \label{eq:SFboundaryConditions_buoyancy_dimless}
    \\
    \hat{\vec{u}}(\hat{z} &= 0, 1) = \vec{0}.
    \label{eq:SFboundaryConditions_velocity_dimless}
\end{align}
Nondimensionalized variables are denoted by a hat, and the dimensionless parameters are defined by:
\begin{align}
    \Ra
    \coloneqq
    \frac{\Delta B \cdot H^3}{\kappa\ \nu},
    \qquad \qquad
    \Pr
    \coloneqq
    \frac{\nu}{\kappa}.
    \label{eq:definitionOfRaAndPr}
\end{align}
The Rayleigh number, $\Ra$, is the ratio of buoyancy forcing ($\Delta B$) to viscous diffusion ($\flatfrac{\kappa \nu}{H^3}$); and the Prandtl number, $\Pr$, is the ratio of the diffusion of momentum ($\nu$) to the diffusion of buoyancy ($\kappa$). 
The former can thus be seen as measure of the applied forcing in RBC, whereas the latter is an intrinsic property of the fluid. 
This nondimensionalization shows that any two RBC systems with the same $\Ra$ and $\Pr$ support the same solutions, i.e. are self-similar.\footnote{A third parameter, the aspect ratio of the domain, $\Gamma \coloneqq L/H$, enters via the lateral boundary conditions; however, the dependence upon the aspect ratio is generally weak so long as $\Gamma > 1$ --- see \cite{ar:AhlersEtAl2009}, section~~3E; also \cite{ar:JohnstonDoering2009,ar:Bailon-CubaEtAl2010,ar:ZhouEtAl2012} --- and the dependence is weaker for periodic boundaries than for rigid boundaries.}

It is worth noting that this nondimensionalization specifically singles out the diffusive regime as the regime of interest, relevant for considerations of stability. 
For consideration of the convective solutions, a nondimensionalization based on the buoyancy forcing is more useful. 
This ``free fall'' or ``free convective'' scaling gives velocity and time scales $U_B \coloneqq \sqrt{\Delta B\ H}, T_B \coloneqq \sqrt{H / \Delta B}$, and is ubiquitous in the CBL literature (where $U_B$ is denoted $w^*$, see, e.g.,\textcite{bk:Garratt}. 
Such a scaling also gives an a priori estimate for the Reynolds number, $\Re \propto \Ra^{1/2}\Pr^{-1/2}$.\footnote{The a priori scaling for the Nusselt number that this predicts, $\Nu \propto \Ra^{1/2}$ --- the so-called ``ultimate scaling'' --- is steeper than observed to date in experimental or numerical dry RBC, because the non-turbulent surface layers next to the boundaries prevent a thermal shortcut.} 
This approximate $\Re(\Ra)$ scaling is observed for the regimes applicable to this paper.

The equation set \eqref{eq:SFboussinesq_momentum_dimless}-\eqref{eq:SFboundaryConditions_velocity_dimless} has a unique stationary zero-flow solution, with a linear buoyancy gradient between the plates and a quadratic pressure profile:
\begin{align}
    \vec{u} 
    = 
    \vec{0},
    \qquad
    b 
    = 
    \frac{1}{2}(1 - z),
    \qquad
    P 
    = 
    P_0 
  + \frac{z}{2} (1 - \frac{z}{2}).
    \label{eq:SFinitialConditions_dimless} 
\end{align}
This solution is both linearly and nonlinearly unstable to perturbations if and only if the Rayleigh number exceeds a critical value, $\Ra_\text{c}$; importantly, the stability does not depend on the Prandtl number \parencite[see, for instance,][]{bk:ChandrasekharStability,ar:Joseph1966,ar:LindsayStraughan1990}. 
Below $\Ra_\text{c}$, solutions are purely diffusive; above $\Ra_\text{c}$, a circulation develops which increases the heat transport. 
This circulation can either be steady, periodic, quasi-periodic, or turbulent, depending on the governing parameters $(\Ra, \Pr)$. 
The precise value of $\Ra_\text{c}$ depends on the velocity boundary conditions at the top and bottom boundaries, but not on the dimensionality of the domain; for our chosen no-slip conditions, $\Ra_\text{c} \approx 1708$, and the wavelength of the most unstable mode is $\lambda_c \approx 2.02 H$ \parencite[table~3]{bk:ChandrasekharStability}.

The domain- and time-averaged dimensionless buoyancy flux is given by the Nusselt number: 
\begin{align}
    \Nu 
    \coloneqq
    \left<
        \vec{k} \vdot 
        \left(
            \hat{\vec{u}} \hat{b}
          - \hat{\grad}{\hat{b}}
        \right)
    \right>_{V,t}
    =
    \left<
        \hat{w}\hat{b} - \pdv{\hat{b}}{\hat{z}}
    \right>_{A,t},
\end{align}
which is the ratio of the actual buoyancy flux to the buoyancy flux of the purely diffusive solution. 
Averaging the buoyancy equation \eqref{eq:SFboussinesq_buoyancy_dimless} over a horizontal plane and over time (denoted $\langle\dots\rangle_{A,t}$) shows that the Nusselt number is independent of height in a statistically stationary flow.

Exact results for the domain- and time-averaged kinetic and thermal dissipation rates, $\varepsilon_{\vec{u}}$ and $\varepsilon_b$, are given by \parencite[appendix~1]{ar:Siggia1994,bk:ChandrasekharStability}:
\begin{align}
	\varepsilon_{\vec{u}}
	&\coloneqq
    \langle
        \grad{\vec{u}} \contract \grad{\vec{u}}
    \rangle_{V,t}
    =
    \Ra (\Nu - 1),
    \label{eq:SFboussinesq_dissipationEnergy}
    \\
    \varepsilon_{b}
    &\coloneqq
    \langle
        \grad b \vdot \grad b
    \rangle_{V,t}
    =
    \Nu.
    \label{eq:SFboussinesq_dissipationBuoyancy}
\end{align}
Here the ``double dot product'' denotes the complete contraction of two rank-two tensors, following the convention $\tensor{A} \contract \tensor{B} \coloneqq \sum_{a,b} A_{ab}B^{ab}$. 
Thus the vertical buoyancy flux is the only quantity that characterizes the stationary-state global energetic response of the system to the applied forcing ($\Ra, \Pr$)\footnote{It is worth noting that these results \eqref{eq:SFboussinesq_dissipationEnergy}-\eqref{eq:SFboussinesq_dissipationBuoyancy} are quite general; in particular they do not rely on the plates being smooth and flat, and they apply equally well also to the cases of stress-free velocity or constant buoyancy flux boundary conditions.}. 
The statistically steady-state Rayleigh-B\'{e}nard problem can then be framed as asking the question: if we apply a buoyancy forcing $\Ra$ to a Boussinesq fluid characterized by $\Pr$, what is the resulting $\Nu$? 
Scaling theories for $\Nu$ as a function of $\Ra$ and $\Pr$ are well-developed, and there is good agreement between the theory and numerical and experimental results until at least $\Ra = 10^{11}$ for $\Pr = \mathcal{O}(1)$ \parencite{ar:AhlersEtAl2009,ar:ChillaSchumacher2012}.
{} 
It is therefore a strong test of any dynamical low-order model of RBC to reproduce these scalings.

\subsection{2D direct numerical simulation of RBC}\label{sec:DNS}
To provide a reference ``truth'' for later sections in the paper, results from two dimensional direct numerical simulations of Rayleigh-B\'{e}nard convection over a wide range of $\Ra$ are presented. 
These simulations also serve to illustrate the phenomenology of RBC, and to indirectly validate the numerical methods via comparison with reference results.

While the restriction to two dimensions may seem like too great a simplification, global and large-scale results of Rayleigh-B\'{e}nard convection in two and three dimensions are remarkably similar so long as the Prandtl number is not too small. 
{}
The classical results regarding the critical Rayleigh number, critical wavelength, and onset of convection are unaffected (see \cite[ch.~2]{bk:ChandrasekharStability}; though not explicitly stated, the stability analysis does not depend on the dimensionality of the domain). 
After the onset of convection, for $\mathcal{O}(1) \Pr$ and greater, the scalings of global parameters such as the Nusselt and Reynolds numbers, as well as the boundary layer depths, are virtually the same in 2D as in 3D (although the magnitudes differ slightly) --- see \textcite{ar:SchmalzlEtAl2004}.
Many theoretical analyses of the problem have either included two dimensions as a special case, or actually assumed only two dimensions, the successful \textcite{ar:GrossmannLohse2000} scaling theory for the Nusselt and Rayleigh numbers being a prime example of the latter. 
Therefore we choose to perform 2D simulations, given the similarity between 2D and 3D results and the vastly reduced computational requirements for 2D calculations.

Our simulation suite runs from fully diffusive ($\Ra \simeq 10^2$) to well into the turbulent regime ($\Ra \simeq 10^{10}$). 
Rayleigh numbers have been chosen such that there is at least one simulation per factor of ten of $\Ra$, with extra simulations run in the vicinity of $\Ra_\text{c}$. 
The Prandtl number is fixed to be $\Pr = 0.707$, the value for dry air at STP. 
Reviews of RBC suggest that qualitative results remain similar so long as the asymptotic range of $\Pr$ is the same, i.e. $\Pr = \mathcal{O}(1)$ rather than $\Pr \to 0$ or $\infty$ \parencite{ar:AhlersEtAl2009,ar:ChillaSchumacher2012}. 
In particular, the scaling exponent $\Nu \propto \Ra^{\beta}$ is not strongly Prandtl-number dependent.

\subsubsection{Choice of resolution}\label{sec:resolution}
\begin{itemize}
    \item By ``resolution'', $\Delta_r$ , we mean the smallest length scale at which structures of the flow are well captured by the model.
    \item By ``filter scale'', $\Delta_f$, we mean the length scale(s) associated with any filter applied to the flow, whether to the solutions or to the governing equations.
    \item By ``grid scale'' (alternatively, ``grid length'' or ``grid spacing''), $\Delta_g$, we mean the actual distance between points (or cell centres) within a discretized model.
\end{itemize}
A direct numerical simulation of a fluid (``DNS'') must ``resolve'' all dynamically relevant scales of the fluid flow in order to justify the assumption that no small-scale processes need to be parametrized. 
But there are various metrics by which we can test whether a flow is ``resolved''. To \textit{fully} resolve a turbulent flow, the grid spacing must resolve at least a factor of ten into the viscous subrange \parencite{ar:Kerr1985}, which is very computationally expensive. 
However, to get the majority of the statistics right the requirements are less extreme: the Kolmogorov dissipation length, $\eta \coloneqq H (\Pr^2 / \varepsilon_{\vec{u}})^{1/4}$, must be resolved \parencite{ar:Groetzbach1983}. 
Within fully-developed turbulence in the bulk of the fluid the exact result for the global kinetic energy dissipation rate, \eqref{eq:SFboussinesq_dissipationEnergy}, may be used to estimate the smallest dynamically relevant scale:
\begin{align}
    \frac{\eta}{H}
    &=
    \left(
    	\frac{\Pr^2}{(\Nu - 1)\Ra}
    \right)^{\frac{1}{4}}.
    \label{eq:kolmogorovLength}
\end{align}

Towards the boundaries, the kinetic and thermal boundary layers must be resolved --- dissipation is typically higher in these regions, reducing the smallest dynamically relevant length scale. 
\textcite{ar:ShishkinaEtAl2010} estimated local dissipation lengths based on dissipation rates defined within the boundary layers, using these to estimate the minimum number of points $N_u, N_b$ required within each boundary layer (thickness $\delta_{\vec{u}}, \delta_b$) in order to adequately resolve the flow. 
This estimate is for $10^6 < \Ra < 10^{10}$, so for $\Ra \leq 10^6$ we use the values of $N_u, N_b$ estimated for $\Ra = 10^6$. 
Note that this extra resolution is only required in the vertical direction.

At any point in the flow the smallest of $\{\eta, \delta_b/N_b, \delta_{\vec{u}}/N_u \}$ must be resolved. 
Collecting the results of \textcite{ar:Groetzbach1983} and \textcite{ar:ShishkinaEtAl2010}, the grid spacing is required to satisfy $\Delta x_{\eta} < 2 \eta, \Delta x_b < \flatfrac{\delta_b}{0.35 \Ra^{0.15}}, \Delta x_{\vec{u}} < \flatfrac{\delta_{\vec{u}}}{0.31 \Ra^{0.15}} $ to be adequate to resolve each respective scale.

To make use of the resolution requirements, the boundary layer thicknesses must be estimated. 
Since the centre of the domain will be statistically well-mixed after the onset of convection, we must have $\delta_b / H \sim \flatfrac{1}{2\Nu}$. 
For the parameter regimes of this study, $\Nu \sim \Ra^{2/7}$ and so $\delta_b / H \sim \Ra^{-2/7}$ \parencite{ar:CastaingEtAl1989,ar:ShraimanSiggia1990,ar:AhlersEtAl2009}. 
Prandtl-Blasius boundary layer theory suggests that the kinetic boundary layer thickness should scale as $\delta_{\vec{u}} / H \sim \Ra^{-1/4}$, and $\delta_{\vec{u}} < \delta_b$ is expected over the entire Rayleigh number range here considered \parencite[][fig.~3]{ar:AhlersEtAl2009}. 
To estimate the prefactors, an over-resolved simulation with $\Delta x / H = \Delta z / H = 0.01$ was run at $\Ra = 10^5$, finding $\delta_{\vec{u}} \approx 0.56 \Ra^{-1/4}, \delta_b \approx 2.8 \Ra^{-2/7}$; these prefactors do indeed ensure that $\delta_{\vec{u}} < \delta_b$ for the Rayleigh number regime of the study. 

For each $\Ra$ we construct an orthogonal, rectangular grid such that the grid spacing is always smaller than the smallest of these length scales. 
This grid consists of, in the $z$-direction: a uniform grid with spacing $\Delta z^{(0)} = \Delta x_{\vec{u}}$ for $0 \leq \abs{z - z_{boundary}} \leq \delta_{\vec{u}}$; a uniform grid with spacing $\Delta z^{(1)}: \Delta x_{\vec{u}} < \Delta z^{(1)} < \Delta x_b$ for $\delta_{\vec{u}} < \abs{z - z_{boundary}} \leq \delta_b$; a nonuniform grid expanding linearly from $\Delta z^{(1)} \to \Delta z^{(2)}$ over the range $\delta_b < \abs{z - z_{boundary}} \leq 2\delta_b$; a uniform grid with spacing $\Delta z^{(2)} = 2 \eta$ for $2\delta_b < z < H - 2\delta_b$. 
In the horizontal direction, grid spacing is uniformly equal to $2 \eta$ throughout the domain.
Details of the grid for each simulation are given in table~\ref{tab:DNSdetails}.

In principle, we could directly check that the resolution is sufficient post-hoc by refining the grid and re-computing all of the statistics; if they do not change as the resolution increases, then the lower resolution ``fully resolves'' the flow. 
In practice, for this paper we note that the grid spacings of our simulations are comparable to those in similar DNS of 2D RBC \parencite[e.g.][]{ar:JohnstonDoering2009}. 
Details of the numerical method are given in section~\ref{sec:numericalMethod} as a special case of the multi-fluid solver.

\begin{table}[hbt!]
    \centering
    \begin{tabular}{ c || c | c | c | c | c | c }
     $\Ra$ & $T_\text{tot} / 4T_B$ & $\Delta t / 4T_B$ & $\Delta z_c / H = \Delta x / H$ & $\Delta z_w / H$ & $\eta / H$ & $\delta_{\vec{u}} / H$ \\ 
     \hline
     \hline
     $10^2$ & $25$ & $6.393\times10^{-5}$ & $0.04$ & $0.04$ & N/A & N/A \\ 
     $10^3$ & $25$ & $1.599\times10^{-3}$ & $0.04$ & $0.04$ & N/A & N/A \\ 
     $1.6\times10^3$ & $51$ & $6.393\times10^{-4}$ & $0.02$ & $0.02$ & N/A & N/A \\ 
     $1.7\times10^3$ & $51$ & $6.393\times10^{-4}$ & $0.02$ & $0.02$ & N/A & N/A \\ 
     $1.8\times10^3$ & $127$ & $6.393\times10^{-4}$ & $0.02$ & $0.02$ & N/A & N/A \\
     $2\times10^3$ & $38$ & $7.992\times10^{-4}$ & $0.02$ & $0.02$ & $1.410\times10^{-1}$ & $8.459\times10^{-2}$ \\  
     $10^4$ & $25$ & $1.598\times10^{-3}$ & $0.02$ & $0.01$ & $7.494\times10^{-2}$ & $5.656\times10^{-2}$ \\ 
     $5\times10^4$ & $25$ & $1.998\times10^{-3}$ & $0.02$ & $0.01$ & $4.240\times10^{-2}$ & $3.783\times10^{-2}$ \\ 
     $10^5$ & $38$ & $1.598\times10^{-3}$ & $0.02$ & $0.01$ & $3.346\times10^{-2}$ & $3.181\times10^{-2}$ \\ 
     $5\times10^5$ & $25$ & $9.990\times10^{-4}$ & $0.02$ & $7.067\times10^{-3}$ & $1.951\times10^{-2}$ & $2.127\times10^{-2}$ \\ 
     $10^6$ & $25$ & $9.990\times10^{-4}$ & $0.02$ & $5.963\times10^{-3}$ & $1.551\times10^{-2}$ & $1.789\times10^{-2}$ \\ 
     $5\times10^6$ & $38$ & $9.990\times10^{-4}$ & $1.797\times10^{-2}$ & $2.990\times10^{-3}$ & $9.151\times10^{-3}$ & $1.196\times10^{-2}$ \\ 
     $10^7$ & $60$ & $5.115\times10^{-4}$ & $1.454\times10^{-2}$ & $2.515\times10^{-3}$ & $7.300\times10^{-3}$ & $1.006\times10^{-2}$ \\ 
     $2\times10^7$ & $51$ & $3.996\times10^{-4}$ & $1.165\times10^{-2}$ & $2.114\times10^{-3}$ & $5.827\times10^{-3}$ & $8.459\times10^{-3}$ \\ 
     $10^8$ & $38$ & $3.197\times10^{-4}$ & $6.729\times10^{-4}$ & $1.130\times10^{-3}$ & $3.459\times10^{-3}$ & $5.657\times10^{-3}$ \\ 
     $10^9$ & $22\ (45)$ & $1.279\times10^{-4}$ & $4.543\times10^{-4}$ & $4.544\times10^{-4}$ & $1.645\times10^{-3}$ & $3.181\times10^{-3}$ \\ 
     $10^{10}$ & $20\ (76)$ & $7.992\times10^{-5}$ & $1.563\times10^{-3}$ & $1.789\times10^{-4}$ & $7.832\times10^{-4}$ & $1.789\times10^{-3}$ \\ 
     \hline
    \end{tabular}
    \caption{Details of grid spacing, time-step size, and simulation time for the 2D DNS of RBC (section~\ref{sec:DNS}). Times are nondimensionalized by the (approximate) eddy turnover time, $T_e \approx 4 T_B = 4\sqrt{H / \Delta B}$. The final two columns give the physical length scales used to estimate the required resolution, the (bulk) Kolmogorov dissipation length $\eta / H$ (equation~\eqref{eq:kolmogorovLength}) and the kinetic boundary layer thickness, $\delta_{\vec{u}} / H \approx 0.56\Ra^{-1/4}$. The $\Ra = 10^9$ and $10^{10}$ simulations were spun up on a coarser grid (the $\Ra = 10^8$ grid), then after reaching equilibrium the grid was refined. The simulation time on the finer grid is given, followed by, in parentheses, the total simulation time on both grids for that Rayleigh number.}\label{tab:DNSdetails}
\end{table}

\subsubsection{Calculation of \texorpdfstring{$\Nu$, $\Re$, $\delta_b$}{Nu, Re, deltab}}\label{sec:calculationOfNuReBL}
The Nusselt number, Reynolds number, and boundary layer depths are calculated as follows:
\begin{itemize}
    \item[$\Nu$:] The most direct way of calculating $\Nu$ is to integrate the (dimensionless) heat flux over the entire domain, then take a time average: $\Nu = \langle wb - \pdv*{b}{z} \rangle_{V,t}$. 
    However, if the flow is statistically stationary, then the time-averaged \textit{horizontally} averaged (dimensionless) heat flux is independent of height, so calculating the time-averaged vertical buoyancy gradient averaged over the top and bottom boundaries gives a second estimate, $\Nu_\text{w} \coloneqq \langle -\pdv*{b}{z} \rangle_{A,t;z=0,H}$. 
    The equivalence of these two expressions for $\Nu$ provides an extra check for the statistical steadiness of the numerical solutions. 
    Another check for statistical stationarity is provided via the kinetic and thermal dissipation rates (calculated using equations~\eqref{eq:SFboussinesq_dissipationEnergy}-\eqref{eq:SFboussinesq_dissipationBuoyancy}).
    Thus for a statistically stationary state, convergence of $\Nu = \Nu_\text{w} = \varepsilon_{b} = 1 + \flatfrac{\varepsilon_{\vec{u}}}{\Ra}$ is required.
    
    \item[$\Re$:] The calculation of a Reynolds number based on the definition $\Re \coloneqq \flatfrac{U L}{\nu}$ requires the choice of a velocity scale and a length scale. 
    For RBC, the only length scale we can reasonably choose for a bulk Reynolds number must be the domain height $H$, as this is the only external length scale in the problem. 
    However, what is a reasonable representative velocity scale, $U$? Several possible choices are suggested in \textcite{ar:Kerr1996,ar:AhlersEtAl2009}; we shall consider velocity scales based on the turning points of the velocity variance profile:
    \begin{align}
        U_1 \coloneqq \sqrt{\text{mean}(\text{var}(u)_{x,t})};
        \qquad
        U_2 \coloneqq \sqrt{\max(\text{var}(u)_{x,t})};
        \qquad
        U_3 \coloneqq \sqrt{\max(\text{var}(w)_{x,t})}
    \end{align}
    An a priori estimate of $\Re$ can be found by assuming free-convective scaling, $U = U_B \coloneqq \sqrt{\Delta B\ H}$, implying $\Re = \sqrt{\flatfrac{\Delta B\ H^3}{\nu^2}} = \Ra^{1/2}\Pr^{-1/2}$.
    
    \item[$\delta_b$:] If the flow is statistically stationary, the buoyancy will be well-mixed in the interior of the domain, the time-averaged buoyancy profile must be approximately constant outside of the boundary layers, and approximately linear within due to the fixed buoyancy boundary conditions.
    Thus one measure of the thermal boundary layer thickness is
    \begin{equation}
        \delta_b^{(1)}
        \coloneqq
      - \frac{\Delta B}{2 \dv{\left<b\right>_{x,t}}{z}|_\text{wall}}.
    \end{equation}
    Following \textcite{ar:Kerr1996}, we also estimate the thermal boundary layer thickness from the locations of the maxima of the buoyancy variance profile:
    \begin{equation}
        \delta_b^{(2)}
        \coloneqq
        \abs{z(\max(\text{var}(b)_{x,t})) - z(\text{wall})}.
    \end{equation}
    Both the upper and lower boundary layer thicknesses should be the same.
\end{itemize}
The above time averages are calculated over at least 5 eddy turnover times ($T_e \approx 4\ T_B$). 
Time-averages are also calculated over twice and three times this minimum averaging time, and all simulations show convergence between the averages taken over these three different times. 
The total simulation time for each Rayleigh number is given in table~\ref{tab:DNSdetails}.

\subsection{The relevance of RBC to atmospheric flows}
While RBC is a valuable test problem in its own right, it is worth considering similarities with and differences from atmospheric flows, in particular the dry atmospheric convective boundary layer (CBL).

Besides the complexities of moisture, the dry RBC problem differs from even dry atmospheric convection in a few important ways. 
Firstly, the Boussinesq approximation is of questionable validity even on the scale of the atmospheric boundary layer; in practice  however, it has long been used in the LES community with excellent results \parencite[e.g. ]{ar:SullivanEtAl2011}. 
Furthermore, the Boussinesq form has been used to facilitate analysis; experiments using a non-Boussinesq (fully compressible) version of the same code show little qualitative or quantitative differences from their Boussinesq counterparts. 

Secondly, the lower boundary in the CBL is neither smooth, nor uniformly heated. 
Recent results show that neither nonuniform heating \parencite{ar:BakhuisEtAl2018} nor rough boundaries \parencite{ar:ZhuEtAl2019,ar:ToppaladoddiEtAl2021} drastically change the dynamics of RBC, though the latter does tend to increase the heat flux towards the so-called ``ultimate regime'', equivalent to the free-convective regime which dominates discussion of scaling in the atmospheric convective boundary layer. 

Thirdly, the fixed buoyancy boundary conditions are quite different to CBL conditions, where the lower boundary is closer to (and is often modelled as) a fixed buoyancy flux, and there is no fixed upper boundary for the convection (instead there is a stable atmospheric layer). 
In practice, solutions of RBC with fixed flux vs. fixed value boundary conditions are similar, especially in 2D \parencite{ar:VerziccoSreenivasan2008,ar:JohnstonDoering2009} (as are LES simulations of the CBL). 
It is thus only the upper boundary that introduces a major difference between RBC and the CBL. 
Even in that case there has been recent progress on studying modified Rayleigh-B\'{e}nard convection with the compensating heat flux provided by radiation in a layer of finite thickness \parencite{ar:LepotEtAl2018,ar:Doering2019}, which the first authors note ``spontaneously achieves the ``ultimate'' regime of thermal convection''. 

We thus consider the classical Rayleigh-B\'{e}nard problem to be sufficiently close to atmospheric convection to provide a reasonable testbed for investigating the behaviour of a multi-fluid model of turbulent convection. 
There remains the question of the applicable parameter regime, discussed in the next section.

\subsubsection{An analogy between constant-viscosity RBC and large-eddy simulation of higher-Ra RBC.}\label{sec:LESanalogy}
Atmospheric flows generally involve very high Reynolds number; for example, the CBL might have depth $\approx 1000~\textrm{m}$, and (even without a mean wind) velocities in convective updraughts $\approx 1~\textrm{m}~\textrm{s}^{-1}$. 
With kinematic viscosity of air $\approx 10^{-5} ~\textrm{m}^2~\textrm{s}^{-1}$ we have $\Re \approx 10^8$. 
In the context of RBC, this would lead to $\Ra\approx10^{16}$, i.e. much larger than in our simulations.
Given the above considerations of resolution, a DNS of this problem is computationally impossible in 3D with current computing power and would be a challenge even in 2D.

The atmospheric science community address this problem using `Large-Eddy Simulation' (LES) as reviewed by \textcite{ar:Mason1994}.
LES is based upon spatially low-pass filtering the equations of motion with a filter with characteristic scale $\Delta_f$ chosen such that the unfiltered flow remains well within the self-similar Kolmogorov `inertial sub-range' (ISR) of scales. 
In this case, the sub-filter contribution to turbulent fluxes is small and can be represented by a simple eddy-viscosity.
In practice, the eddy-viscosity proposed by \textcite{ar:Smagorinsky1963} has been found to give good results provided the simulation actually is well within the ISR.

In fact, \textcite{ar:Mason1994} points out that acceptable results are obtained from a simulation of the CBL in which a constant viscosity is used at each level based upon the horizontal average of the Smagorinsky value.
One might go further and suggest that the height-dependence is required primarily close to the surface and boundary-layer top where eddy length-scales are restricted.
In this case, a simple view of LES is as follows.
With a well-developed ISR the flow is essentially independent of $\Re$. 
For sufficiently large $\Re$ we can choose an artificial larger viscosity such that the range of scales in the flow is smaller as the Kolmogorov microscale (i.e. the eddy scale at which $\Re=1$) is larger.
The turbulent kinetic energy dissipation rate ($\varepsilon$) remains the same as does the flow at larger scales. 
Essentially the same argument applies to use of wind-tunnels with scale models.

Smagorinsky provides us with a method to estimate this artificially large viscosity, but let us take a more basic view. 
In the ISR the energy spectrum $E(k)=K_0 \varepsilon^\frac{2}{3}k^{-\frac{5}{3}}$, with $K_0$ a constant and $k$ the wave number.
Suppose we choose a filter scale with wave number $k_f$, then the `turbulent' kinetic energy (TKE) in the subfilter flow has a velocity scale $U_f$ given by
\begin{equation}
U_f^2 = \int_{k_f}^\infty E(k)\,\mathrm{d}k = \frac{2}{3}K_0 \varepsilon^\frac{2}{3}k_f^{-\frac{2}{3}} 
\label{eq:U_f}
\end{equation}
Prandtl argues, by analogy with the kinetic theory of gases, that the eddy diffusivity is a product of the turbulent velocity scale and the `mean free path' or `mixing length'. 
It is possible to show this more rigorously using the dynamical equation for stress.
We assume that the mixing length scales with $k_f^{-1}$ and hence the eddy viscosity is given by  $\nu_f\propto U_f k_f^{-1}$.
This assumption is precisely the same as stating that the eddy Reynolds number, $\Re_f\equiv U_f k_f^{-1}/\nu_f\approx 1$, i.e. the filter scale is proportional to the Kolmogorov microscale of the filtered flow, $\eta_f \propto k_f^{-1}$.
Indeed, if we absorb the constants in eq. \eqref{eq:U_f} into $\eta_f$, the Kolmogorov microscale for the filter, then $U_f=\varepsilon^\frac{1}{3}\eta_f^\frac{1}{3}$,  $\Re_f\equiv U_f \eta_f/\nu_f = 1$ and all of the Kolmogorov scales apply.

Note that with this viscosity the kinematic deviatoric stress is given by $\vec{\tau}=U_f\eta_f \left(\grad{\vec{u}}+\grad{\vec{u}}\trans\right)$. 
The spatially filtered equation for the TKE, in steady state and ignoring the transport term (both assumptions being appropriate for the homogeneous isotropic turbulence the ISR is considered to represent) leads to a simple balance between shear production and dissipation:
\begin{equation}
\vec{\tau}  \contract   \left(\grad{\vec{u}}+\grad{\vec{u}}\trans\right)  = 2\varepsilon
\end{equation}
The TKE is given by $\frac{1}{2}U_f^2$.
The scaling above gives $\varepsilon=U_f^3/\eta_f$ (so the timescale for dissipation is $\eta_f/(2U_f)$) then this balance becomes:
\begin{equation}
U_f\eta_f \left(\grad{\vec{u}}+\grad{\vec{u}}\trans\right)  \contract   \left(\grad{\vec{u}}+\grad{\vec{u}}\trans\right)  = 2\frac{U_f^3}{\eta_f} 
\end{equation}
from which $U_f = \eta_f\left[ \frac{1}{2}\left(\grad{\vec{u}}+\grad{\vec{u}}\trans\right)  \contract   \left(\grad{\vec{u}}+\grad{\vec{u}}\trans\right) \right]^\frac{1}{2}$, leading to the \textcite{ar:Smagorinsky1963} formulation of viscosity.

To give a simple example, suppose we have a convective boundary layer with $H$ the depth of the layer, say 1000 m and convective velocity scale $U=2~\mathrm{m~s}^{-1}$, corresponding to $\Delta B \approx 4\times10^{-3}~\mathrm{m~s^{-2}}$.
Then $\Re = UH/\nu = 1.33 \times 10^8$ and $\Ra=1.8\times 10^{16}$.
The `outer' mixing length is often taken to be $L=0.15H$. 
Then a crude estimate of $\varepsilon$ is $\varepsilon = U^3/L = 8/150~\mathrm{m^2~s^{-3}}=5.3\times 10^{-2}~\mathrm{m^2~s^{-3}}$. 
The Kolmogorov microscale is thus $\eta = (\nu^3/\varepsilon)^\frac{1}{4} \approx 0.5$ mm.
If we choose a filter scale such that $\eta_f=1$ m, then $U_f=0.376~\mathrm{m~s}^{-1}$, the eddy viscosity is 0.376 m$^2$ s$^{-1}$ and the Reynolds number of the whole flow is reduced to $\Re \approx 5300$.
This should still be turbulent and is likely to be within the Reynolds number independent regime. 
In fact, the convective boundary layer is very amenable to LES because the large coherent structures with scales of order $H$ dominate, and  \textcite{ar:SullivanEtAl2011} show that ``the majority of the low-order moment statistics (means, variances, and fluxes) become grid independent when the ratio $z_i/(Cs \Delta_f)> 310$''.
Here $z_i$ is the inversion depth (i.e. $H$), $Cs$ is the Smagorinsky constant ($\approx 0.2$), and $\Delta_f$ essentially the grid length (so the actual filter scale is a multiple of this). 
This implies $\Delta_f<z_i/(310Cs)\approx16$ m in this case.
Hence our notional 1 m resolution should be very well-converged LES.

Thus, provided the solution remains in the $\Re$-independent turbulent regime, the relatively low $\Ra$ runs ($\Ra\geq 10^8$) may be interpreted as reasonable approximations to LES of much higher $\Ra$ (and hence $\Re$) flows encountered in the atmosphere. 
Indeed, a similar argument is made by \textcite{ar:MelladoEtAl2018} in a study of Stratocumulus convection, except that they seem to have reversed the semantics of the conclusion by describing their artificially large viscosity runs as DNS. 
This `DNS in a Re-independent regime' is, we would argue, more correctly described as a form of LES as its basis is precisely the same.

A slight note of caution may arise from consideration of the boundary conditions, as the turbulence length scale collapses as one approaches the boundary and buoyancy effects on turbulence become more dominant. 
(The same concerns apply to LES).
With a fixed heat-flux boundary condition, the concern is less as the surface exchange serves merely to transport the given surface flux into the fluid where large eddies can start to transport it. 
In practice, our results are similar for fixed temperature and fixed heat flux boundary conditions, suggesting that, so long as the thermal boundary-layer is adequately resolved the solutions remain applicable to higher $\Re$.

\subsection{Phenomenology of RBC}
\begin{figure}[t!hb]
    \centering
    \begin{subfigure}{0.49\textwidth}
        \caption{$\Ra \simeq 10^5$}
        \includegraphics[trim = 0 0 0 25, clip, width=\textwidth]{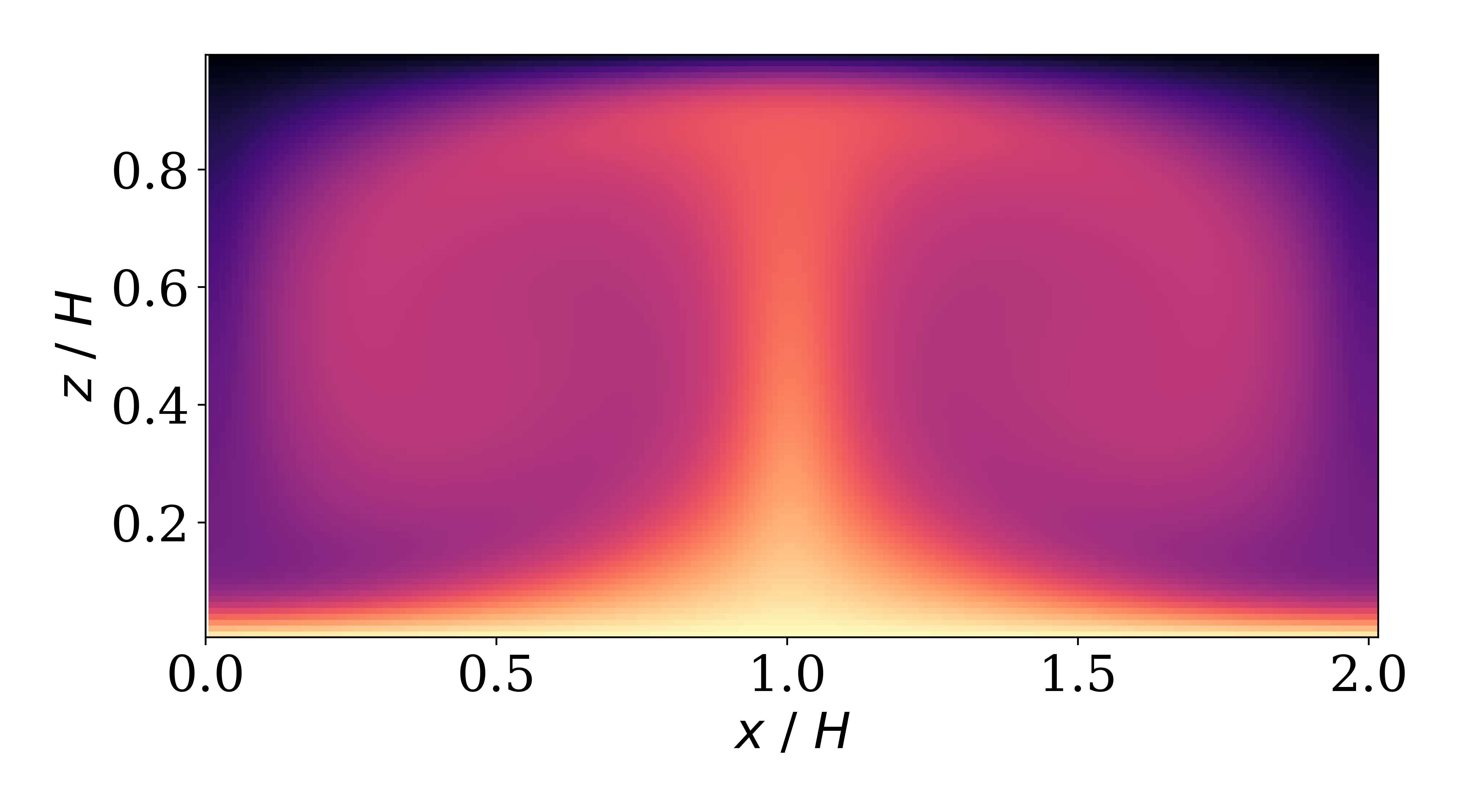}
    \end{subfigure}
    \begin{subfigure}{0.49\textwidth}
        \caption{$\Ra \simeq 10^8$}
        \includegraphics[trim = 0 0 0 25, clip, width=\textwidth]{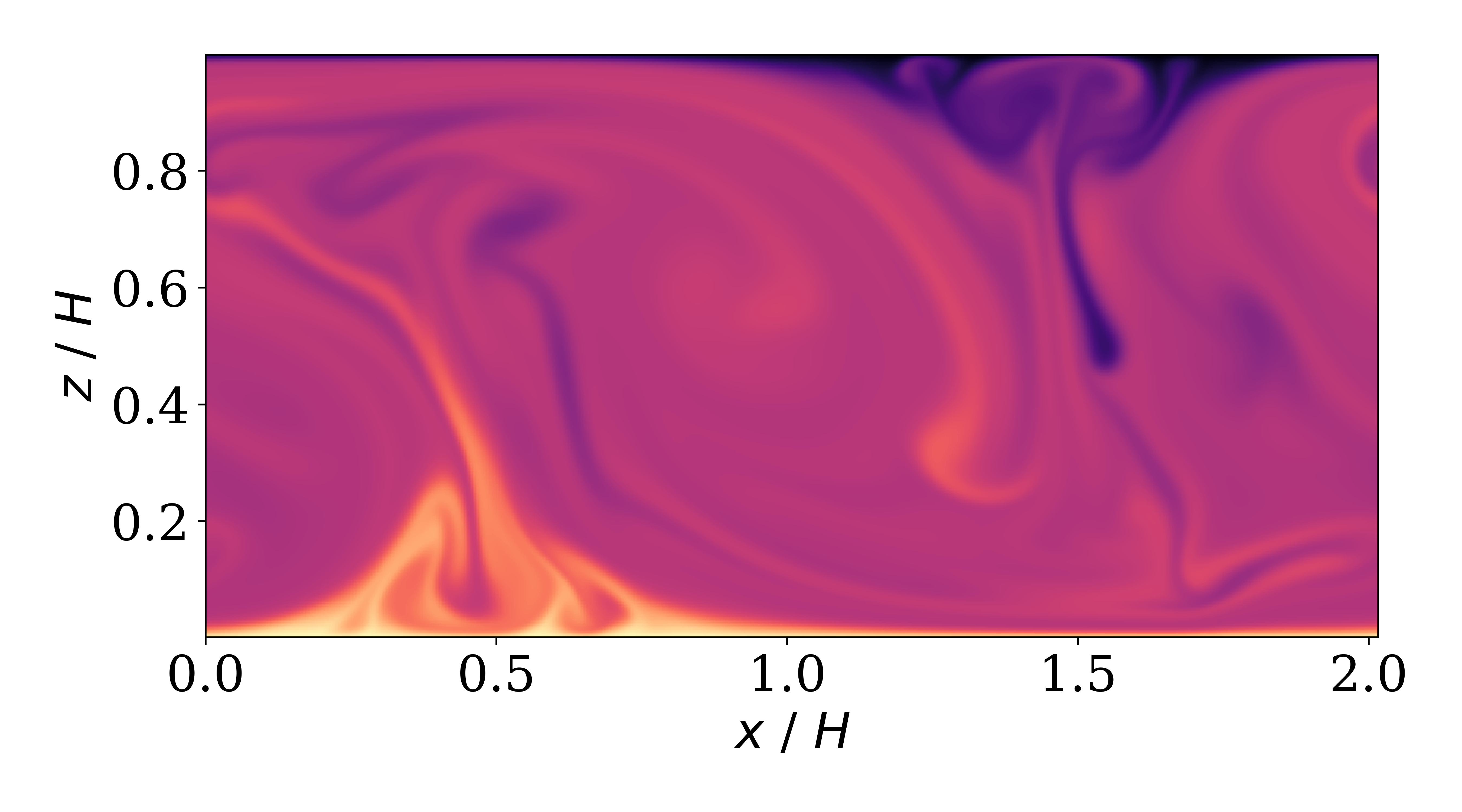}
    \end{subfigure}
    \begin{subfigure}{0.49\textwidth}
        \caption{$\Ra \simeq 10^{10}$}
        \includegraphics[trim = 0 0 0 25, clip, width=\textwidth]{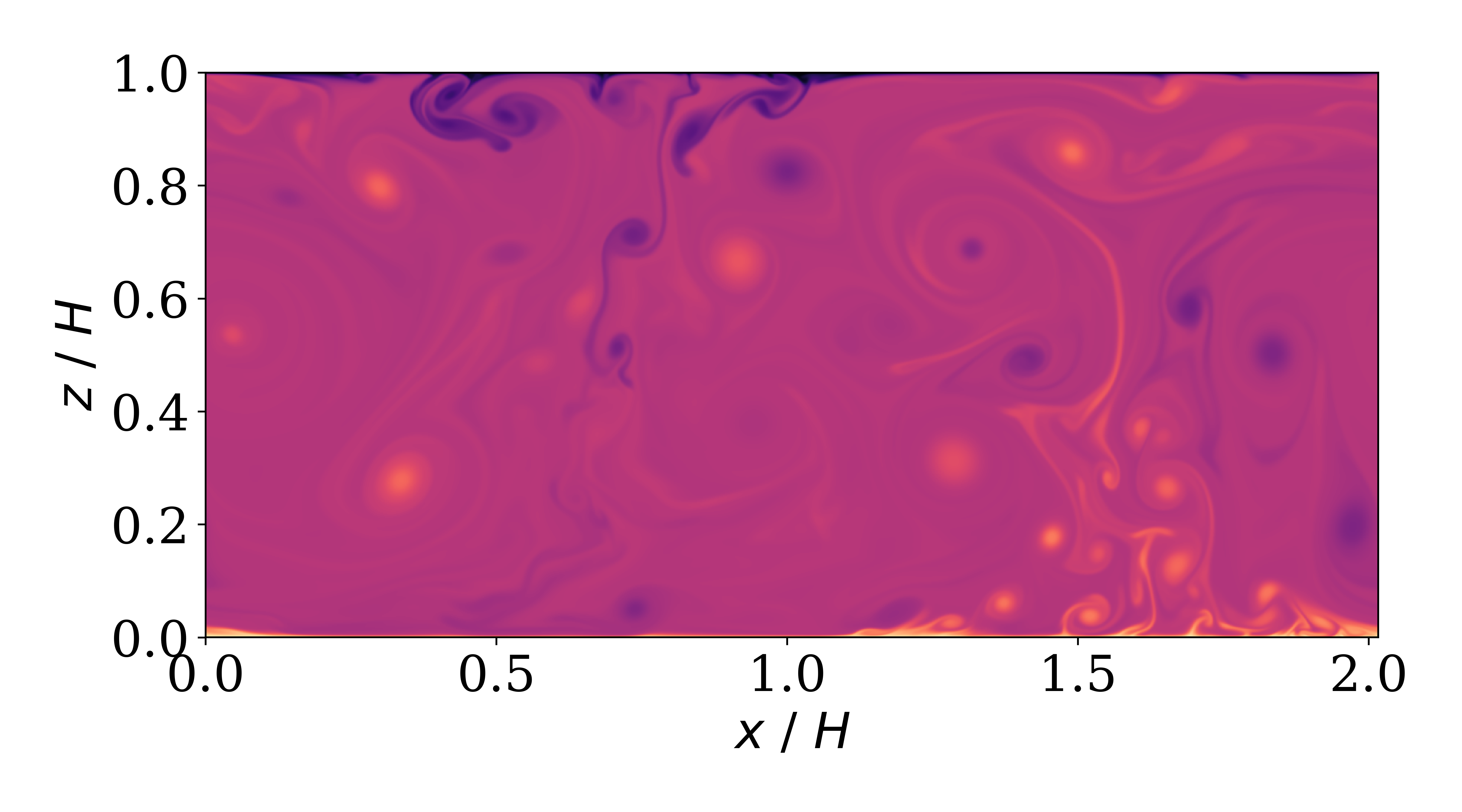}
    \end{subfigure}
    \begin{subfigure}{0.49\textwidth}
        \caption{$\Ra \simeq 10^{10}$ (zoom)}
        \includegraphics[trim = 0 0 0 25, clip, width=\textwidth]{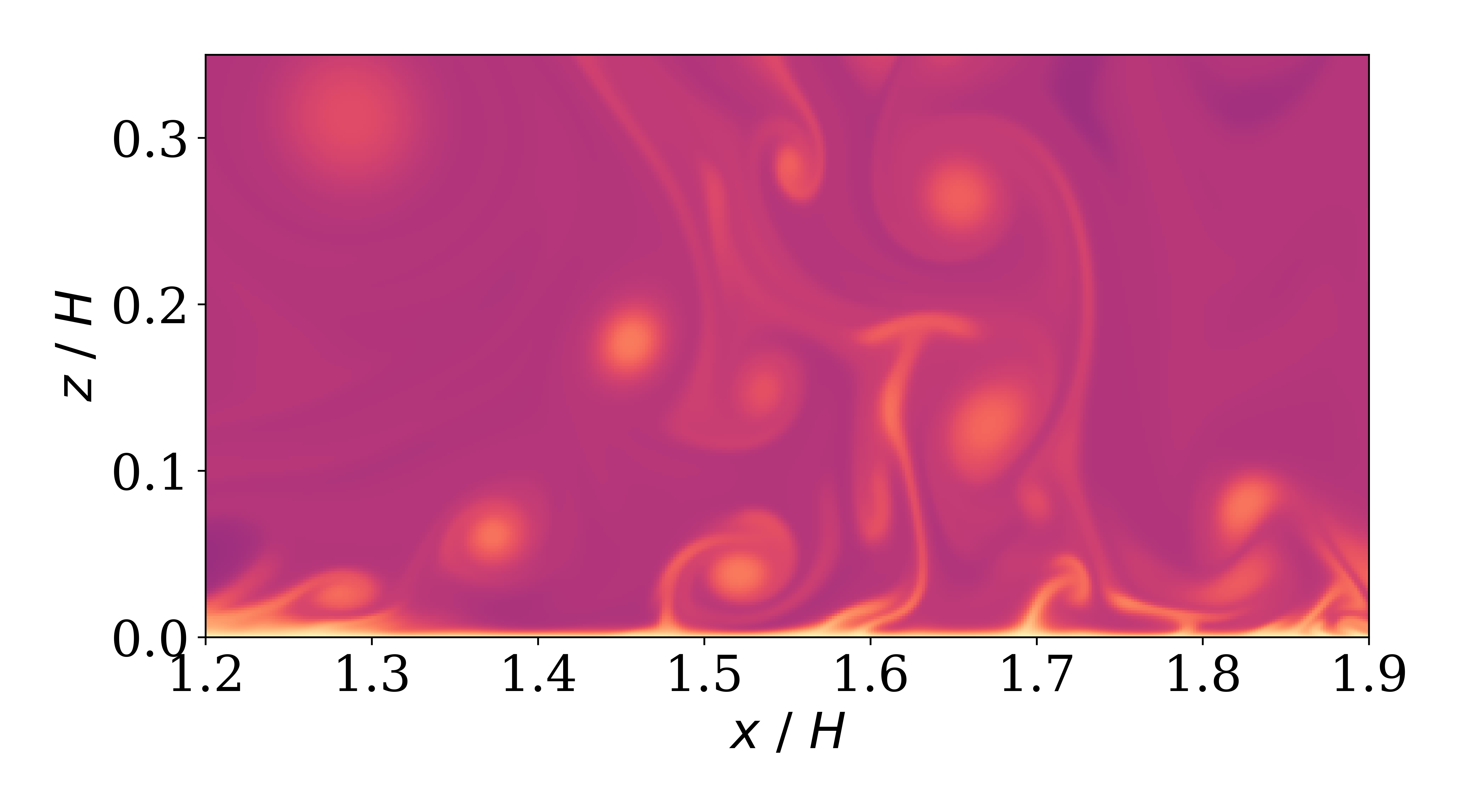}
    \end{subfigure}
    \caption{Snapshots of buoyancy fields in 2D Rayleigh-B\'{e}nard convection at varying Rayleigh number. In (a), the flow is convective but steady; in (b), the flow is turbulent, but only just, with $\Re \simeq 5000$; in (c), the flow is highly turbulent and exhibits many small scale features; (d) is the same flow as (c) but zoomed in to show small-scale features close to the lower boundary layer, and also to demonstrate the resolution. }\label{fig:DNS_RBCexamples}
\end{figure}
\begin{figure}[t!hb]
    \centering
    \begin{tabular}[c]{cc}
    \begin{subfigure}[c]{0.48\textwidth}
        \caption{}
        \includegraphics[trim = 0 0 0 10, clip, width=\textwidth]{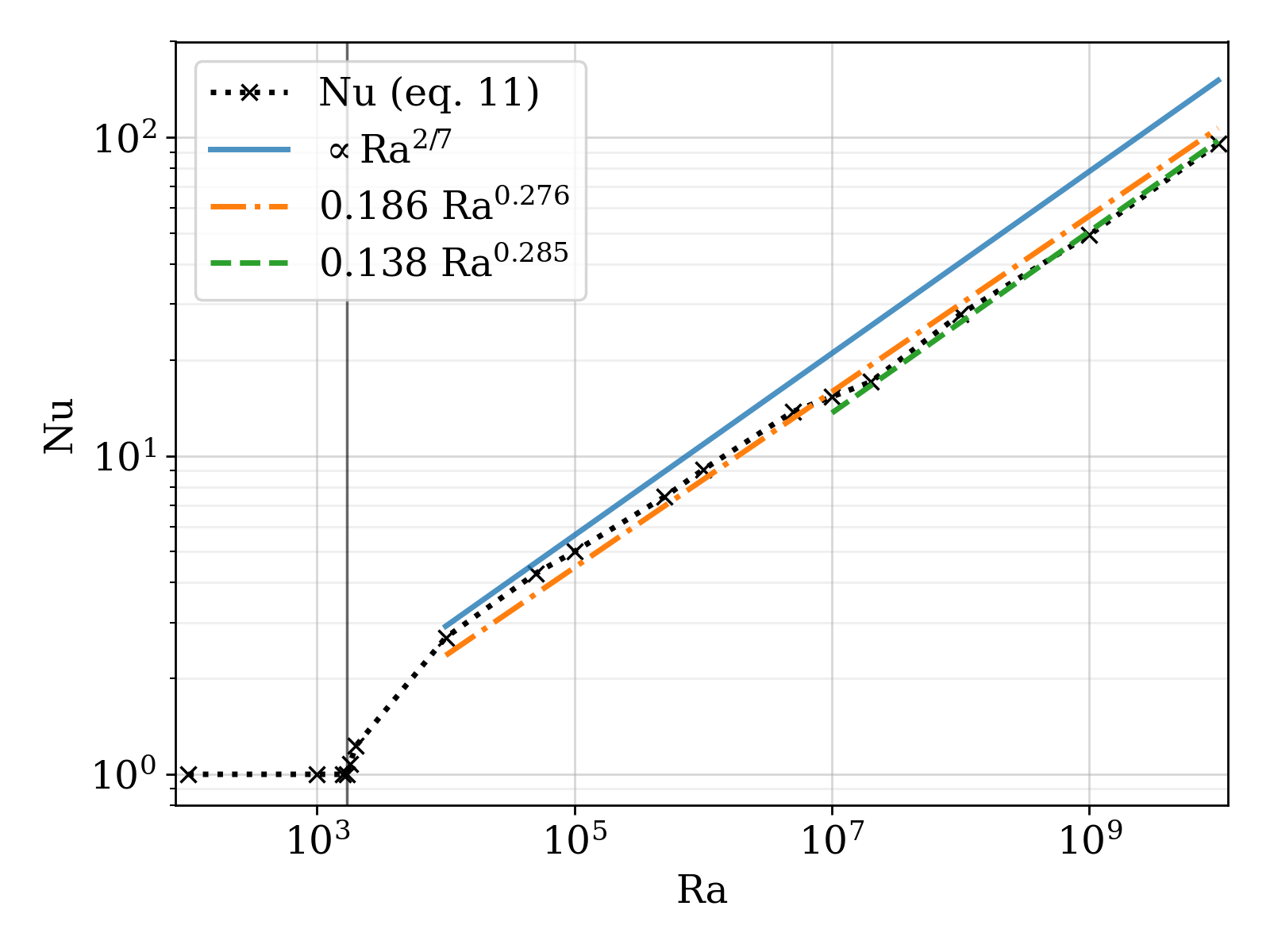}
    \end{subfigure}&
    \begin{subfigure}[c]{0.48\textwidth}
        \caption{}
        \includegraphics[trim = 0 0 0 10, clip, width=\textwidth]{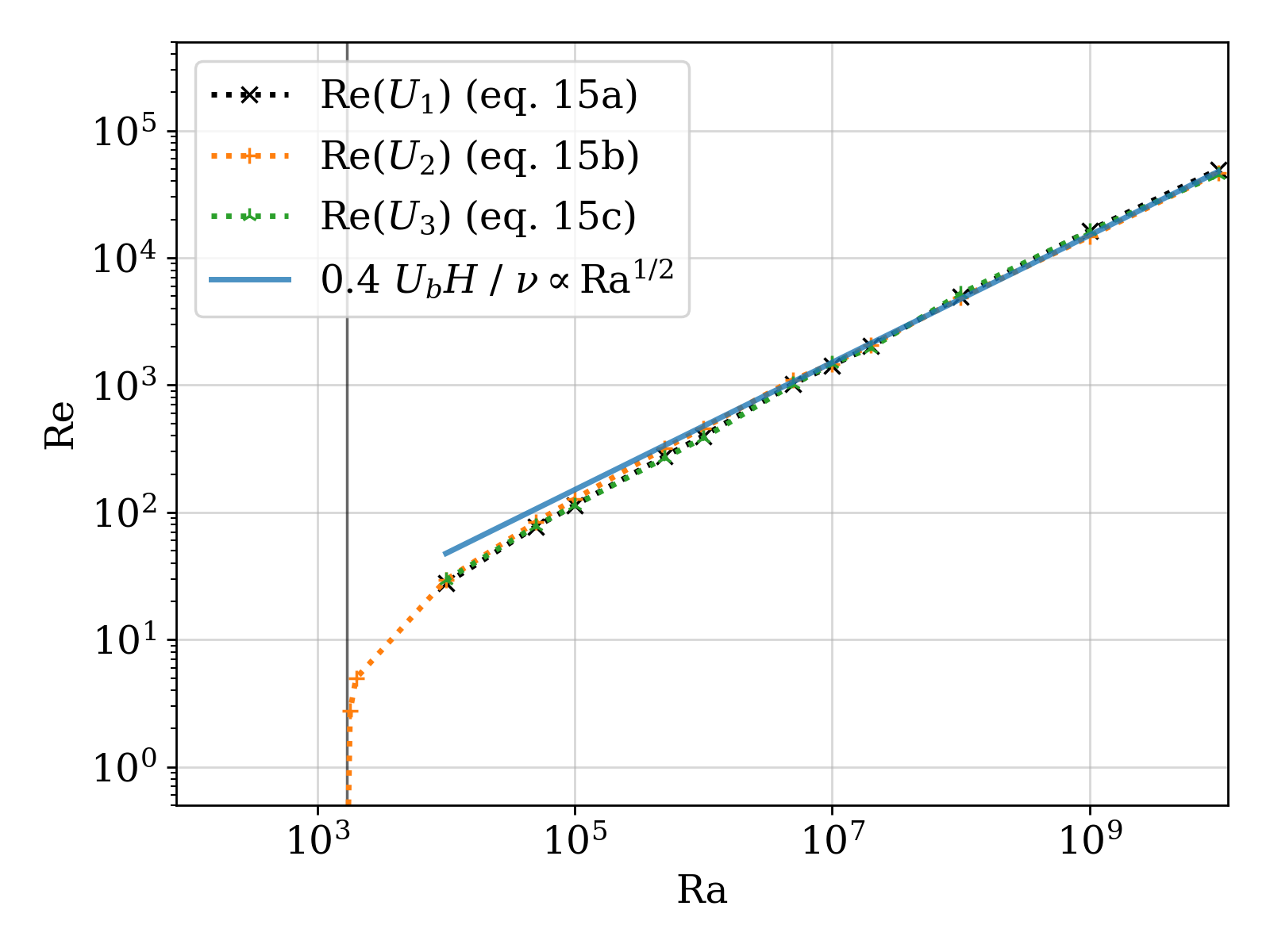}
    \end{subfigure}\\
    \begin{subfigure}[c]{0.48\textwidth}
        \caption{}
        \includegraphics[trim = 0 0 0 10, clip, width=\textwidth]{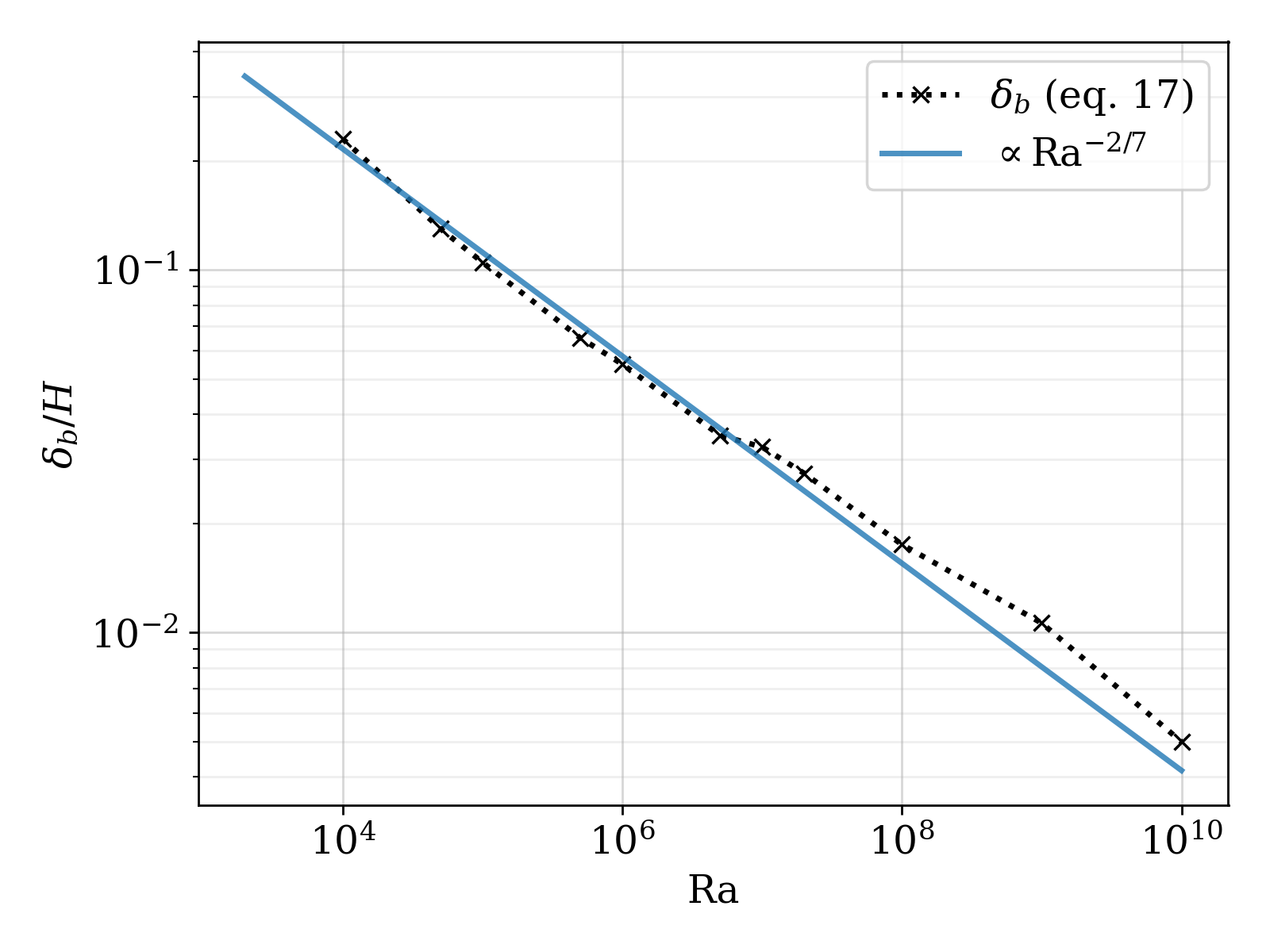}
    \end{subfigure}&
    \begin{subfigure}[c]{0.48\textwidth}
    \end{subfigure}\\
    \end{tabular}
    \caption{Validation of 2D Rayleigh-B\'{e}nard direct numerical simulations, showing scaling with applied buoyancy forcing, $\Ra$, of: (a) heat transport ($\Nu$); (b) momentum transport ($\Re$); and (c) thermal boundary layer thickness ($\delta_b$). 
    In (a)-(c), the black crosses joined by a dotted line denote our main results.
    In (a) and (b), the solid black vertical line marks the theoretical critical Rayleigh number, $\Ra_\text{c} \simeq 1708$.
    In (a) the solid blue line follows the theoretical $\Nu \propto \Ra^{2/7}$ scaling; the orange dash-dotted line follows the best fit line of \textcite{ar:Kerr1996}, $\Nu = 0.186 \Ra^{0.276}$ (3D); the green dashed line follows the best fit of \textcite{ar:JohnstonDoering2009}, $\Nu = 0.138 \Ra^{0.285}$ (2D), valid above $\Ra \simeq 10^7$. 
    In (b) the three dotted lines show Reynolds numbers calculated from the alternative definitions in equation~(15); the theoretical scaling, $\Re \propto \Ra^{1/2}$, is shown as a solid blue line. 
    In c), the solid blue line shows the theoretical scaling, $\delta_b \propto \Ra^{-2/7}$.}\label{fig:DNS_scalingVerification}
\end{figure}
Direct numerical simulations of 2D, dry, Boussinesq Rayleigh-B\'{e}nard convection were performed for the range $10^2 \leq \Ra \leq 10^{10}$ for a fluid with Prandtl number $0.707$ (the value for dry air at standard temperature and pressure). 
For each $\Ra$, the fluid was initialized from the hydrostatically-balanced resting state \eqref{eq:SFinitialConditions_dimless}, with small random perturbations to the buoyancy field $\abs{\delta b_\text{pert}} \leq 0.01 \Delta B$ drawn from a uniform distribution. 
The aspect ratio of the domain was set equal to the critical wavelength: $\Gamma = L_x/L_z = \lambda_c / H \approx 2.02$ (for $10^2 \leq \Ra \leq 10^8$, simulations were also run with $\Gamma = 10$, which gave the same results for $\Nu$, $\Re$ etc.; therefore only the smaller aspect ratio results are reported).
Each simulation was run until a statistically-steady equilibrium was reached, determined by the convergence of the time-mean values of $\Nu, \Re, \delta_b$, and the equivalence of the four methods of estimating $\Nu$.

Since $\Nu \approx \Nu_\text{w} \approx \varepsilon_b \approx 1 + \flatfrac{\varepsilon_{\vec{u}}}{\Ra}$ for all simulations (not shown), verifying statistical steadiness, only $\Nu$ is discussed hereafter. 
All three methods of estimating the Reynolds number also produce very similar results (fig.~\ref{fig:DNS_scalingVerification}b), and the free-convective scaling (with proportionality factor $\approx 0.4$) gives good agreement with the observed scaling, especially for $\Ra \gtrsim 10^6$.

Figure~\ref{fig:DNS_RBCexamples} shows single-time snapshots of the 2D buoyancy field in fully developed RBC at various Rayleigh numbers. 
The solutions show several characteristic regimes. 
For $\Ra < \Ra_\text{c}$, diffusion damps out any motion and the solution is entirely diffusive (not shown). 
As $\Ra$ increases above $\Ra_\text{crit}$ the solution exhibits first steady convection (a), then transitional turbulence (b), and finally fully-developed convective turbulence (c-d). 
This broad phenomenology is valid in both 2D and 3D, so for the remainder of the paper we restrict to 2D. 
Reproducing this phenomenology serves both to demonstrate the usefulness of RBC as a model of convection, and to validate the chosen numerical method.

The scalings of $\Nu, \Re,$ and $\delta_b$ are shown in Fig.~\ref{fig:DNS_scalingVerification}, along with snapshots of buoyancy fields from representative simulations in each phenomenological regime in figure~\ref{fig:DNS_RBCexamples}. 
A transition from diffusive to convective behaviour is observed in both the Nusselt (figure~\ref{fig:DNS_scalingVerification}a) and Reynolds (figure~\ref{fig:DNS_scalingVerification}b) numbers at $\Ra \simeq 1700$, in agreement with the prediction $\Ra_{c} = 1708$. 
A transition to turbulence follows between $10^7 \lesssim \Ra \lesssim 10^8$, as expected given that $\Re \approx 2000$ for $\Ra \approx 2\times10^7$. 
This can be seen in the qualitative nature of the flow: figure~\ref{fig:DNS_RBCexamples}a is steady, representative of all flows with $\Ra_{c} \lesssim \Ra \lesssim 10^7$; while above $\Ra \gtrsim 2\times 10^7$ the flow is intermittent and exhibits patterns on multiple scales, characteristic of turbulence, as seen in Figs.~\ref{fig:DNS_RBCexamples}b-d.

The Nusselt number obeys a power law close to $\Ra^{2/7}$, and the Reynolds number a power law close to $\Ra^{1/2}$, from shortly after the onset of convection to the highest Rayleigh number considered. 
These are the expected exponents within this parameter regime \parencite{ar:AhlersEtAl2009,ar:ChillaSchumacher2012}. 
The three different possibilities for the velocity scale in the Reynolds number calculation give similar results. 
A reduction in the prefactor of the power law for $\Nu$ is observed between $10^7 < \Ra < 10^8$, which coincides with the onset of turbulence. 
A similar transition is seen in the results of \textcite{ar:JohnstonDoering2009} for finite-difference DNS of 2D dry RBC with $\Pr = 1$. 
Above $\Ra \simeq 10^7$ they observe a power law relationship between $\Ra$ and $\Nu$ of $\Nu = 0.138 \Ra^{0.285}$, which our data are in excellent agreement with.

Any two-fluid parametrization of RBC should therefore aim to capture the described scaling behaviour of $\Nu$ and $\Re$ with $\Ra$.
 
\section{Multi-fluid equation set and closure choices}\label{sec:multiFluidDerivation}
As a first step towards building a multi-fluid parametrization of convective turbulence, we motivate and present a two-fluid single-column model of Rayleigh-B\'{e}nard convection. 
The full viscous multi-fluid Boussinesq equation set is \parencite{ar:ShipleyEtAl2021b_inPrep}:
\begin{align}
    \pdv{\sigma_i}{t}
  + \div(\sigma_i \vec{u}_i)
    &=
    \overline{\mathcal{S}_i^+}
  - \overline{\mathcal{S}_i^-},
    \label{eq:MFBoussinesq_sigma_noApprox}
    \\
\begin{split}
    \pdv{\sigma_i \vec{u}_i}{t}
  + \div(\sigma_i \vec{u}_i \otimes \vec{u}_i)
    &=
    \sigma_i b_i \unit{k}
  - \sigma_i \grad{\overline{P}}
  - \grad(\sigma_i p_i)
  - \left[
        \overline{P} \grad{\sigma_i}
      - \overline{P \grad{I_i}}
    \right]
    \\
    &\quad
  + \nu \laplacian{\sigma_i \vec{u}_i}
  - \nu \div{\overline{(\vec{u} \otimes \grad{I_i})\trans}}
  - \nu \overline{\grad{I_i} \vdot (\grad{\vec{u}})\trans}
    \\
    &\quad
  + \overline{\vec{u} \mathcal{S}_i^+}
  - \overline{\vec{u} \mathcal{S}_i^-}
  - \div(
        \overline{I_i \vec{u} \otimes \vec{u}}
      - \sigma_i \vec{u}_i \otimes \vec{u}_i
    ),
    \label{eq:MFBoussinesq_momentum_noApprox}
\end{split}
    \\
\begin{split}
    \pdv{\sigma_i b_i}{t}
  + \div(\sigma_i \vec{u}_i b_i)
    &=
    \kappa \laplacian{\sigma_i b_i}
  - \kappa \overline{\grad{I_i} \vdot \grad{b}}
  - \kappa \div{\overline{b \grad{I_i}}}
    \\
    &\quad
  + \overline{b \mathcal{S}_i^+}
  - \overline{b \mathcal{S}_i^-}
  - \div(
        \overline{I_i \vec{u} b}
      - \sigma_i \vec{u}_i b_i
    ),
    \label{eq:MFBoussinesq_buoyancy_noApprox}
\end{split}
    \\
    \sum_i \div(\sigma_i \vec{u}_i)
    &=
    0.
    \label{eq:MFBoussinesq_mass_sumConstraint}
    \\
    \sum_i \sigma_i
    &=
    1
    \label{eq:MFBoussinesq_sigma_sumConstraint}.
\end{align}
Here an overbar denotes a spatial filter \parencite{ar:Germano1992}; $i \in \{0,1,\dots,n\}$ indexes the fluid partitions; $I_i$ is an indicator function for fluid $i$; $\sigma_i \coloneqq \overline{I_i}$ is the fraction of fluid $i$ contained within a characteristic filter volume; $\vec{u}_i \coloneqq \flatfrac{\overline{I_i \vec{u}}}{\sigma_i}$ and $b_i \coloneqq \flatfrac{\overline{I_i b}}{\sigma_i}$ are the velocity and buoyancy fields of fluid $i$; $p_i \coloneqq \flatfrac{\overline{I_i P}}{\sigma_i} - \overline{P}$ is the difference between the conditionally-filtered pressure in fluid $i$ and the unconditionally filtered pressure $\overline{P}$; $\overline{\mathcal{S}_i^{\pm}}, \overline{\vec{u}\mathcal{S}_i^{\pm}}, \overline{b\mathcal{S}_i^{\pm}}$ are respectively sources and sinks of fluid fraction, momentum, and buoyancy in fluid $i$ arising from the relabelling of fluid. 
The unconditionally-filtered pressure, $\overline{P}$, ensures the incompressibility of the mean flow, equation~\eqref{eq:MFBoussinesq_mass_sumConstraint}.

Equations~\eqref{eq:MFBoussinesq_sigma_noApprox}-\eqref{eq:MFBoussinesq_sigma_sumConstraint} are derived by conditionally spatially filtering the Boussinesq equations~\eqref{eq:SFboussinesq_momentum}-\eqref{eq:SFboussinesq_mass} in the manner set out by \textcite{ar:ThuburnEtAl2018}; however, here viscous terms and sources and sinks of fluid fraction are retained from the outset. 
The only terms neglected here are those arising from possible non-commutation of the spatial filter with the partial derivatives. 
For a full derivation and discussion of the terms requiring closure, see \textcite{ar:ShipleyEtAl2021b_inPrep}.

\subsection{Closures}\label{sec:closures}
The terms in equations~\eqref{eq:MFBoussinesq_sigma_noApprox}-\eqref{eq:MFBoussinesq_buoyancy_noApprox} which require closure can be split into: 
\begin{itemize}
    \item $p_i$, the difference between the conditionally-filtered pressure in fluid $i$ and the unconditionally filtered pressure;
    \item $\overline{P}\grad{\sigma_i} - \overline{P \grad{I_i}}$, $- \nu \div{\overline{(\vec{u} \otimes \grad{I_i})\trans}} - \nu \overline{\grad{I_i} \vdot (\grad{\vec{u}})\trans}$, and $- \kappa \overline{\grad{I_i} \vdot \grad{b}} - \kappa \div{\overline{b \grad{I_i}}}$, which arise from conditionally-filtering the pressure gradient, viscous diffusion, and buoyancy diffusion terms;
    \item $- \div( \overline{I_i \vec{u} \otimes \vec{u}} - \sigma_i \vec{u}_i \otimes \vec{u}_i )$, and $- \div(\overline{I_i \vec{u} b} - \sigma_i \vec{u}_i b_i)$, which are often termed ``subfilter fluxes'' and are akin to the Reynolds stress and subfilter buoyancy flux, respectively, in normal higher-order modelling of turbulence;
    \item $\overline{\mathcal{S}_i^{\pm}}, \overline{\vec{u}\mathcal{S}_i^{\pm}}, \overline{b\mathcal{S}_i^{\pm}}$, which arise from filtering the re-labelling of fluid parcels.
\end{itemize}
We present closures that attempt to model the dominant coherent overturning structures of RBC (seen in the DNS, figure~\ref{fig:DNS_RBCexamples}).

For this study, differences between conditionally-filtered and unconditionally-filtered pressures are parametrized as $p_i = \left( \sum_j \sigma_j \gamma_i \div{\vec{u}_j} \right) - \gamma_i \div{\vec{u}_i}$, where $\gamma_i$ is a volume (or ``bulk'') viscosity. 
This has successfully been used by \textcite{ar:WellerEtAl2020}, where it was argued that such a form is plausible since in the underlying Boussinesq flow, the pressure is simply a Lagrange multiplier to enforce the divergence-free condition. 
It is also possible to derive this form by analogy with the ``bulk viscous pressure'' which arises in compressible fluid dynamics, as in \textcite{bk:BatchelorFluids} --- see \textcite{ar:ShipleyEtAl2021b_inPrep} for details.

Residual terms arising from conditionally-filtering the pressure gradient and diffusion terms are closed via a mean-field approximation: 
\begin{itemize}
    \item $\overline{P}\grad{\sigma_i} - \overline{P \grad{I_i}} \to \overline{P}\grad{\sigma_i} - \overline{P}\grad{\sigma_i} = 0$;
    \item $- \nu \div{\overline{(\vec{u} \otimes \grad{I_i})\trans}} - \nu \overline{\grad{I_i} \vdot (\grad{\vec{u}})\trans} \to - \nu \div{(\overline{\vec{u}} \otimes \grad{\sigma_i})\trans} - \nu \grad{\sigma_i} \vdot (\grad{\overline{\vec{u}})\trans}$; 
    \item $ - \kappa \overline{\grad{I_i} \vdot \grad{b}} - \kappa \div{\overline{b \grad{I_i}}} \to - \kappa \grad{\sigma_i} \vdot \grad{\overline{b}} - \kappa \div{\overline{b} \grad{\sigma_i}} $.
\end{itemize} 
These choices retain the correct sum over all fluids for the entire pressure, viscous, and diffusive terms, respectively. 
They also cause the fluid fractions to behave passively in the case of two fluids, and in the absence of transfers: the Eulerian derivatives for $\vec{u}_i$ and $b_i$ do not depend on $\sigma_i$ if the two fluids have the same $\vec{u}_i$ and $b_i$.

The resolved velocities and buoyancies of the multi-fluid split are assumed to dominate the single-fluid subfilter fluxes, such that $\overline{I_i \vec{u} \otimes \vec{u}} \approx \sigma_i \vec{u}_i \otimes \vec{u}_i$ and $\overline{I_i \vec{u} b} \approx \sigma_i b_i \vec{u}_i$.
This is the same as assuming that the multi-fluid split captures \textit{all} of the subfilter variability in the momentum and buoyancy fluxes, i.e. neglecting the residual subfilter fluxes of momentum and buoyancy, $\div( \overline{I_i \vec{u} \otimes \vec{u}} - \sigma_i \vec{u}_i \otimes \vec{u}_i )$ and $\div(\overline{I_i \vec{u} b}- \sigma_i \vec{u}_i b_i)$.
While this will never be exactly true, it is instructive to see how well a multi-fluid model with no extra subfilter modelling can perform when simulating a fully turbulent flow.
In the single column context this requires the vertical grid to adequately resolve the boundary layers, as in the DNS.

To proceed further, we must decide what the labels $I_i$ represent. 
The simplest choice is to restrict to two fluids; the symmetries of the Rayleigh-B\'{e}nard problem suggest choosing one falling and the other rising: let $i = 0$ denote fluid with $w \leq 0$, and $i = 1$ denote fluid with $w > 0$ \parencite[as in][]{ar:WellerEtAl2020}. 
Then fluid $1$ represents ``updrafts'' while fluid $0$ represents ``downdrafts''. 
This choice of definitions for the two fluids, coupled with the discrete symmetry of the unfiltered equations under the simultaneous transformations $z \to \tfrac{H}{2} - z$, $b \to -b$, forces $\int_\mathcal{D} \sigma_i \diff{}{V} = \tfrac{1}{2}$. 
This constraint can be used as a ``sanity check'' for both the initial conditions and the transfer terms $\mathcal{S}_i^\pm$. 
The discrete symmetry of the fluids under exchange also forces $\gamma_i = \gamma_j$ if $\gamma$ is not a function of $z$.

Specializing to two fluids allows the sources of fluid fraction $i$ to be written as $\overline{\mathcal{S}_i^+} = \sigma_j S_{ji}$, where $\sigma_j S_{ji}$ is the rate of transfer of fluid fraction from $j$ to $i$. 
A similar relation follows for the sinks. 
We choose to model the exchanges of momentum and buoyancy from fluid $i$ to $j$ as a characteristic value, $\vec{u}^T_{ij}$ or $b^T_{ij}$, times the rate of transfer of fluid fraction from $i$ to $j$, $\sigma_i S_{ij}$. 
This aligns with the modelling approach taken in other recent works on multi-fluid modelling \parencite{ar:ThuburnEtAl2018,ar:ThuburnEtAl2019,ar:WellerMcIntyre2019,ar:WellerEtAl2020,ar:McIntyreEtAl2020}.
    
Partitioning the flow based on the sign of $w$ forces $w_{ij}^T = 0$. 
For a single-column model, it remains only to specify the form of the fluid fraction transfer rate, $S_{ij}$, and the transferred buoyancy, $b^{T}_{ij}$ (for a 2D or 3D model, the horizontal components of the transferred velocity would also need to be specified). 
For the fluid fraction transfer rate we choose
\begin{align}
    S_{ij}
    &=
    \max(-\div{\vec{u}_i},0),
\end{align}
which in 1D is similar to dynamical entrainment, and follows the successful implementation of the same divergence-based transfer in \textcite{ar:WellerEtAl2020}.
This aims to capture the large-scale overturning circulation, and is exactly correct for the first normal mode of RBC with stress-free boundaries \parencite{ar:ShipleyEtAl2021b_inPrep}. 
\textcite[][chapter~2]{phd:McIntyre2020} also shows that this choice of transfer rate removes the problematic Kelvin-Helmholtz--like instability for a two-fluid Boussinesq system \parencite{ar:ThuburnEtAl2019}.

The transferred buoyancy must depend on the distribution of buoyancy within each fluid, and on the detailed dynamics of the relabelling. 
In the absence of this information, we choose a simple model:
\begin{align}
    b^{T}_{ij}
    =
    b_i
  + (-1)^i C \abs{b_i},
\end{align}
with some dimensionless constant $C \geq 0$. 
That is, the buoyancy of fluid parcels relabelled from $i$ to $j$ is modelled as the mean buoyancy within the fluid $i$ plus or minus some constant times the magnitude of the buoyancy, to crudely approximate the subfilter buoyancy variability. 
The signs are chosen to model the fact that the fluid transferred from the falling (0) to the rising (1) fluid is expected to be more buoyant than the average falling fluid parcel for that height, while the reverse should be true for transfers from the rising (1) to the falling (0) fluid. 
This is a similar formulation to that used by \textcite{ar:ThuburnEtAl2019}, though in theirs the transferred value depends on both the initial and destination fluids, rather than just the initial fluid.

Making these closure assumptions reduces the equation set to:
\begin{align}
    \pdv{\sigma_i}{t}
  + \div(\sigma_i \vec{u}_i)
    &=
    \sigma_j S_{ji}
  - \sigma_i S_{ij},
    \label{eq:2FBoussinesq_sigma_closed}
    \\
\begin{split}
    \pdv{\sigma_i \vec{u}_i}{t}
  + \div(\sigma_i \vec{u}_i \otimes \vec{u}_i)
    &=
    \sigma_i b_i \unit{k}
  - \sigma_i \grad{\overline{P}}
  - \grad(\sigma_i p_i)
    \\
    &\quad
  + \nu \laplacian{\sigma_i \vec{u}_i}
  - \nu \div(\overline{\vec{u}} \otimes \grad{\sigma_i})\trans
  - \nu \grad{\sigma_i} \vdot (\grad{\overline{\vec{u}})\trans}
    \\
    &\quad
  + \sigma_j \vec{u}^T_{ji} S_{ji}
  - \sigma_i \vec{u}^T_{ij} S_{ij},
    \label{eq:2FBoussinesq_momentum_closed}
\end{split}
    \\
\begin{split}
    \pdv{\sigma_i b_i}{t}
  + \div(\sigma_i \vec{u}_i b_i)
    &=
    \kappa \laplacian{\sigma_i b_i}
  - \kappa \grad{\sigma_i} \vdot \grad{\overline{b}}
  - \kappa \div{\overline{b} \grad{\sigma_i}}
    \\
    &\quad
  + \sigma_j b^T_{ji} S_{ji}
  - \sigma_i b^T_{ij} S_{ij},
    \label{eq:2FBoussinesq_buoyancy_closed}
\end{split}
\end{align}
with $i \in \{0,1\}$, and the specific parametrization choices:
\begin{align}
    S_{ij} 
    &=
    \max(- \div{\vec{u}_i}, 0),
    \label{eq:2Fclosure_divTransfer}
    \\
    w^{T}_{ij}
    &=
    0,
    \label{eq:2Fclosure_wTransfer}
    \\
    b^{T}_{ij}
    &=
    b_i
  + (-1)^i C \abs{b_i}
    \label{eq:2Fclosure_bTransfer}
    \\
    p_i
    &=
    \left(
        \sum_j \sigma_j \gamma \div{\vec{u}_j}
    \right)
  - \gamma \div{\vec{u}_i}.
    \label{eq:2Fclosure_gamma}
\end{align}
The equations are given in vector form because of the desire to eventually create a 3D grey-zone convection parametrization; to that end the subsequent numerical method is also three-dimensional. 
Note, however, that in the form \eqref{eq:2FBoussinesq_sigma_closed}-\eqref{eq:2Fclosure_gamma}, the horizontal components of the transferred velocity still require closure.

\subsubsection{Boundary conditions}\label{sec:BCs}
Conditionally filtering the boundary conditions for RBC gives $\vec{u}_i(z = 0,H) = \vec{0}$, $b_i(z = 0,H) = \pm \tfrac{\Delta B}{2}$. 
The Neumann boundary condition for the unconditionally filtered pressure (required for the numerical solution, which solves elliptic equations for the pressures) is hydrostatic, $\dv{\overline{P}}{z}(z=0,H) = \overline{b}(z=0,H)$.
Boundary conditions on the perturbation pressures are chosen to be zero-gradient, $\dv{\overline{p_i}}{z}(z=0,H) = 0$.

Because the $\sigma_i$ equation is a transport equation with no diffusion, boundary values of $\sigma_i$ are not in the domain of dependence of its solution.
The asymptotic boundary behaviour of $\sigma_i$ is thus entirely dependent on the asymptotic behaviour of the transfer terms as the boundaries are approached.
Boundary values of $\sigma_i$ are however required for the momentum and buoyancy equations, which do contain second derivatives of $\sigma_i$.
These boundary values should be set by extrapolated values of $\sigma_i$ from the interior of the domain.
However, for this study we choose zero-gradient conditions for $\sigma_i$ for better numerical behaviour.
Heuristically this means that we are imposing no creation of fluid in either partition at the boundary.

\subsection{Scaling of pressure differences between fluids}\label{sec:scaling}

In single-column form, equations \eqref{eq:2FBoussinesq_sigma_closed}-\eqref{eq:2Fclosure_gamma} contain two free parameters: $\gamma$ and $C$. 
$C$ is dimensionless and should be $\lesssim \mathcal{O}(1)$, but $\gamma$ has the dimensions of (bulk) viscosity and does not have an obvious magnitude. 
In this section we present a scaling argument for $\gamma$ with the external dimensionless control parameters $\Ra, \Pr$, thus reducing the model to the choice of two dimensionless constants which should both be $\mathcal{O}(1)$.

In convection, a distinction is often made between filamentary plumes and a well-mixed environment; this distinction is clearly seen in the example RBC buoyancy fields of figure~\ref{fig:DNS_RBCexamples}, and is the basis of the conceptual ``updraft''-``environment'' partition. 
We assume that such a plume has a length $\mathcal{O}(H)$, a width $\delta$, and the along-plume flow scales with the large-scale circulation $U \sim U_B = \sqrt{\Delta B\ H}$.
Orienting a local Cartesian co-ordinate system such that $\unit{x}$ points parallel to the plume and $\unit{z}$ points normal to it, the scaled continuity equation gives:
\begin{align}
    \frac{U}{H}\pdv{\tilde{u}}{\tilde{x}}
    &=
  - \frac{W}{\delta}\pdv{\tilde{w}}{\tilde{z}}
    \quad
    \implies
    \quad
    W
    =
    U \frac{\delta}{H}.
\end{align}
Splitting the buoyancy equation similarly into its plume-parallel and -normal parts gives:
\begin{align}
    \left(
        \pdv{\tilde{b}}{\tilde{t}}
      + \pdv{\tilde{u}\tilde{b}}{\tilde{x}}
      + \pdv{\tilde{w}\tilde{b}}{\tilde{z}}
    \right)
    &=
    \frac{\kappa T_b}{\delta^2}
    \left(
        \frac{\delta^2}{H^2} \pdv[2]{\tilde{b}}{\tilde{x}}
      + \pdv[2]{\tilde{b}}{\tilde{z}}
    \right).
\end{align}
Note the buoyancy scaling cancels here.
The simplest choice of the time scale is $T_b = \flatfrac{\delta^2}{\kappa}$, which makes the coefficient of the final term on the RHS one, consistent with filamentary plumes being diffusion-limited in well-developed turbulent flows.
Scaling the plume-parallel momentum equation with time scale $ T_m = T_b / \Pr$,  buoyancy with $\Delta B$ and pressure with  
$P \sim U^2$ (Bernoulli scaling), leads to:
\begin{align}
    \pdv{\tilde{u}}{\tilde{t}}
  + \pdv{\tilde{u}\tilde{u}}{\tilde{x}}
  + \pdv{\tilde{u}\tilde{w}}{\tilde{z}}
    &=
    \Re \frac{\delta^2}{H^2}
    \left(
      - \pdv{\tilde{p}}{\tilde{x}}
      + \tilde{b}\ \unit{g} \vdot \unit{x}
    \right)
  + \left(
        \frac{\delta^2}{H^2} \pdv[2]{\tilde{u}}{\tilde{x}}
      + \pdv[2]{\tilde{u}}{\tilde{z}}
    \right),
\end{align}
where $\Re = \flatfrac{U H}{\nu}= \Pr^{-1/2}\Ra^{1/2}$.
The pressure gradient and buoyancy terms are assumed to drive the flow, and so $\Re\flatfrac{\delta^2}{H^2} = \mathcal{O}(1)$ and:
\begin{align}
    \frac{\delta}{H}
    &=
    \Re^{-1/2}.
\end{align}
Hence the across-plume pressure contrast --- i.e. the difference in pressure between the plume and the bulk --- may be scaled as $P_z = P \delta / H = \Delta B \delta$. 

These results are the standard Prandtl-Blasius results with $\delta$ the boundary-layer depth, consistent with the presumption that plumes in RBC are simply detached from the boundary layers. 
This is a standard assumption for the kinetic boundary layer depth in scaling analysis of RBC, for example in the successful theory of \textcite{ar:GrossmannLohse2000} for the Nusselt and Reynolds number scalings. 
The $\Re \propto \Ra^{1/2}$ result is also expected for RBC in the parameter regimes under study in this paper \parencite[][table 2]{ar:AhlersEtAl2009}.

We wish to parametrize the difference between the conditionally-filtered pressure in partition $i$, and the unconditionally-filtered pressure, as a bulk viscous stress: $p_i = - \gamma(\div{\vec{u}_i} - \sum_j{\sigma_j \div{\vec{u}_j}})$, equation~\eqref{eq:2Fclosure_gamma}. 
Assuming that the multi-fluid split is dominated by a plume vs. bulk contrast, then $\vec{u}_i$ scales with the velocity of the plumes, $\sqrt{\Delta B\ H}$, and the divergence within each fluid should then scale as $\div{\vec{u}_i} = (U/H)\divtilde{\tilde{\vec{u}}_i}$ (so long as the filter width is $\gtrsim\mathcal{O}(H)$). 
Collecting the nondimensionalized expressions for the pressure and the bulk viscous stress gives:
\begin{align}
    \gamma \frac{U}{H} \divtilde{\tilde{\vec{u}}_i}
    &=
    \Delta B\ \delta \pdv{\tilde{p}}{\tilde{z}}
    \nonumber
    \\
    \implies
    \frac{\gamma}{\nu}
    &=
    \mathcal{O}(1)
    \times \frac{\Delta B H}{\nu U}\delta
    =
    \mathcal{O}(1)
    \times
    \frac{U^2 H}{\nu U}\frac{\delta}{H}
    =  \mathcal{O}(1)\times \Re^\frac{1}{2}
    \nonumber
    \\
    \implies
    \frac{\gamma}{\nu}
    &=
    \hat{\gamma}_0 \Ra^{1/4} \Pr^{-1/4},
    \label{eq:gammaScaling}
\end{align}
introducing the $\mathcal{O}(1)$, dimensionless constant $\hat{\gamma}_0$.

This scaling law for $\gamma(\Ra,\Pr)$ reduces the model for the pressure perturbation to the specification of an $\mathcal{O}(1)$ constant, $\hat{\gamma}_0$. 
Although $\hat{\gamma}_0$ must be determined empirically, this determination need only be performed at one Rayleigh number. 
Since $\Pr = 0.707$ is constant throughout our experiments, we choose to subsume the factor of $\Pr^{-1/2} \approx 1.19$ into the definition of $\hat{\gamma}_0$ from now on.

\section{Numerical methods}\label{sec:numericalMethod}
\subsection{Single-fluid solver}\label{sec:numericalMethod_singleFluid}
Single-fluid reference solutions (section~\ref{sec:DNS}) were computed using the single-fluid Boussinesq finite volume code \texttt{boussinesqFoam} (available at \href{https://www.github.com/AtmosFOAM/AtmosFOAM}{www.github.com/AtmosFOAM/AtmosFOAM}). 
This solves the single-fluid Boussinesq equation set \eqref{eq:SFboussinesq_buoyancy_dimless}-\eqref{eq:SFboussinesq_mass_dimless} using precisely the same numerical method as detailed below for the multi-fluid equation set, but with only one fluid. 
This single-fluid solver gives statistically identical results to the multi-fluid solver when the latter is run with no coupling terms between the fluids, such that the $\sigma_i$ are simply passive tracers.

\subsection{Two-fluid solver}\label{sec:numericalMethod_twoFluid}

The two-fluid Boussinesq equation set \eqref{eq:2FBoussinesq_buoyancy_closed}-\eqref{eq:2Fclosure_gamma} is solved in advective form using the finite volume solver \texttt{multiFluidBoussinesqFoam}; this is part of the \texttt{AtmosFOAM} library of CFD codes for atmospheric fluid dynamics, based on the \texttt{OpenFOAM} open-source CFD library. 
The code is available at \href{http://www.github.com/AtmosFOAM/AtmosFOAM-multiFluid}{www.github.com/AtmosFOAM/AtmosFOAM-multiFluid}. 
The method is similar to that detailed in section~3 of \textcite{ar:WellerEtAl2020}; an overview, and choices specific to this paper, are presented below.

The spatial discretization uses Arakawa C-grid staggering in the horizontal and Lorenz staggering in the vertical. 
Temporal discretization is Crank-Nicolson with off-centring coefficient $\alpha = 0.55$.

Prognostic variables are $b_i$ and $\sigma_i$ at cell centres, and the volume flux $\phi_i \coloneqq \vec{u}_i \vdot \vec{S}_f$ at cell faces, where $\vec{S}_f$ is the outward-pointing area vector of face $f$. 
Advection of $b_i$ and $\sigma_i$ is total variation-diminishing (with a van Leer limiter) to preserve boundedness, while advection of $\phi_i$ is linear upwind. 
Thus the spatial discretization is (almost) second-order accurate.

The transfer terms $S_{ij}$ are handled explicitly, while the momentum and buoyancy transfers are implicit and operator-split, as in \parencite{ar:WellerMcIntyre2019,ar:McIntyreEtAl2020,ar:WellerEtAl2020}.

Diagnostic variables are the pressures $P$ and $p_i$ at cell centres. 
Solutions for both $P$ and $p_i$ are implicit but not simultaneous: first a Poisson equation is solved for $P$, which maintains a divergence-free mean velocity field (i.e. it ensures eq.~\eqref{eq:MFBoussinesq_mass_sumConstraint} is satisfied), followed by a Helmholtz equation for each $p_i$. 
These solutions are then iterated to convergence. 
The generalized Geometric-Algebraic MultiGrid (GAMG) method is used for the implicit pressure solves, with an absolute tolerance of $10^{-6}$.

Two outer iterations (for the whole of the above method) and two inner iterations (for the implicit pressure solves) are performed per time-step.

Apart from the transfer terms, this method is suitable for an arbitrary number of fluids, in up to 3 spatial dimensions. 
However, the transfer terms, and their inclusion into the algorithm, are currently specific to two fluids.

\section{Two-fluid single-column model results}\label{sec:results}

\begin{table}[hbt!]
    \centering
    \begin{tabular}{ c | c | c  }
     $\Ra$ & $T_\text{tot} / 4T_B$ & $\Delta t / 4T_B$ \\ 
     \hline
     \hline
     $10^2$ & $19$ & $1.998\times10^{-4}$ \\ 
     $10^3$ & $63$ & $3.197\times10^{-4}$ \\ 
     $2\times10^3$ & $38$ & $2.557\times10^{-4}$ \\ 
     $10^4$ & $19$ & $1.279\times10^{-3}$ \\ 
     $10^5$ & $19$ & $1.279\times10^{-3}$ \\ 
     $10^6$ & $19$ & $7.992\times10^{-4}$ \\ 
     $10^7$ & $19$ & $3.197\times10^{-4}$ \\ 
     $2\times10^7$ & $19$ & $5.115\times10^{-4}$ \\ 
     $10^8$ & $19$ & $3.197\times10^{-4}$ \\ 
     $10^9$ & $19$ & $1.598\times10^{-4}$ \\ 
     $10^{10}$ & $19$ & $5.115\times10^{-5}$ \\ 
     \hline
    \end{tabular}
    \caption{Details of time-step size and total simulation time for the two-fluid single-column results (section~\ref{sec:results}). 
    Resolutions are the same as the vertical resolution of the DNS, explained in section~\ref{sec:resolution} and given in table~\ref{tab:DNSdetails}. 
    All two-fluid single-column simulations at a given $\Ra$ required similar time-steps regardless of $\hat{\gamma}_0$ and $C$, therefore only the values for $\hat{\gamma}_0 = 1.861$, $C=0.5$ ($\Ra \leq 10^7$), $\hat{\gamma}_0 = 1.861$, $C=0$ ($\Ra > 10^7$) are given.}
    \label{tab:singleColumnDetails}
\end{table}

For $10^2 \leq \Ra \leq 10^{10}$, single-column two-fluid simulations were run with the same vertical resolution as the reference DNS (see table~\ref{tab:DNSdetails}) for various values of $\hat{\gamma}_0$ and $C$. 
The qualitative nature of the solutions is described in  section~\ref{sec:twoFluidPhenomenology}, followed by an analysis of sensitivity to the choice of $\hat{\gamma}_0$ and $C$ in section~\ref{sec:sensitivityToGammaAndC}.
In section~\ref{sec:scalingOfNuWithRa} the global buoyancy and momentum transport, $\Nu$ and $\Re$, is examined as a function of the buoyancy forcing $\Ra$.

For all simulations, the initial state was constructed from a resting hydrostatically-balanced solution with a linear buoyancy profile and uniform $\sigma_i = 0.5$ in both fluids.
Small non-zero velocities equal to $\pm 10^{-3}\ U_B$ were added to ensure correct labeling, and random perturbations of magnitude $\abs{\delta b} \leq 0.0008\ \Delta B$ drawn from a uniform distribution were added to the initial linear profile to seed instability\footnote{This value was chosen in order to approximate the same initial available potential energy in both the DNS and the single-column simulations. However, the (linear) growth rate of instabilities in a single fluid is not dependent on the size of the initial perturbation, and so the exact magnitude of the initial perturbations does not matter so long as it is small.}. 
Simulations were run until a steady state was reached ($9-12 T_e$); the steady-state profiles of buoyancy, pressure, vertical velocity, and fluid fraction, were then compared with the corresponding statistically steady-state time-mean conditionally horizontally averaged DNS profiles.
Resolutions, time-step size, and total simulation run time for each simulation are given in table~\ref{tab:singleColumnDetails}.

\begin{figure}[t!hb]
    \begin{subfigure}[t]{0.493\textwidth}
        \caption{DNS, $\Ra = 10^5$}
        \includegraphics[trim = 0 0 0 18, clip, width=\textwidth]{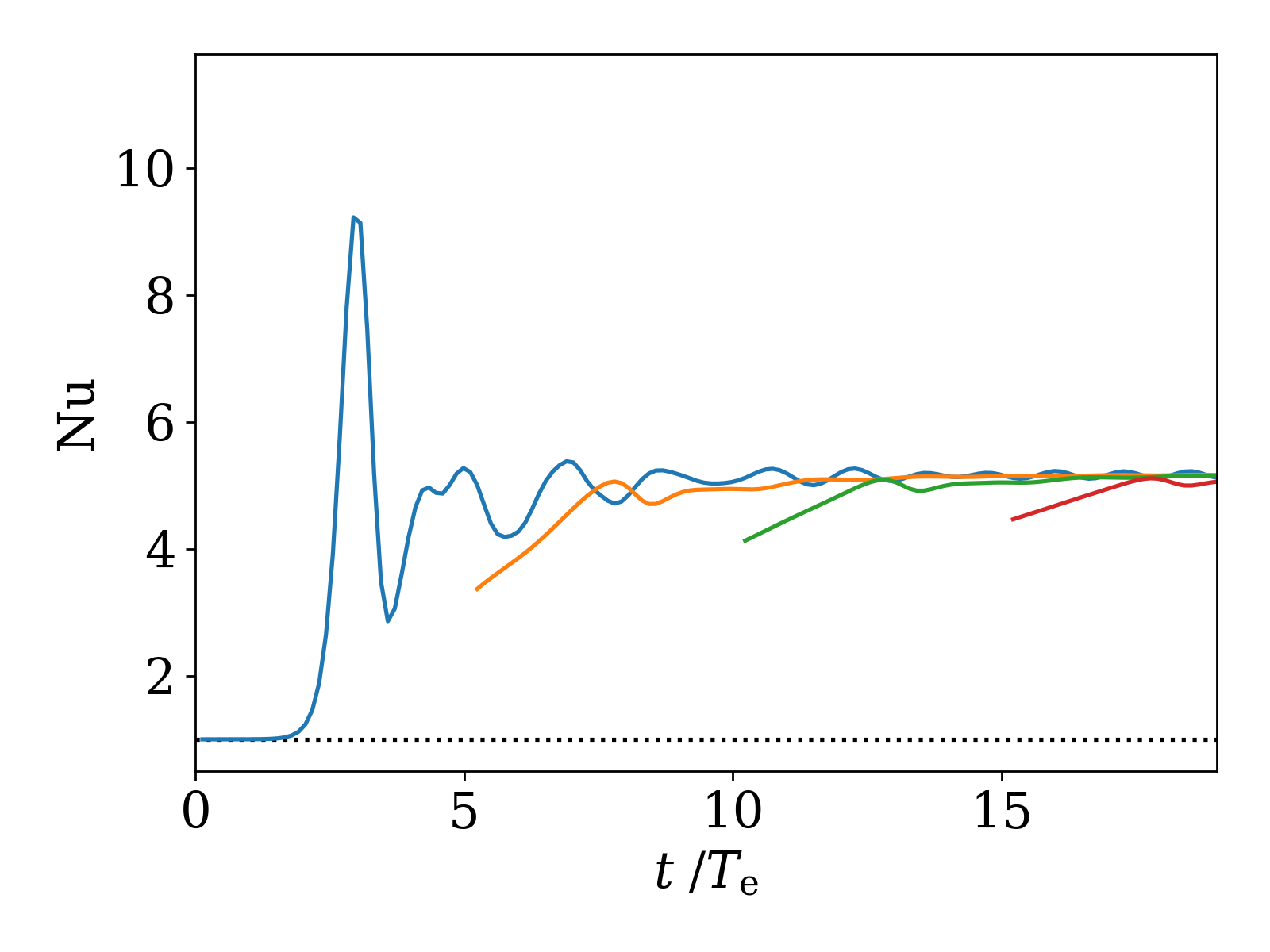}
    \end{subfigure}
    \begin{subfigure}[t]{0.493\textwidth}
        \caption{DNS, $\Ra = 10^8$}
        \includegraphics[trim = 0 0 0 18, clip, width=\textwidth]{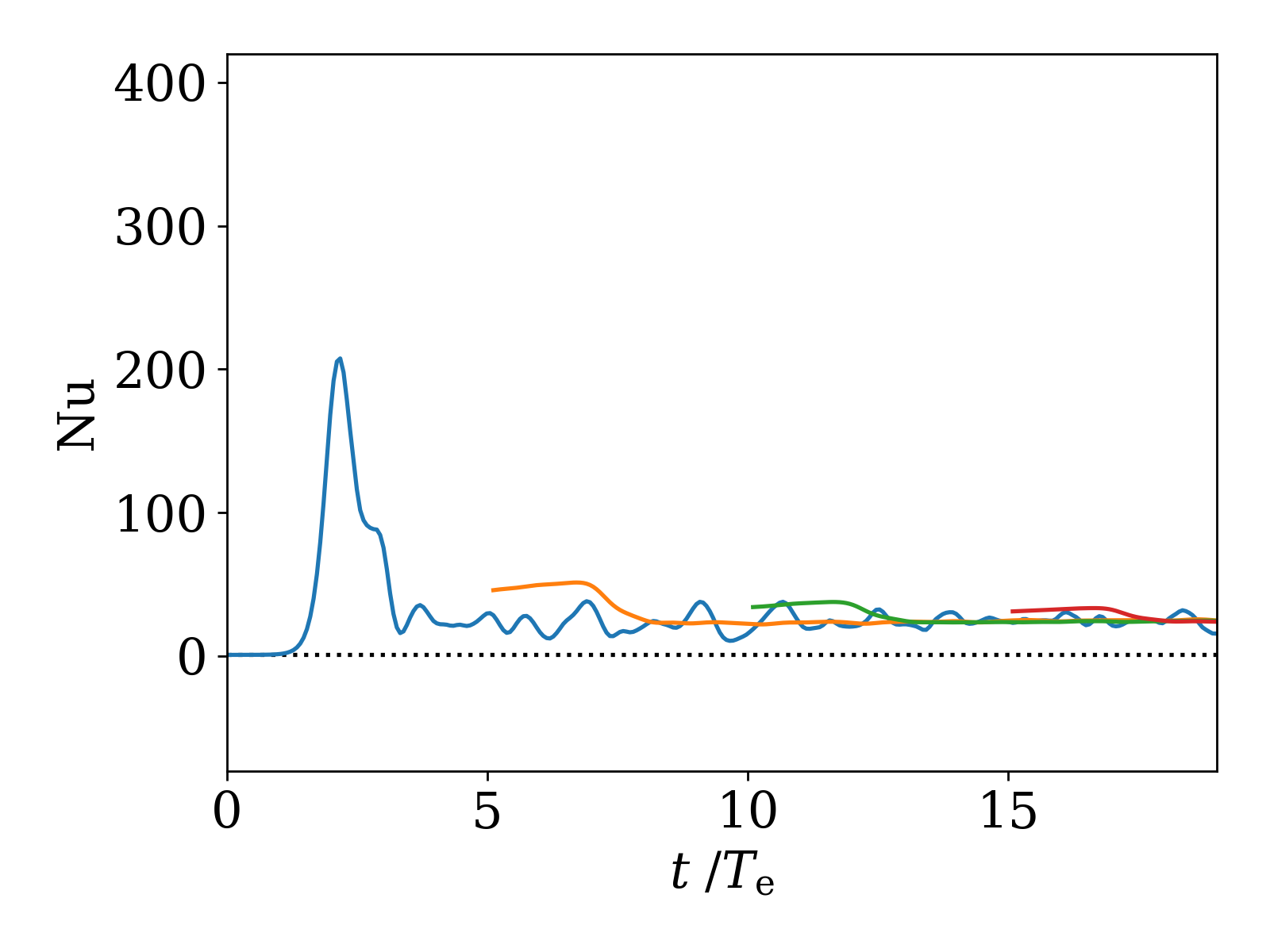}
    \end{subfigure}
    \begin{subfigure}[t]{0.493\textwidth}
        \caption{Single-column, $\Ra = 10^5$}
        \includegraphics[trim = 0 0 0 18, clip, width=\textwidth]{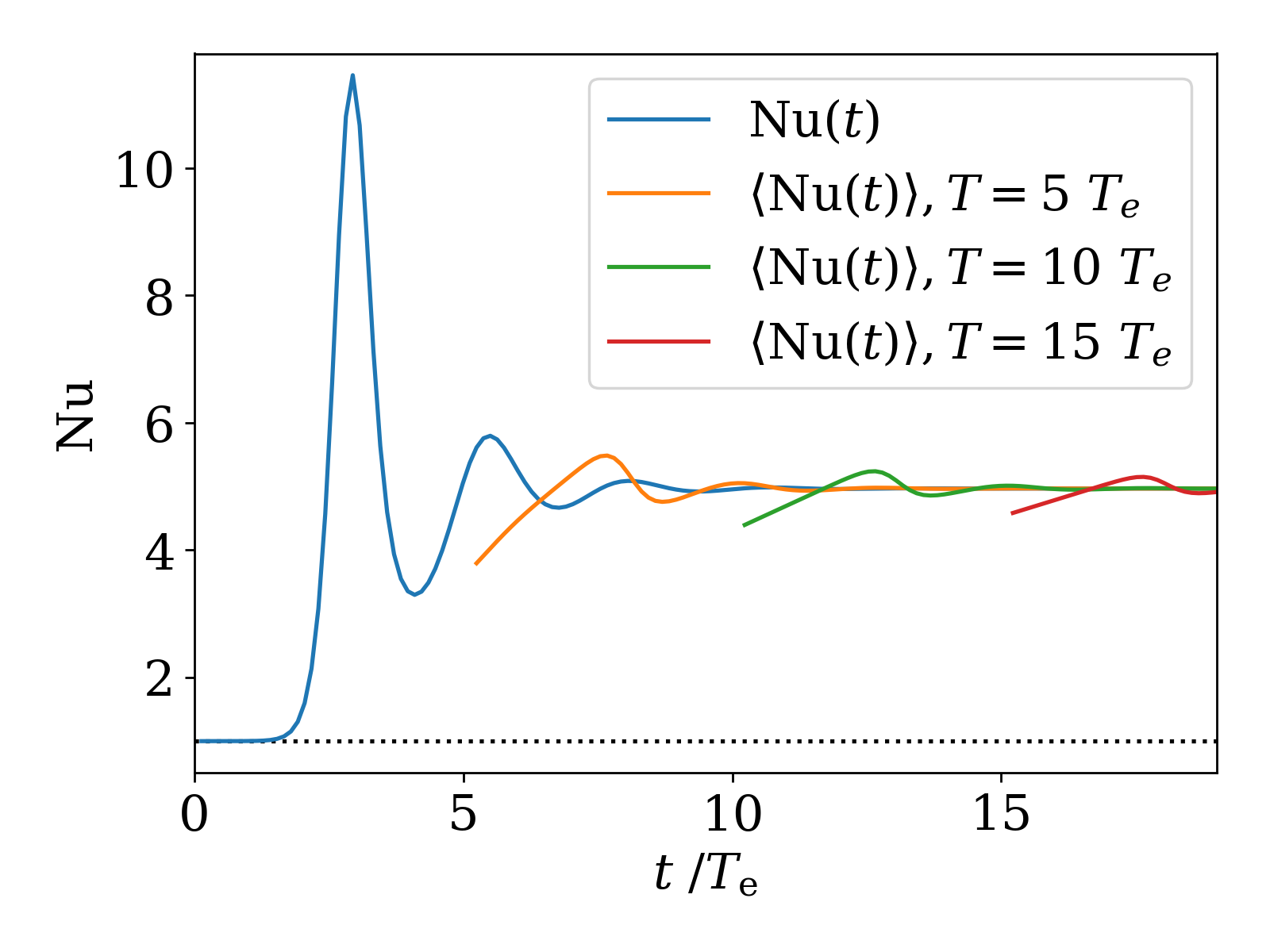}
    \end{subfigure}
    \begin{subfigure}[t]{0.493\textwidth}
        \caption{Single-column, $\Ra = 10^8$}
        \includegraphics[trim = 0 0 0 18, clip, width=\textwidth]{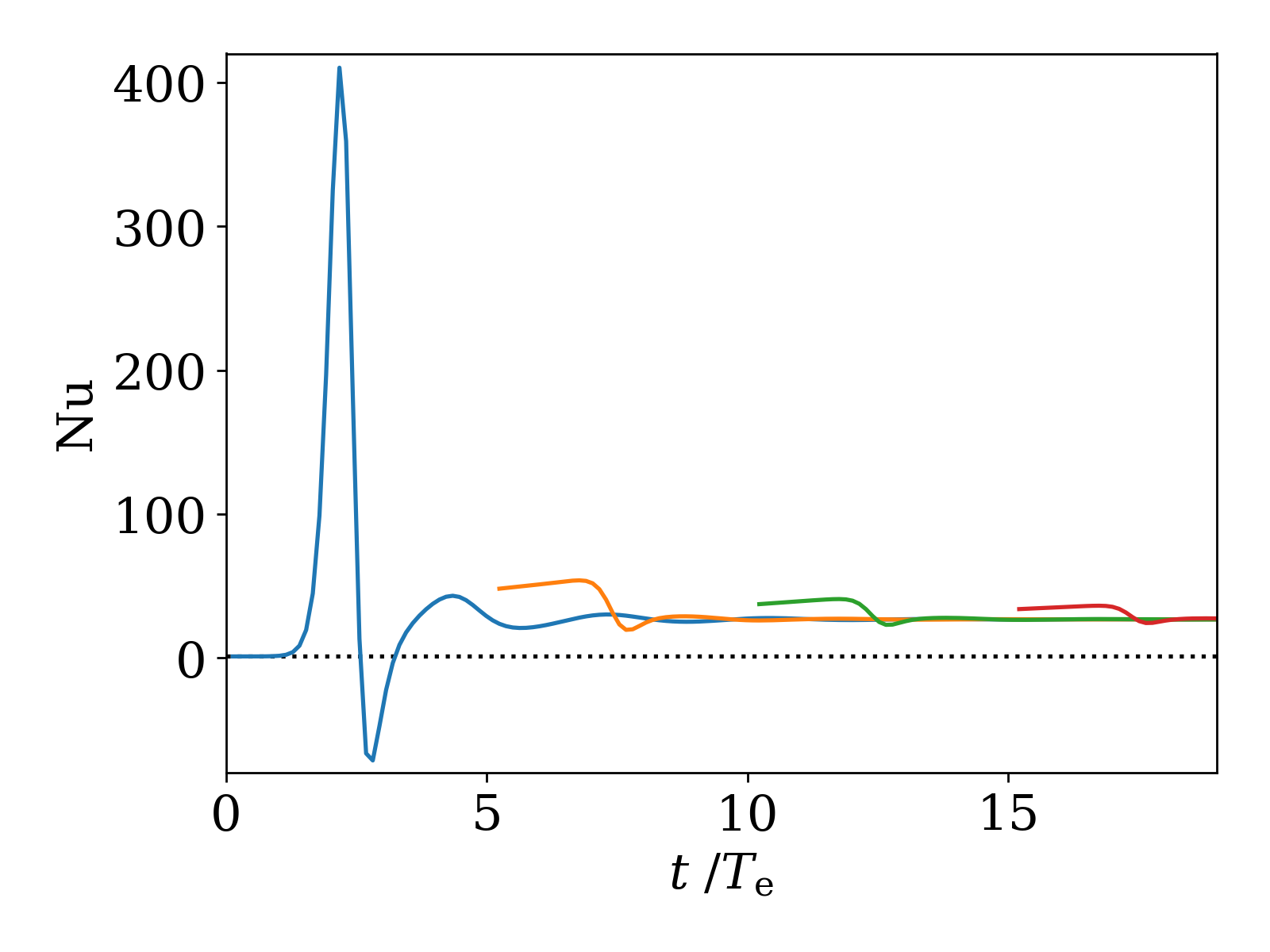}
    \end{subfigure}
    \caption{Nondimensionalized vertical heat flux vs. time for (a,b) DNS and (c,d) single-column models. 
    In each subfigure, the blue curve shows the instantaneous nondimensionalized vertical buoyancy flux, $H\times\flatfrac{(wb - \kappa \pdv{b}{z})}{\kappa \Delta B}$, while the orange, green, and red curves show Nusselt numbers (domain- and time-averaged nondimensionalized buoyancy flux) for different averaging times. 
    In each plot, $\Nu = 1$ is shown as a black dotted line.
    In (c) and (d) $\flatfrac{\gamma}{\nu} = 1.861\times\Ra^{1/4}$, with $C=0.5$ for (c) and $C=0$ for (d) (see figure~\ref{fig:2F1C_NuVsRa}). 
    }\label{fig:NuVsTime}
\end{figure}

The single column model spins up to equilibrium in a remarkably similar manner to the horizontally-averaged DNS; this is demonstrated in figure~\ref{fig:NuVsTime}, which shows the Nusselt number vs. time for both DNS and single-column simulations at $\Ra = 10^5$ and $10^8$.
For these simulations, $\hat{\gamma}_0 = 1.861$, and $C = 0.5, 0$ for $\Ra = 10^5, 10^8$, respectively (see section~\ref{sec:scalingOfNuWithRa}).
At each $\Ra$, convection initiates at a similar time ($\approx 2 T_e$) in both the single-column and DNS flows, seen in the sharp increase in $\Nu$ above the purely diffusive value of $1$. 
This initial convective surge causes a strong peak in the Nusselt number (slightly overestimated by the single-column model), before the system gradually settles down towards equilibrium with decaying Nusselt number under- and overshoots. 
The under- and overshoots appear stochastic for the DNS, whereas they are periodic for the single-column model; that the single-column model appears less chaotic than the DNS is unsurprising.

The same steady state was reached when initializing from other initial conditions (e.g. initializing from the DNS reference profiles), provided the identities of the fluids were initialized correctly and the initial column-integrated fraction of fluid in each fluid was equal to $0.5$. 
This suggests that the steady state is robust. 
Similar qualitative spin-up behaviour is also observed with different values of $\hat{\gamma}_0$ and $C$. 
Thus, for the remainder of the paper we consider only the steady state, and not the spin-up.

We begin our study of the two-fluid single-column model steady-state by looking at the qualitative behaviour of the equilibrium profiles in different Rayleigh number regimes. 
We then investigate the sensitivity of those profiles to the two closure constants, $C$ and $\hat{\gamma}_0$. 
Finally we examine the scaling of the global parameters $\Nu$ and $\Re$ with $\Ra$ produced by the model.

\subsection{Phenomenology}\label{sec:twoFluidPhenomenology}
For each of the characteristic Rayleigh numbers $\Ra = 10^5, 10^8, 10^{10}$ (as in figure~\ref{fig:DNS_RBCexamples}), we present and discuss an example two-fluid single column simulation. 
Rather than use the fixed value used above for discussion of the spin-up, the values of $\hat{\gamma}_0$ and $C$ in the example simulations were chosen to have the best qualitative fit to the conditionally horizontally averaged DNS for all profiles. 
The discussion for each of these examples qualitatively applies to all simulations within the characteristic Rayleigh number regime.

\subsubsection*{Laminar ($\Ra = 10^5$)}
\begin{figure}[t!hb]
    \centering
    \begin{subfigure}[t]{0.49\textwidth}
        \caption{Buoyancy, $b$}
        \includegraphics[trim = 0 0 0 35, clip, width=\textwidth]{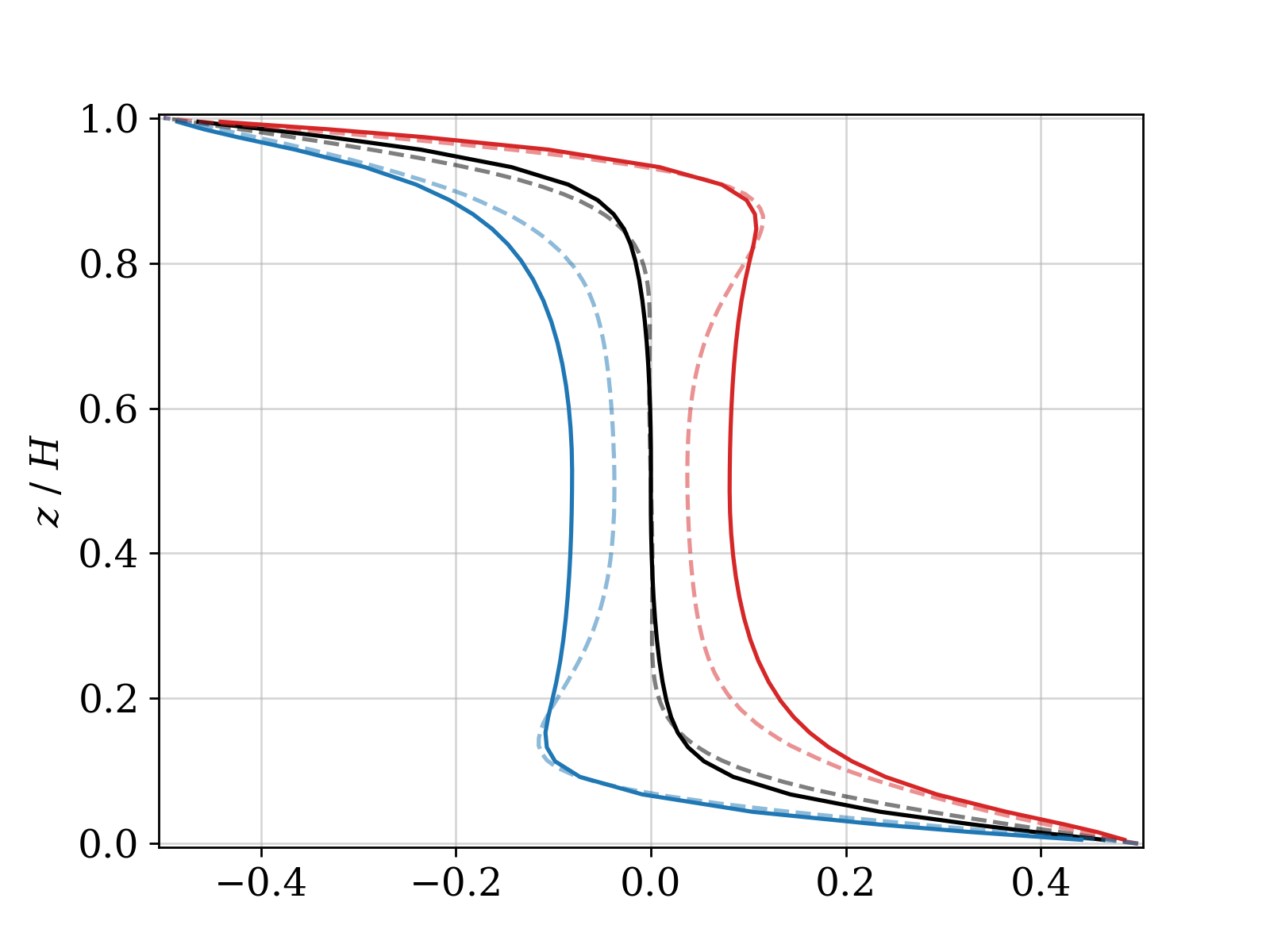}
    \end{subfigure}
    \begin{subfigure}[t]{0.49\textwidth}
        \caption{Pressure, $P$}
        \includegraphics[trim = 0 0 0 35, clip, width=\textwidth]{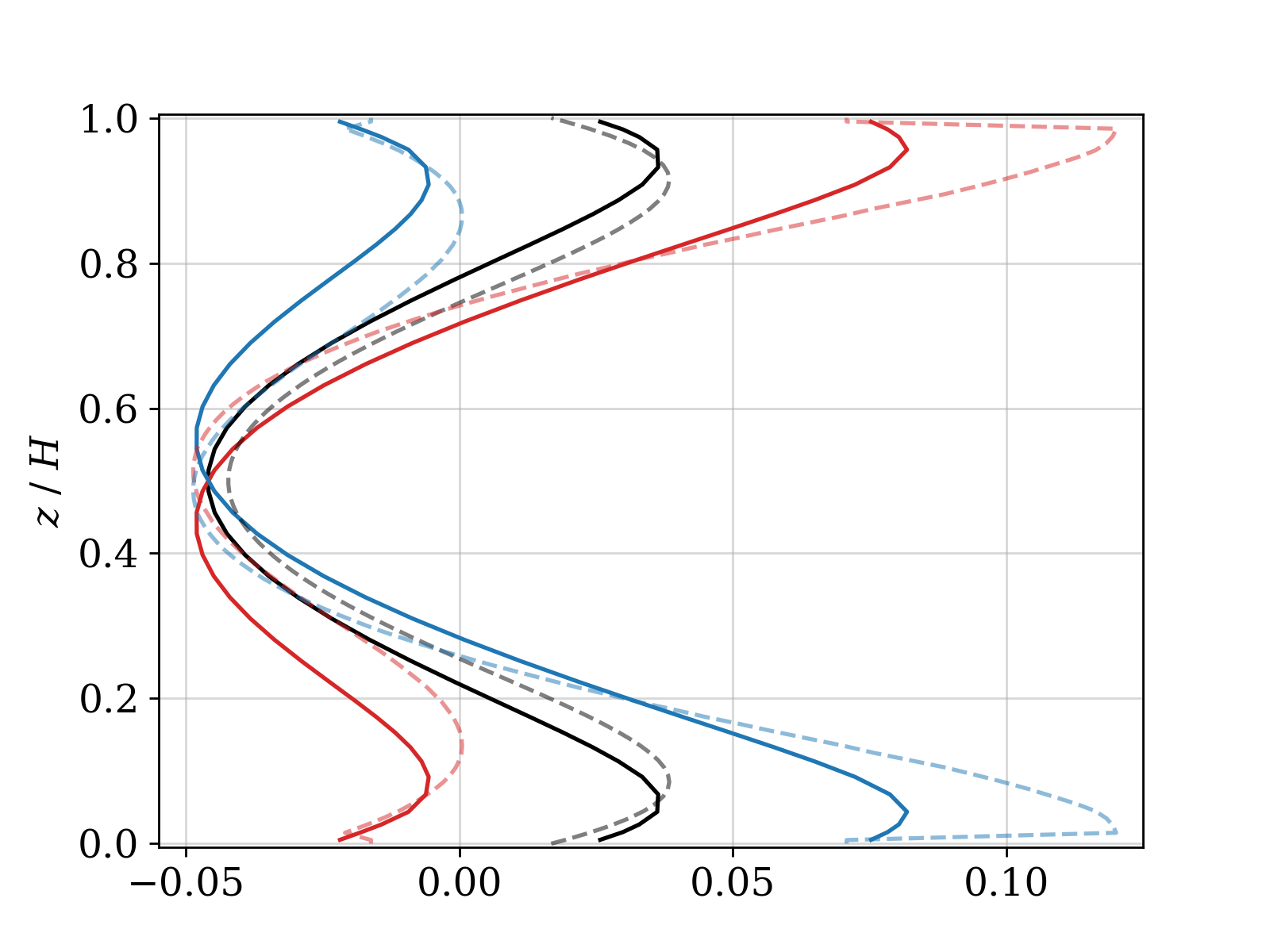}
    \end{subfigure}
    \begin{subfigure}[t]{0.49\textwidth}
        \caption{Vertical velocity, $w$}
        \includegraphics[trim = 0 0 0 35, clip, width=\textwidth]{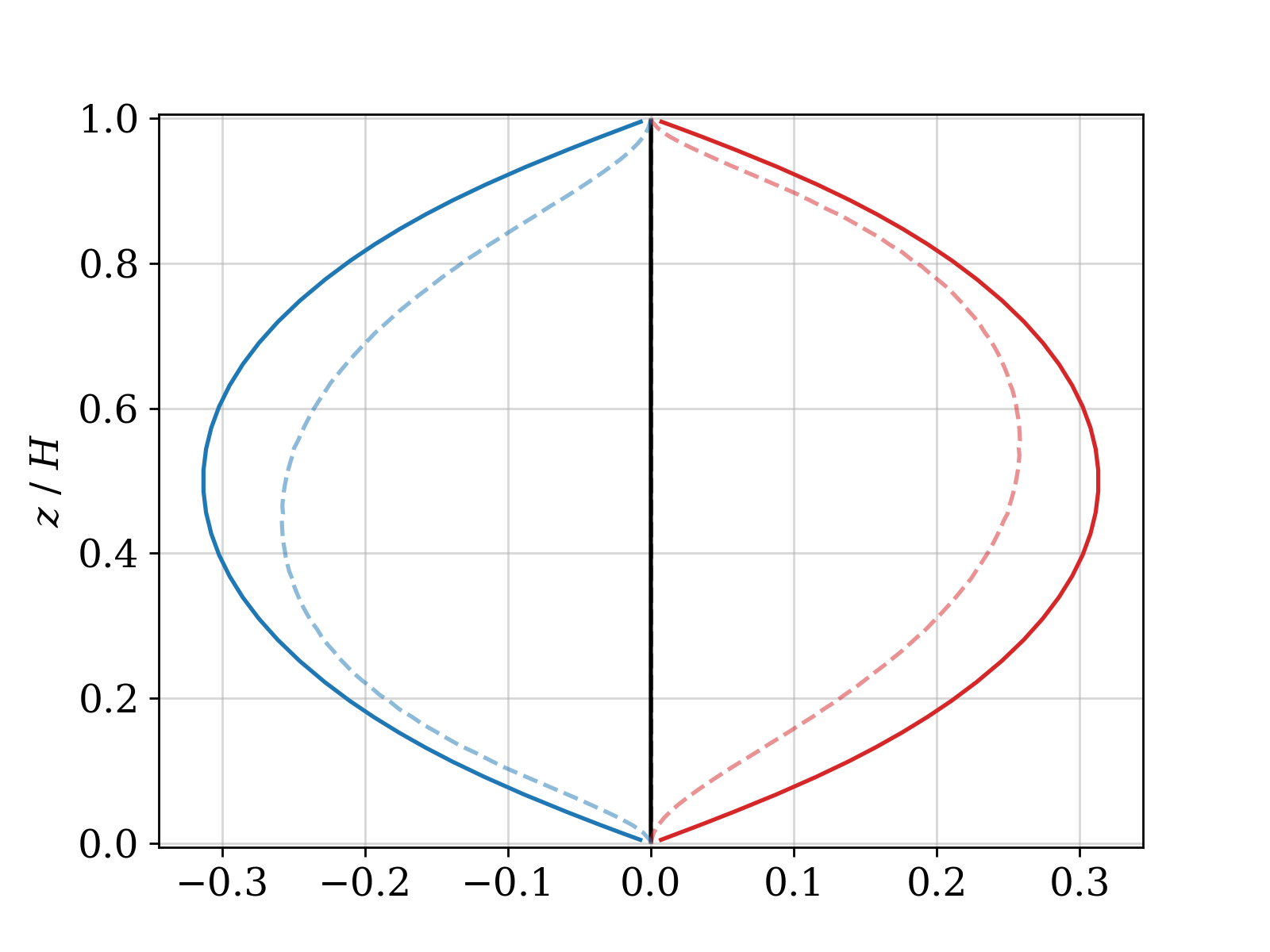}
    \end{subfigure}
    \begin{subfigure}[t]{0.49\textwidth}
        \caption{Fluid fraction, $\sigma_{w > 0}$}
        \includegraphics[trim = 0 0 0 35, clip, width=\textwidth]{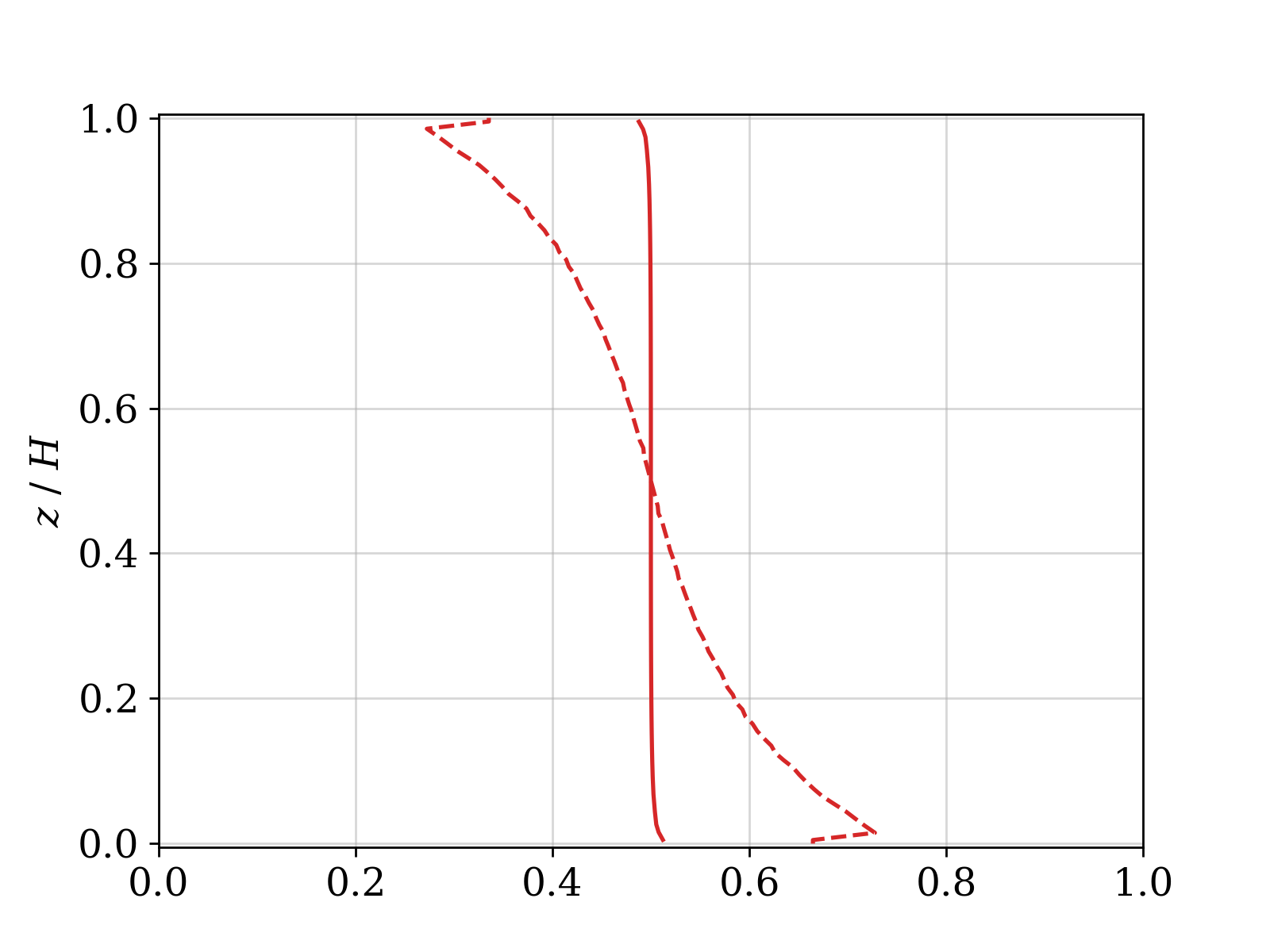}
    \end{subfigure}
    \begin{subfigure}[t]{0.4\textwidth}
        \centering
        \includegraphics[trim = -10 0 20 0, clip, width=\textwidth]{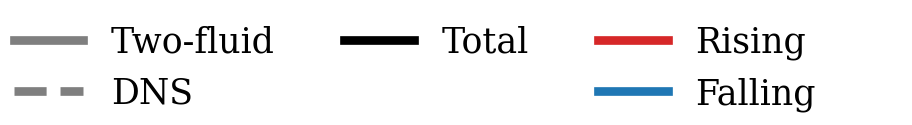}
    \end{subfigure}
    \caption{Two-fluid single-column model of $\Ra = 10^5$ RBC governed by equations \eqref{eq:2FBoussinesq_sigma_closed}-\eqref{eq:2Fclosure_gamma} and \eqref{eq:gammaScaling}, with closure constants $\hat{\gamma}_0 \approx 0.75, C = 0.5$. 
    Conditionally horizontally- and time-averaged profiles from the DNS are shown for reference. 
    $\Nu = 7.1$, reference $\Nu_{\text{DNS}} = 5.0$. 
    }
    \label{fig:Ra10^5_2F1C_results}
\end{figure}

At $\Ra = 10^5$, the DNS exhibits laminar convective rolls (see Fig.~\ref{fig:DNS_RBCexamples}a). 
This solution is qualitatively characteristic of the flow for all laminar $\Ra$, $\Ra_\text{c} < \Ra \lesssim 10^7$.
Steady state results of a two-fluid single-column model governed by equations~\eqref{eq:2FBoussinesq_sigma_closed}-\eqref{eq:2Fclosure_gamma} with $\hat{\gamma}_0 \approx 0.75$, $C = 0.5$ are shown in Fig.~\ref{fig:Ra10^5_2F1C_results}. 
The mean buoyancy (a) and pressure (b) profiles match closely between the DNS and the single column model; in particular the model correctly predicts a well-mixed buoyancy in the fluid interior, with a sharp buoyancy gradient close to the top and bottom boundaries. 
The shape of the pressure profile is also correct, though the maxima are slightly too high close to the boundaries.

Good agreement is also seen between the DNS and two-fluid single column model for the individual fluid buoyancy profiles: the overall shape is correct, though the profiles are too far apart in the middle of the domain, leading to surplus buoyancy transport for a given velocity profile. 
Experiments varying $C$ (see section~\ref{sec:sensitivityToC}) demonstrated $C > 0$ was required to reproduce a buoyancy overshoot at the top (bottom) of the rising (falling) fluid. 
By overshoot, we mean the part of the buoyancy profile at the interface between the bulk and the buoyancy boundary layer where $\dv{b_i}{z}$ changes sign. 
These overshoots can be seen in the 2D buoyancy field of the DNS flow of figure~\ref{fig:DNS_RBCexamples}a and are a general feature of $\mathcal{O}(1)$ Prandtl number laminar RBC. (For $\Pr > 1$, the overshoots become so strong that they begin to be seen even in the mean buoyancy profile; such profiles can be seen in e.g. Fig.~4b of \cite{ar:SchmalzlEtAl2004}.) 
The value $C = 0.6$ gives the best shape for $b_i(z)$ for $\Ra = 10^5$, but $C \approx 0.5$ works for all laminar $\Ra$.  

The individual fluid velocity profiles are roughly the correct shape; the slight asymmetry in the location of the maxima in each fluid in the DNS is due to the gradient of the volume fraction profile in the DNS (i.e. forcing the correct gradient of $\sigma_i$ reproduces the asymmetry in the vertical velocity profiles).

The pressure profiles with each fluid are captured by the scheme, suggesting that to leading order $p_i \propto - \gamma \div{\vec{u}_i}$ is an appropriate model of the pressure differences. 
The model is particularly good close to the boundaries, but the fluids are better mixed in the interior of the domain in the DNS, causing the pressure differences there to be smaller than predicted by the single column model. 
This could possibly be remedied by using a $z$-dependent $\gamma$ parametrization, which would fit well with the discussion of LES in section~\ref{sec:LESanalogy}.

The two-fluid model keeps area fractions, $\sigma_i(z)$, close to $0.5$.
This is expected as the divergence-based transfer is known to keep $\sigma_i(z)$ roughly constant \parencite{ar:WellerEtAl2020}. 
In contrast, the area fractions diagnosed from the DNS diverge from $0.5$ either side of the centre (where symmetry demands equal fractions), reaching a maximum close to the boundaries approaching $0.3$ and $0.7$.  

\subsubsection*{Transition to turbulence ($\Ra = 10^8$)}
Between $10^7 < \Ra \lesssim 5\times10^8$, the DNS solutions transition from laminar flow to fully developed turbulence. 
The buoyancy field of figure~\ref{fig:DNS_RBCexamples}b is characteristic of this transitional regime. 
Besides the solutions becoming intermittent and transient rather than (quasi-)periodic, the plume separation from the boundary layer fundamentally changes: above $\Ra \approx 10^7$, regions of recirculation develop at the base of the plumes.

\begin{figure}[t!hb]
    \centering
    \begin{subfigure}[t]{0.49\textwidth}
        \caption{Buoyancy, $b$}
        \includegraphics[trim = 0 0 0 10, clip, width=\textwidth]{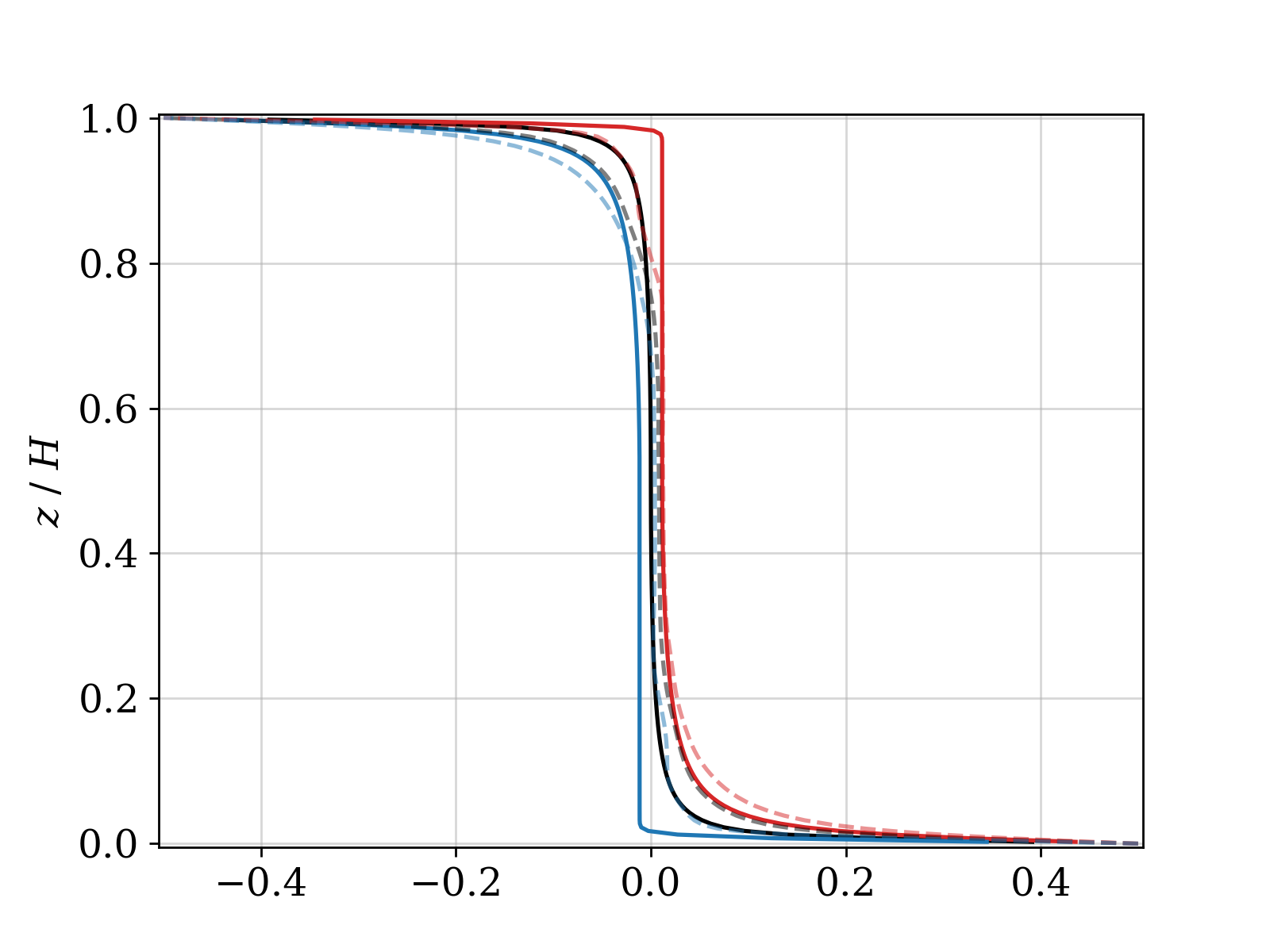}
    \end{subfigure}
    \begin{subfigure}[t]{0.49\textwidth}
        \caption{Pressure, $P$}
        \includegraphics[trim = 0 0 0 10, clip, width=\textwidth]{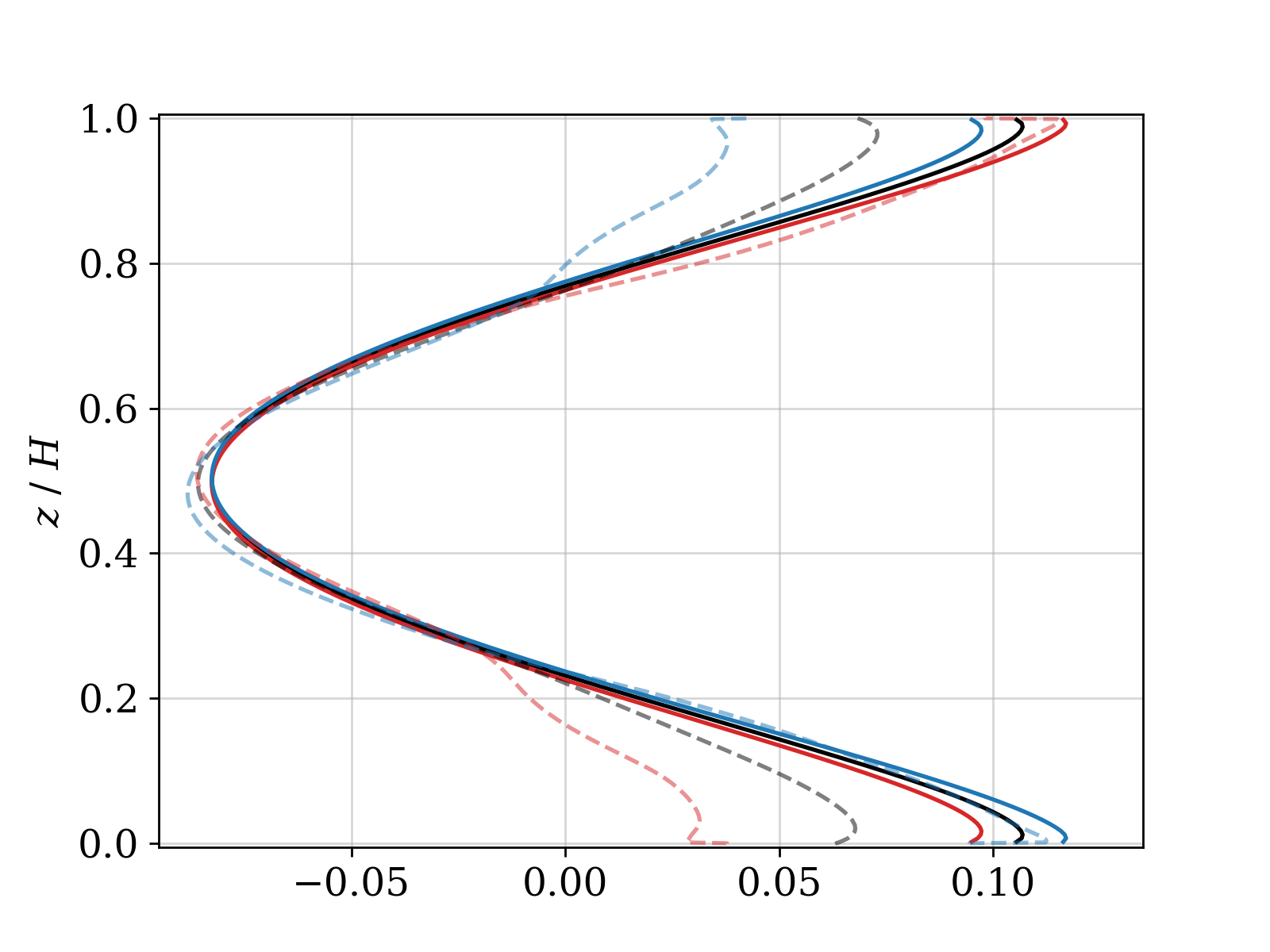}
    \end{subfigure}
    \begin{subfigure}[t]{0.49\textwidth}
        \caption{Vertical velocity, $w$}
        \includegraphics[trim = 0 0 0 10, clip, width=\textwidth]{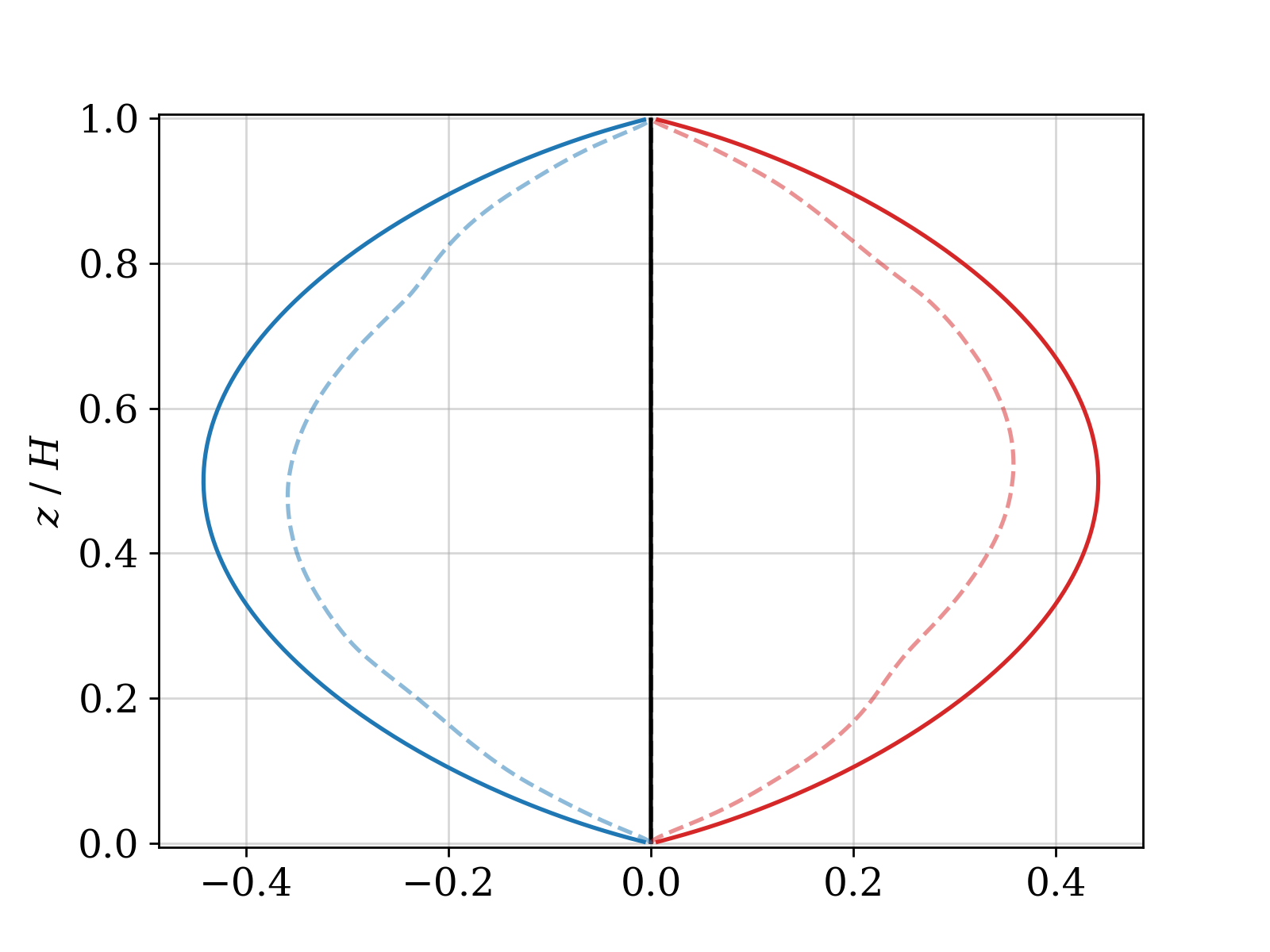}
    \end{subfigure}
    \begin{subfigure}[t]{0.49\textwidth}
        \caption{Fluid fraction, $\sigma_{w > 0}$}
        \includegraphics[trim = 0 0 0 10, clip, width=\textwidth]{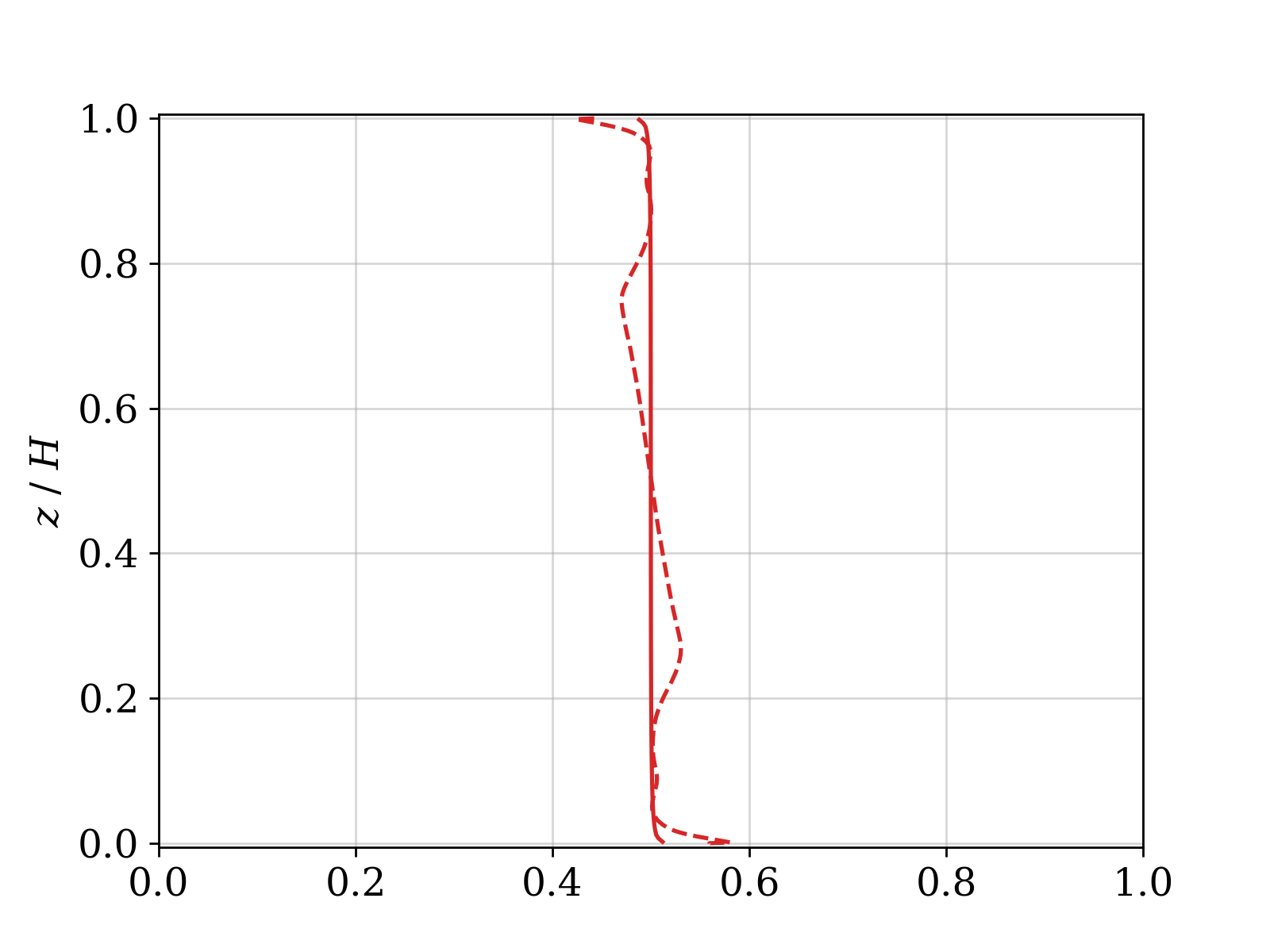}
    \end{subfigure}
    \begin{subfigure}[t]{0.4\textwidth}
        \centering
        \includegraphics[trim = -10 0 20 0, clip, width=\textwidth]{images/legend_2fProfiles_h.png}
    \end{subfigure}
    \caption{Two-fluid single-column model of $\Ra = 10^8$ RBC governed by equations \eqref{eq:2FBoussinesq_sigma_closed}-\eqref{eq:2Fclosure_gamma} and \eqref{eq:gammaScaling}, with closure constants $\hat{\gamma}_0 \approx 0.47$, $C = 0$. 
    Conditionally horizontally- and time-averaged profiles from the DNS are shown for reference. 
    $\Nu = 41.6$; reference $\Nu_\text{DNS} = 27.9$.
    }
    \label{fig:Ra10^8_2F1C_results}
\end{figure}

Results of a two-fluid single-column model with $\hat{\gamma}_0 \approx 0.47$, $C = 0$ are compared with those from the horizontally-averaged DNS in Fig.~\ref{fig:Ra10^8_2F1C_results}. 
As with the $\Ra = 10^5$ results, the values of $\hat{\gamma}_0$ and $C$ were chosen to give the best qualitative agreement for all profiles. 
Better prediction of the pressure differences between the fluids near the boundaries is achieved by increasing $\hat{\gamma}_0$ by a factor of $\approx 2$; however this degrades the agreement of the mean pressure profile with the DNS profile.
This again suggests that $\gamma$ should be a function of $z$, either directly or through dependence on other properties of the flow, for instance the TKE.

Comparisons with the DNS reference profiles are mostly the same as for the laminar case, except that the additional mixing caused by the recirculation regions at the base of the plumes modifies the profiles in the near-boundary regions. 
This has the most obvious effect on the buoyancy profiles within each fluid, which no longer overshoot, and on the volume fraction profile, which is no longer monotonic. 
The lack of overshoots is reproduced by transferring the mean buoyancy, $C=0$, a suitable model for well-mixed turbulent flow. 
The detailed differences to the profiles caused by these recirculation regions are however not reproduced by this simple parametrization: better representation of the mass exchanges $S_{ij}$ is required. 
The recirculation is counter to the large-scale circulation, and hence is not captured either by our arguments for the scaling of $\gamma$, or by the divergence-based mass transfer.

\subsubsection*{Fully developed turbulence ($\Ra = 10^{10}$)}
Above $\Ra \simeq 5 \times 10^8$, the DNS flow is fully turbulent, exhibiting structures on many scales from the domain depth down to the exceptionally thin boundary layers, shown in Figs.~\ref{fig:DNS_RBCexamples}c-d for $\Ra = 10^{10}$. 
The recirculations at plume base first exhibited in the transitional regime divide into multiple small plumes which organize into a larger-scale circulation. 
The bulk of the domain is statistically well-mixed.

\begin{figure}[t!hb]
    \centering
    \begin{subfigure}[t]{0.49\textwidth}
        \caption{Buoyancy, $b / \Delta B$}
        \includegraphics[trim = 0 0 0 35, clip, width=\textwidth]{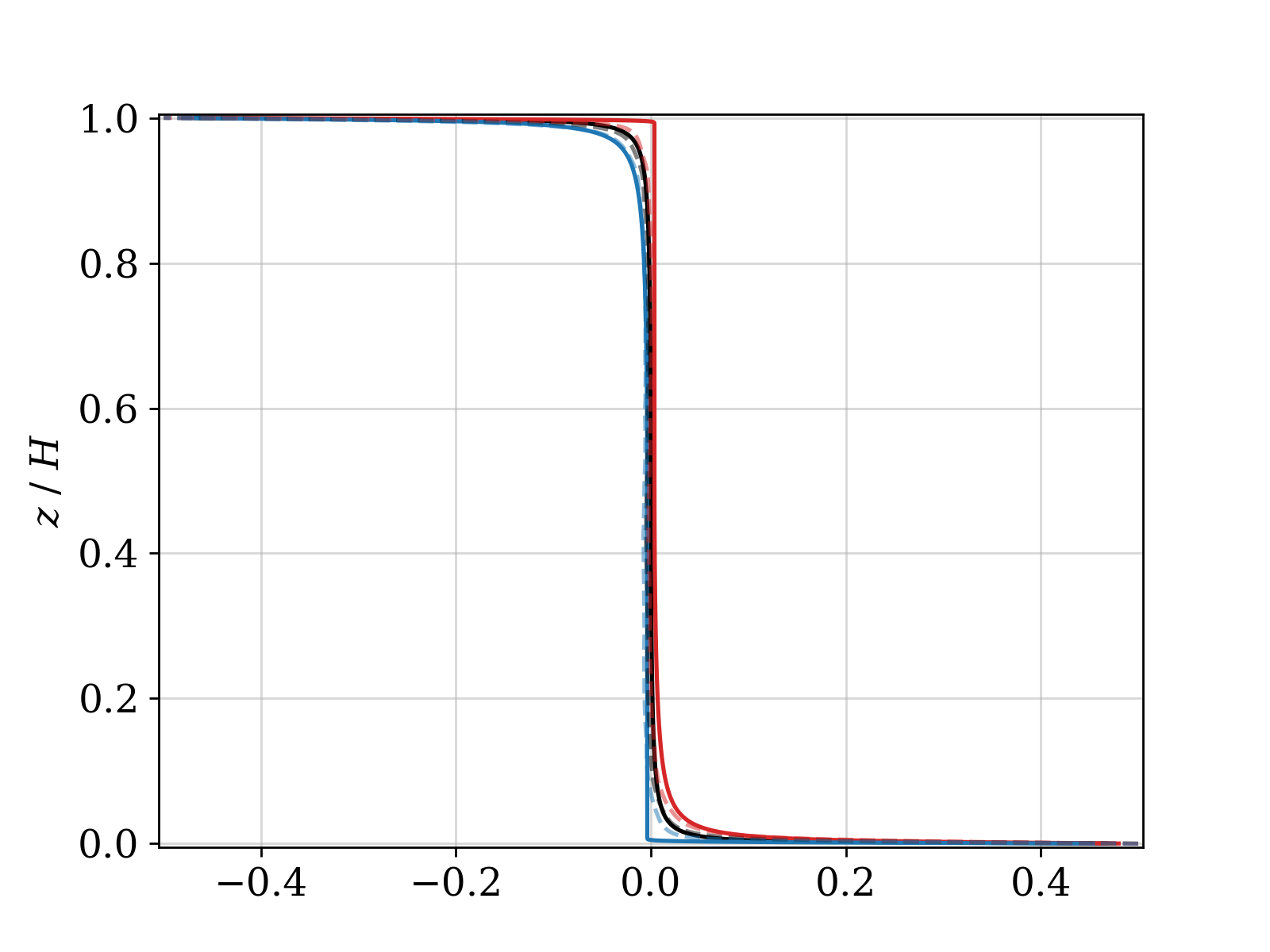}
    \end{subfigure}
    \begin{subfigure}[t]{0.49\textwidth}
        \caption{Pressure, $P / (\Delta B\ H)$}
        \includegraphics[trim = 0 0 0 35, clip, width=\textwidth]{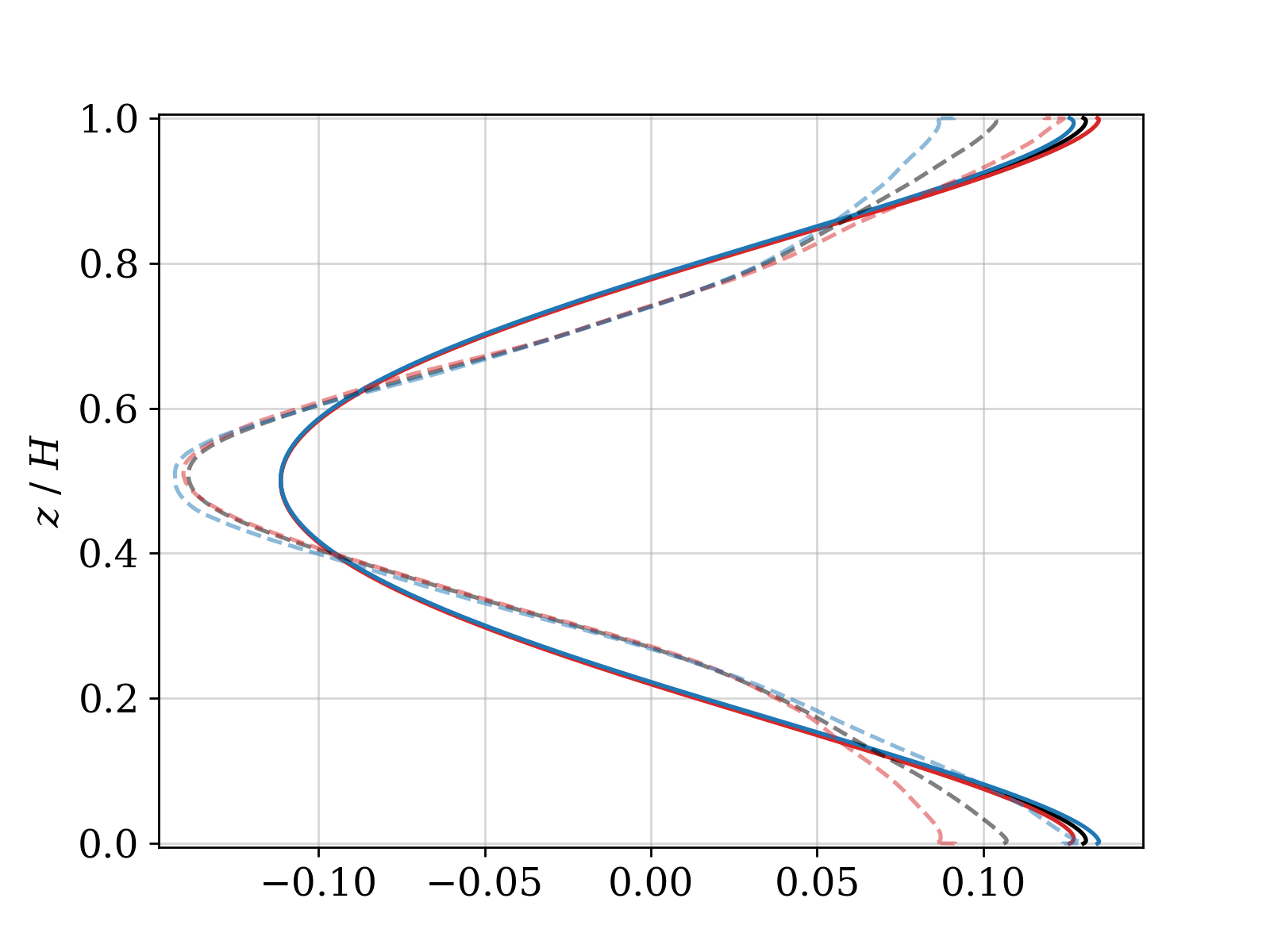}
    \end{subfigure}
    \begin{subfigure}[t]{0.49\textwidth}
        \caption{Vertical velocity, $w / \sqrt{\Delta B\ H}$}
        \includegraphics[trim = 0 0 0 35, clip, width=\textwidth]{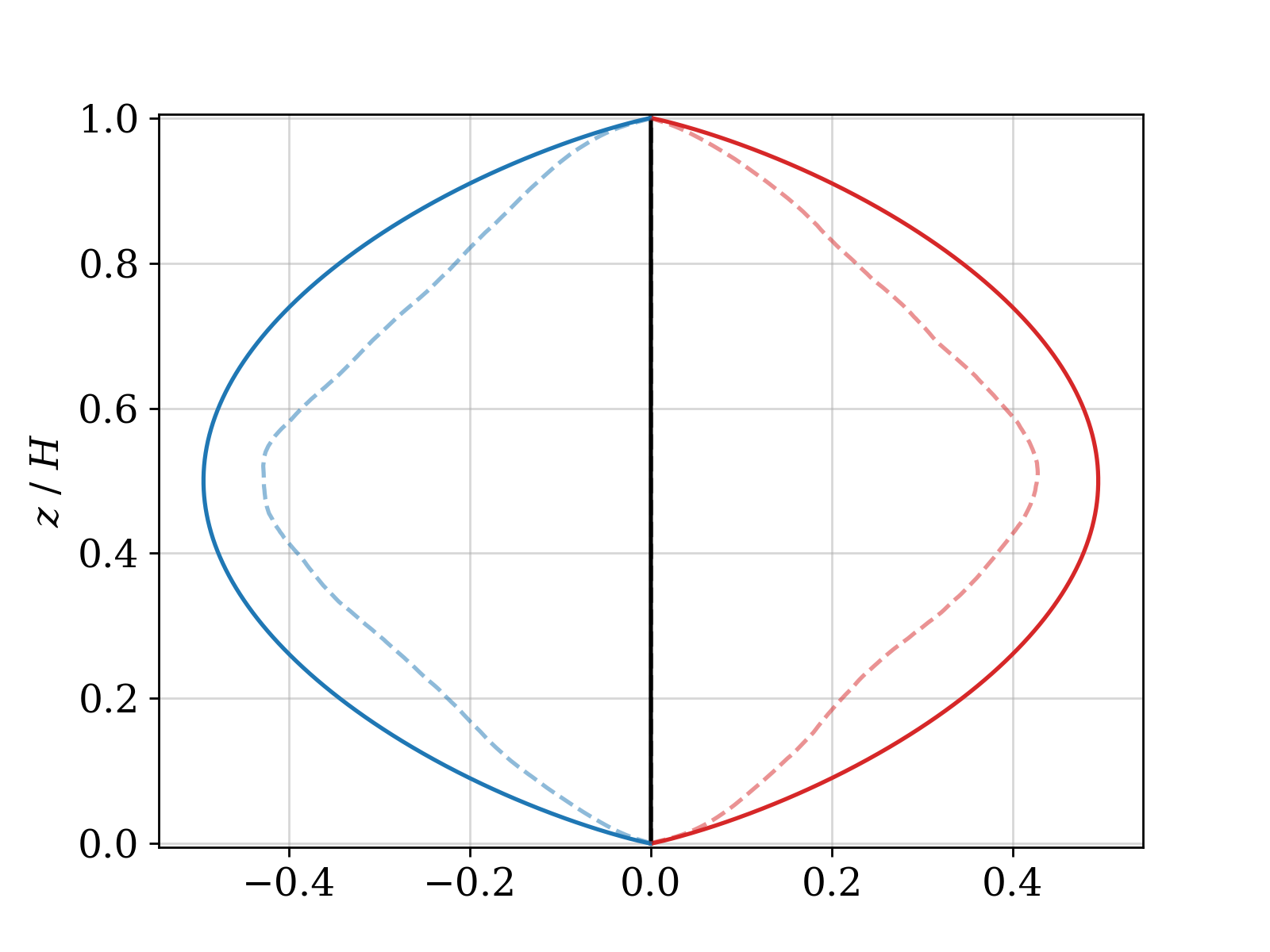}
    \end{subfigure}
    \begin{subfigure}[t]{0.49\textwidth}
        \caption{Fluid fraction, $\sigma_{w > 0}$}
        \includegraphics[trim = 0 0 0 35, clip, width=\textwidth]{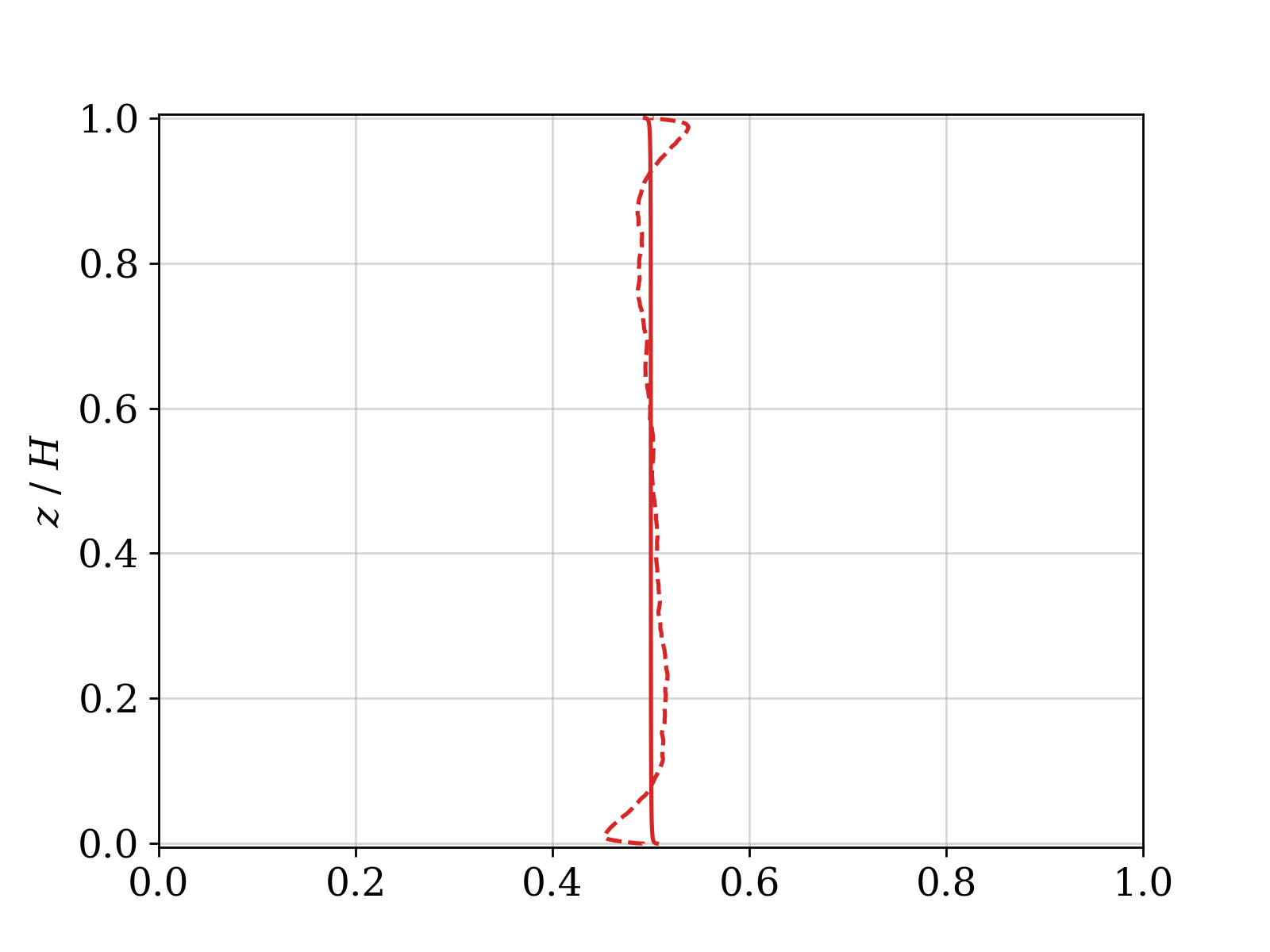}
    \end{subfigure}
    \begin{subfigure}[t]{0.4\textwidth}
        \centering
        \includegraphics[trim = -10 0 20 0, clip, width=\textwidth]{images/legend_2fProfiles_h.png}
    \end{subfigure}
    \caption{Two-fluid single-column model of $\Ra = 10^{10}$ RBC governed by equations \eqref{eq:2FBoussinesq_sigma_closed}-\eqref{eq:2Fclosure_gamma} and \eqref{eq:gammaScaling}, with closure constants $\hat{\gamma}_0 \approx 0.44$, $C = 0$. 
    Conditionally horizontally- and time-averaged profiles from the DNS are shown for reference. 
    $\Nu = 228$; reference $\Nu_\text{DNS} = 94.5$.
    }
    \label{fig:Ra10^10}
\end{figure}

Results from a two-fluid single-column model with $\hat{\gamma}_0 \approx 0.44$, $C=0$ are shown in Fig.~\ref{fig:Ra10^10}. 
Qualitative agreement with the buoyancy and vertical velocity profiles is still good, but the mean pressure profile predicted by the model now has too little curvature in the centre of the domain, and does not get the gradient correct close to the boundaries. 
Again, the complex mixing of the turbulent flow has strong effects on the volume fraction profile, causing the volume fraction of rising (falling) fluid to be less than $0.5$ close to the lower (upper) boundary.

\begin{figure}[t!hb]
    \centering
    \begin{subfigure}[t]{\textwidth}
        \caption{}
        \includegraphics[trim = 0 0 0 20, clip, width=\textwidth]{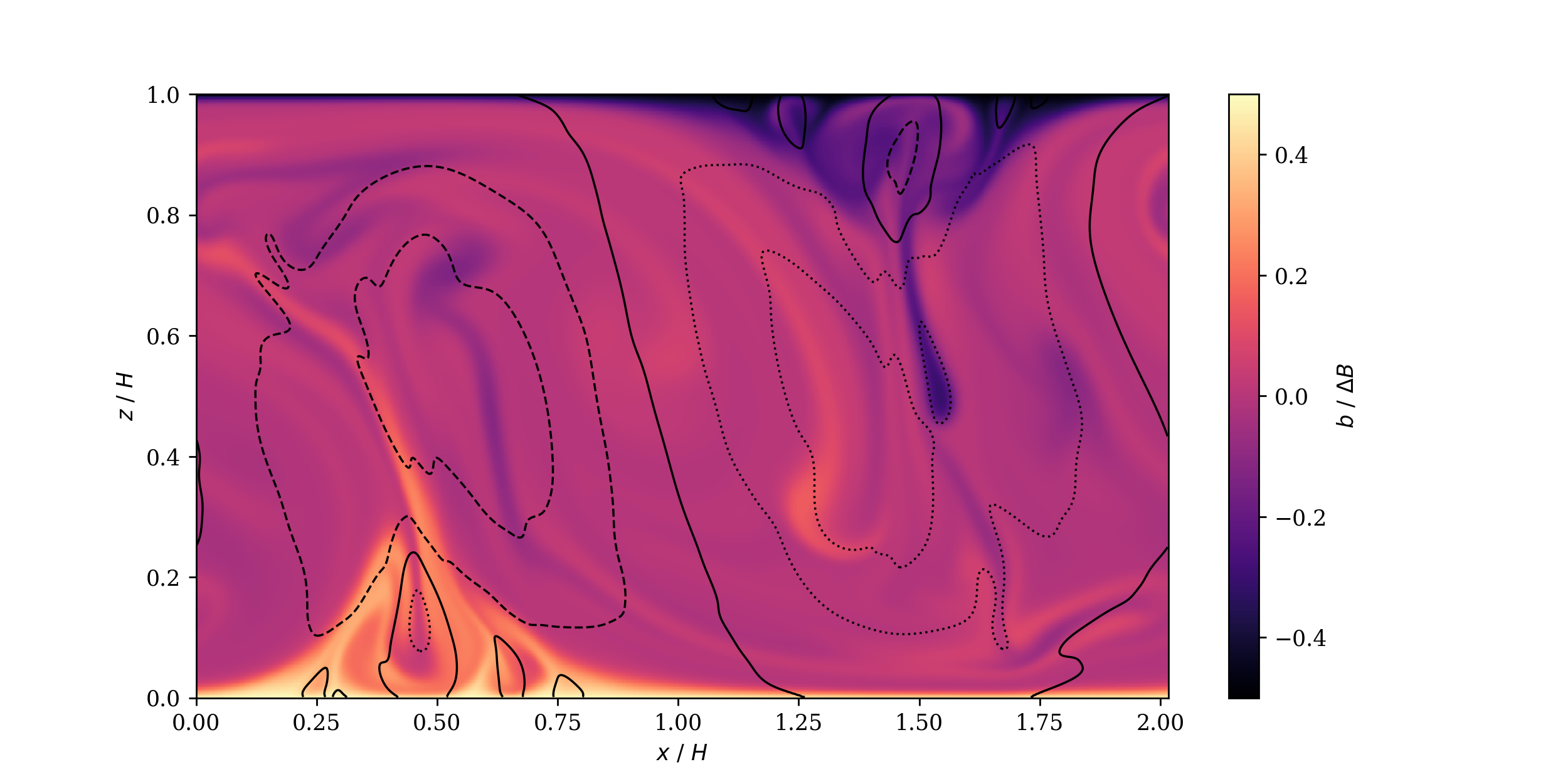}
    \end{subfigure}
    \begin{subfigure}[t]{\textwidth}
        \caption{}
        \includegraphics[trim = 0 0 0 20, clip, width=\textwidth]{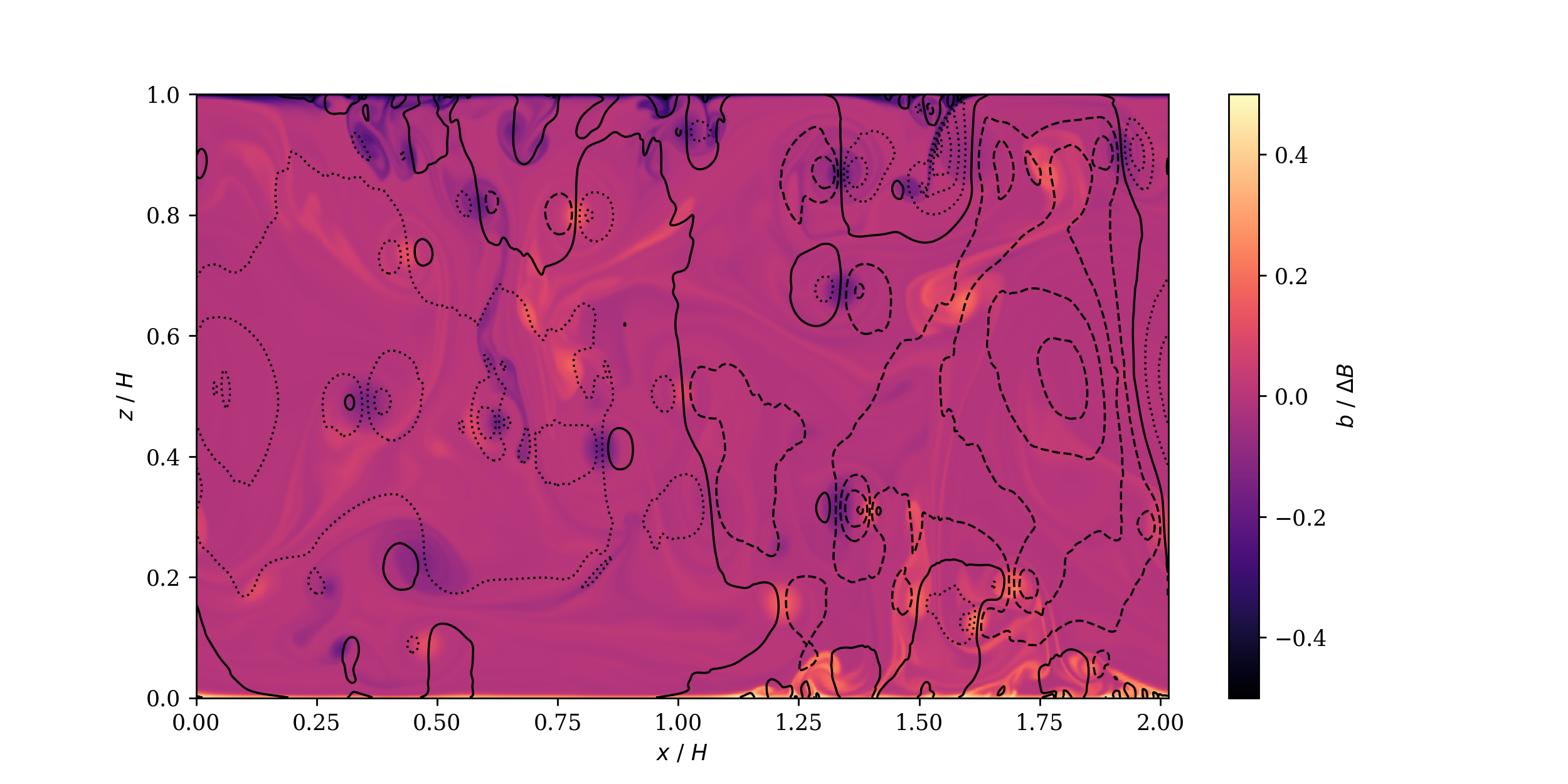}
    \end{subfigure}
    \caption{Snapshots of DNS buoyancy fields with overlaid vertical velocity contours at $\Ra = 10^8$ (a) and $\Ra = 10^{10}$ (b). 
    Dashed contours denote $w > 0$, dotted $w < 0$, and the solid contour denotes $w = 0$. 
    Contours above and below $w=0$ are spaced at intervals of $U_B / 4$.}\label{fig:DNSbuoyancy_wContours}
\end{figure}

These larger discrepancies between the DNS and the two-fluid model model are possibly because the $w = 0$ interface is now very complex. 
Figure~\ref{fig:DNSbuoyancy_wContours} shows the $w=0$ interface superimposed on the DNS buoyancy fields at $\Ra = 10^8$ and $\Ra = 10^{10}$. 
Although the dominant rising/falling two-fluid split is still into columns of falling and rising air with an approximately vertical interface even in the higher $\Ra$ case, the simple split is increasingly complicated by the complex vortical motions in the bulk of the fluid, and especially close to the base of the plumes. 
The intricate dynamics of these interfaces are not accounted for by our single-column model. 

While there are quantitative discrepancies, for all three Rayleigh numbers the overall the agreement between horizontally-averaged DNS and the two-fluid single-column model is good. 
Approximately the correct profiles are captured even in the highly turbulent regime of $\Ra = 10^{10}$.
The model performs remarkably well given it has no representation of sub-filter variability beyond the two-fluid split, showing that the model captures the essential coherent overturning structures of Rayleigh-B\'{e}nard convection in all three characteristic regimes.

\subsection{Sensitivity to \texorpdfstring{$\hat{\gamma}_0$}{gamma0} and \texorpdfstring{$C$}{C}}\label{sec:sensitivityToGammaAndC}
In this section, the sensitivity of the model to the dimensionless closure parameters $\hat{\gamma}_0$ and $C$ is investigated. 
The effects of changing $\hat{\gamma}_0$ and $C$ are similar at all Rayleigh numbers, so for brevity only $\Ra = 10^5$ is presented.

\subsubsection{Sensitivity to \texorpdfstring{$\hat{\gamma}_0$}{gamma0}}\label{sec:sensitivityToGamma}
\begin{figure}[t!hb]
    \centering
    \begin{subfigure}[t]{\textwidth}
        \caption{Buoyancy, $b / \Delta B$}
        \includegraphics[trim = 0 0 0 29, clip, width=\textwidth]{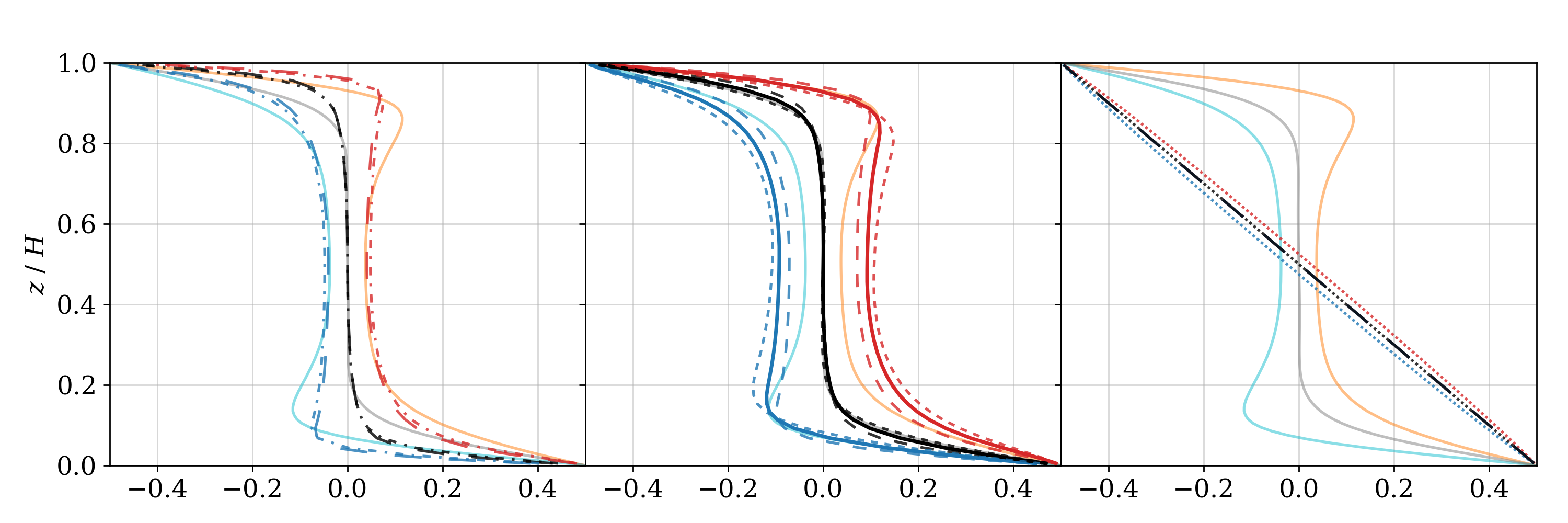}
    \end{subfigure}
    \begin{subfigure}[t]{\textwidth}
        \caption{Pressure, $P / (\Delta B\ H)$}
        \includegraphics[trim = 0 0 0 29, clip, width=\textwidth]{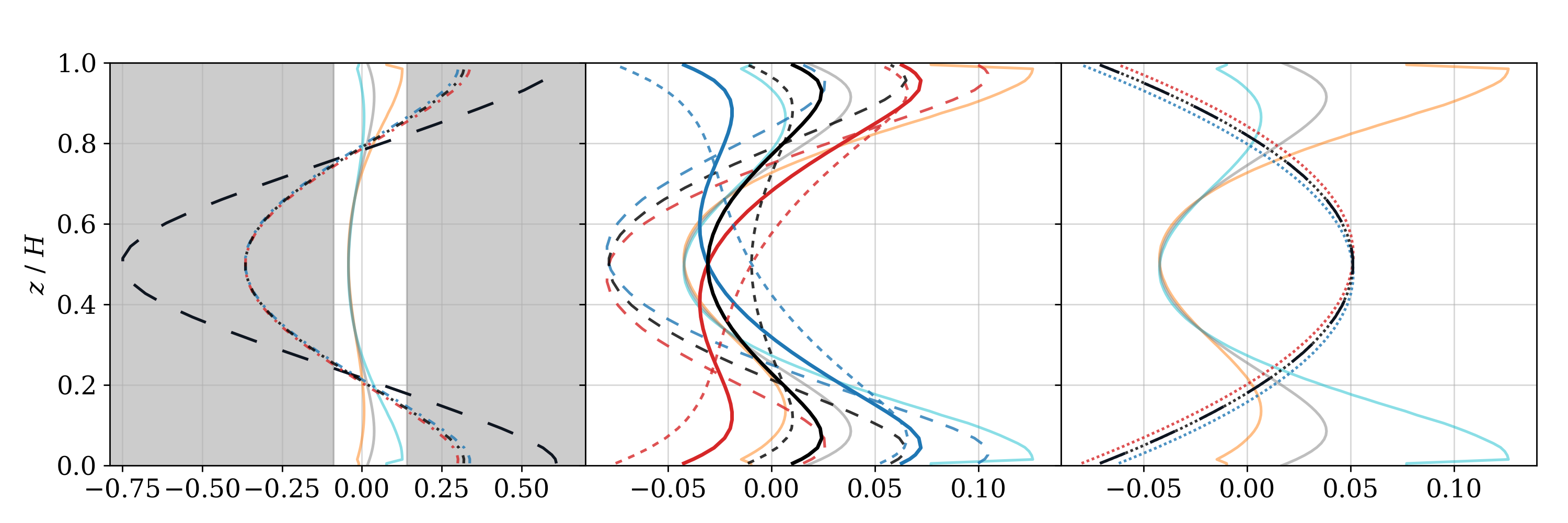}
    \end{subfigure}
    \begin{subfigure}[t]{\textwidth}
        \caption{Vertical velocity, $w / \sqrt{\Delta B\ H}$}
        \includegraphics[trim = 0 0 0 29, clip, width=\textwidth]{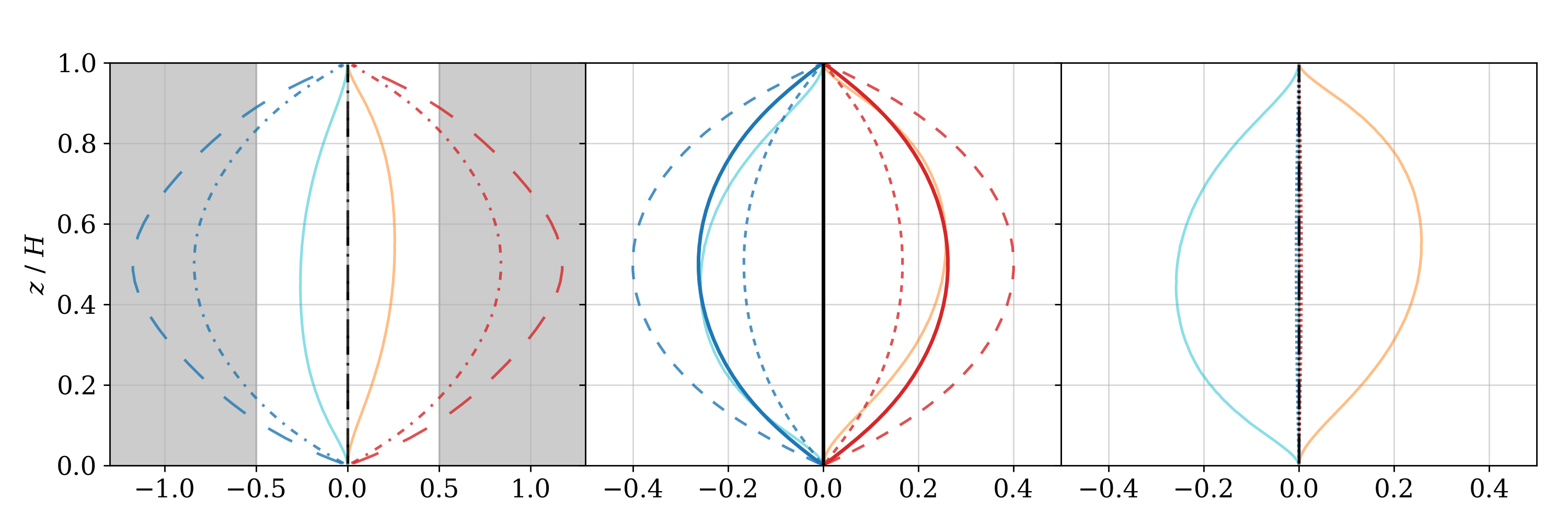}
    \end{subfigure}
    \begin{subfigure}[t]{\textwidth}
        \centering
        \includegraphics[trim = -30 0 30 0, clip,width=\textwidth]{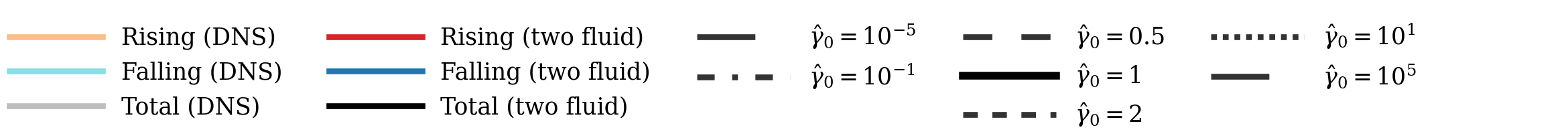}
    \end{subfigure}
    \caption{Two-fluid single-column model of $\Ra = 10^{5}$ RBC governed by equations \eqref{eq:2FBoussinesq_sigma_closed}-\eqref{eq:2Fclosure_gamma} and \eqref{eq:gammaScaling}, with $C = 0.5$, showing sensitivity to $\hat{\gamma}_0$ (defined in eq.~\eqref{eq:gammaScaling}) over the range $10^{-1} \leq \hat{\gamma}_0 \leq 10^{1}$. 
    Profiles in the limit of asymptotically large ($10^5$) and small ($10^{-5}$) $\hat{\gamma}_0$ are also shown for reference. 
    $\hat{\gamma}_0 = \mathcal{O}(1)$ is expected based on the scale analysis of section~\ref{sec:scaling}.
    Small values of $\hat{\gamma}_0$ ($\lesssim \mathcal{O}(10^{-1})$) are shown in the left column, values of order 1 in the middle column, and large magnitudes ($\gtrsim \mathcal{O}(10)$) in the right column.
    Grey shaded regions in plots in the left column highlight areas which are not in the domain of plots in the centre and right columns.
    }
    \label{fig:twoFluid_sensitivity_gamma}
\end{figure}

Figure~\ref{fig:twoFluid_sensitivity_gamma} shows the effect on the two-fluid single-column steady-state of varying $\hat{\gamma}_0$ from $10^{-1} \lesssim \hat{\gamma}_0 \lesssim 10^{1}$, along with examples in the asymptotically-large and -small $\hat{\gamma}_0$ regimes. 
The experiments were performed with $C=0.5$ at fixed $\Ra = 10^5$, but the results are similar for all $\Ra$.

The best qualitative match between the single-column and DNS profiles is found when $\hat{\gamma}_0 \approx 0.75$, as discussed earlier, while the correct heat flux is predicted at $\hat{\gamma}_0 \approx 1.861$. 
These values are both $\mathcal{O}(1)$, as expected. 
Agreement with the reference profiles degrades sharply as $\hat{\gamma}_0$ moves away from this range.

Increasing $\hat{\gamma}_0$ increases the buoyancy difference between the fluids, and damps the vertical velocities --- which makes sense since in 1D this parametrization of $p_i$ \textit{is} similar to diffusion of the vertical velocity within a fluid, even though the sum correction means no extra viscous term is added to the mean momentum budget. 
This effect is already clear at $\hat{\gamma}_0 = 2$, where the vertical velocities are only $\approx 2/3$ of those in the DNS, and the pressure profile is much shallower, though still with the correct number of turning points.
By $\hat{\gamma}_0 = 10$, the pressure profile loses the minimum in the centre of the domain, and the vertical velocities are almost zero. 
At asymptotically large $\hat{\gamma}_0$, the system becomes subcritical and the solution is purely diffusive.

Decreasing $\hat{\gamma}_0$ rapidly increases the pressure gradient, and deepens the minimum of the mean pressure in the centre of the domain. 
This drastically increases the vertical velocities --- by $\hat{\gamma}_0 = 10^{-1}$, the maximum vertical velocities are over three times those of the DNS, and over twice those of the simulations with $\hat{\gamma}_0 = 0.75$ discussed in detail earlier. 
Decreasing $\hat{\gamma}_0$ further only slightly changes these results, as seen for the asymptotically-small case of $\hat{\gamma}_0 = \times10^{-5}$.

\subsubsection{Sensitivity to \texorpdfstring{$C$}{C}}\label{sec:sensitivityToC}
\begin{figure}[t!hb]
    \centering
    \begin{subfigure}[t]{\textwidth}
        \caption{Buoyancy, $b / \Delta B$}
        \includegraphics[trim = 0 0 0 29, clip, width=\textwidth]{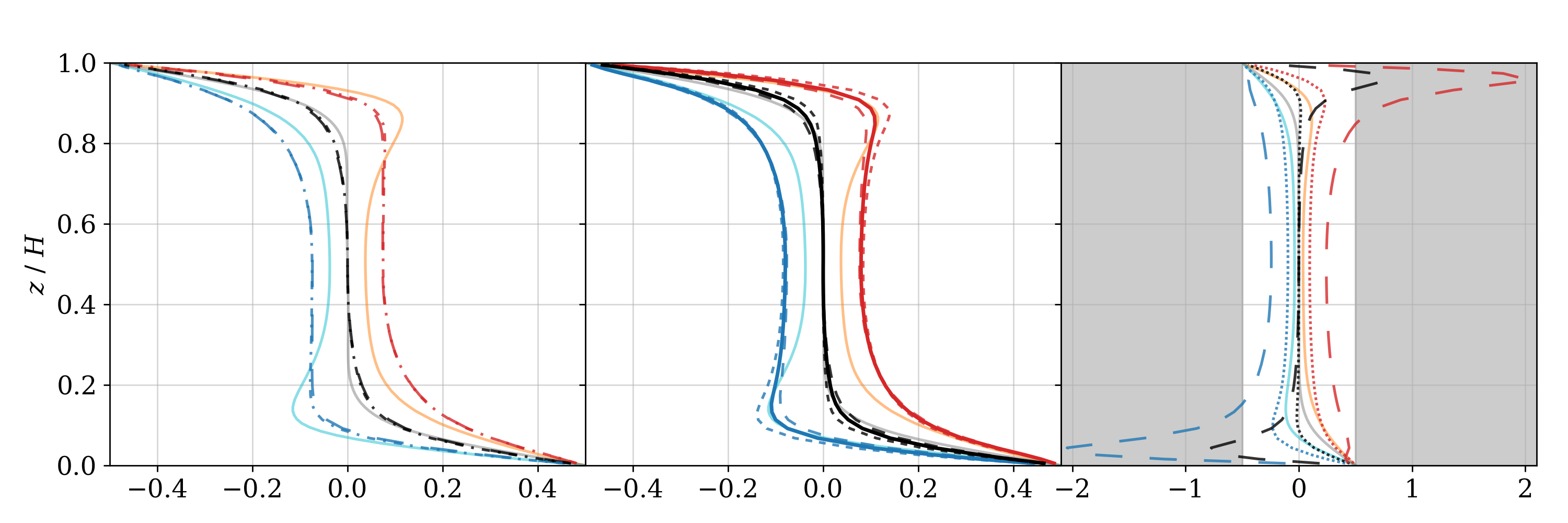}
    \end{subfigure}
    \begin{subfigure}[t]{\textwidth}
        \caption{Pressure, $P / (\Delta B\ H)$}
        \includegraphics[trim = 0 0 0 29, clip, width=\textwidth]{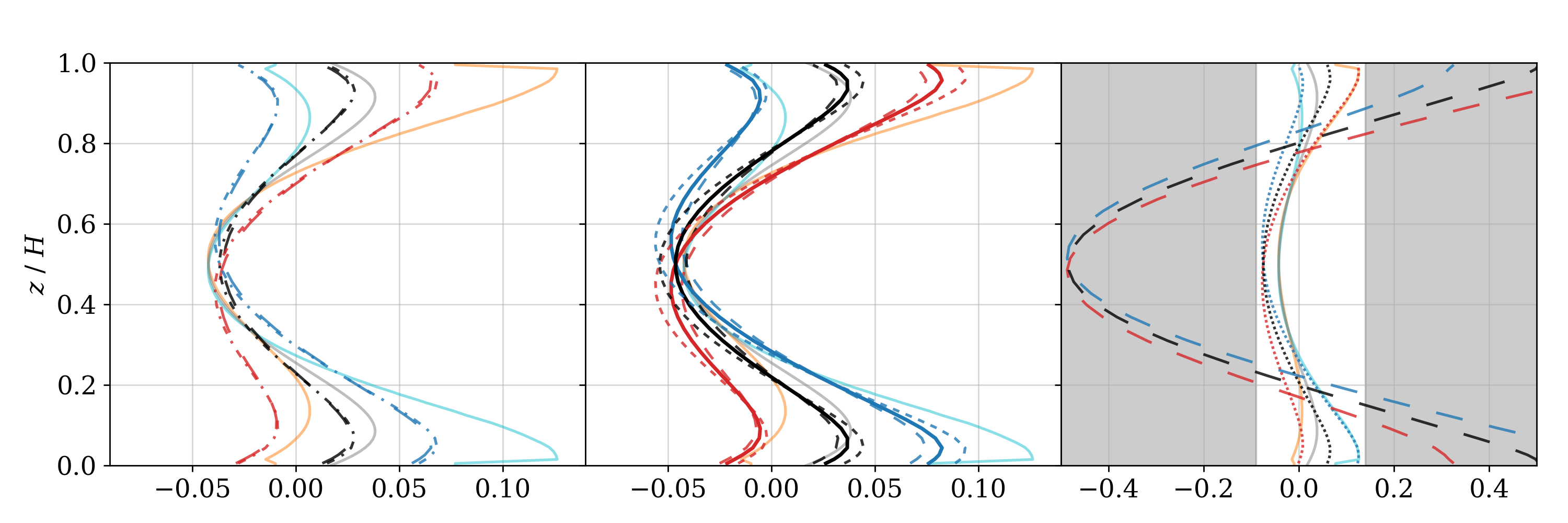}
    \end{subfigure}
    \begin{subfigure}[t]{\textwidth}
        \caption{Vertical velocity, $w / \sqrt{\Delta B\ H}$}
        \includegraphics[trim = 0 0 0 29, clip, width=\textwidth]{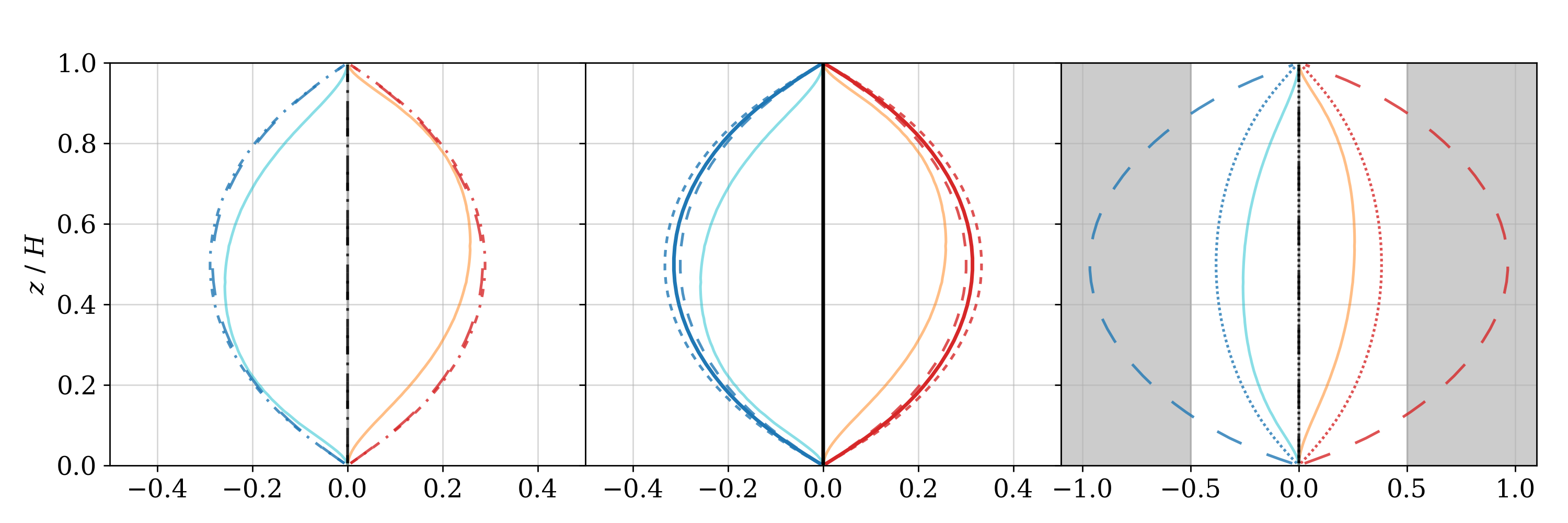}
    \end{subfigure}
    \begin{subfigure}[t]{\textwidth}
        \centering
        \includegraphics[trim = -30 0 30 0, clip,width=\textwidth]{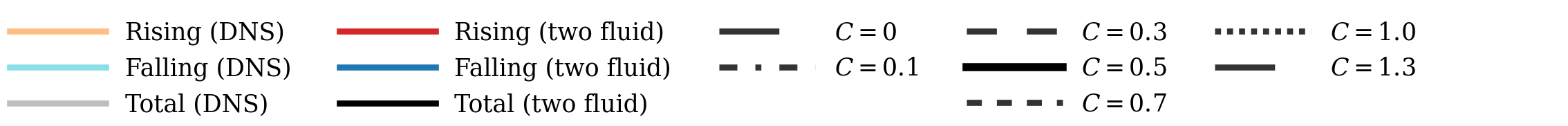}
    \end{subfigure}
    \caption{Two-fluid single-column model of $\Ra = 10^{5}$ RBC governed by equations \eqref{eq:2FBoussinesq_sigma_closed}-\eqref{eq:2Fclosure_gamma} and \eqref{eq:gammaScaling}, with $\hat{\gamma}_0 \approx 0.75$, showing sensitivity to the transferred buoyancy parameter $C$ (defined in eq.~\eqref{eq:2Fclosure_bTransfer}) over the range $0 \leq C \leq 1.3$.
    $C=0.5$ corresponds to the profiles in figure~\ref{fig:Ra10^5_2F1C_results}.
    For $C \gtrsim 1.3$, the solution becomes unstable.
    Results for small values of $C$ ($= 0, 0.1$) are shown in the left column; for values around the central value of $0.5$ in the middle column; and for large values ($\geq 1$) in the right column (see text for interpretation).
    Grey shaded regions in plots in the right column highlight areas which are not in the domain of plots in the left and centre columns. 
    }
    \label{fig:twoFluid_sensitivity_C}
\end{figure}

Figure~\ref{fig:twoFluid_sensitivity_C} shows the steady-state effect of varying $C$ from $0$ (mean buoyancy is transferred: $b_{ij}^T = b_i$) to $1$ (zero buoyancy is transferred over most of the domain: $b_{ij}^T = 0$ wherever $b_i = \abs{b_i}$).
Transfers with $C>1$ amount to transferring buoyancies with magnitude greater than $\Delta B$ close to the boundaries, which causes the solution to become unstable at $C \approx 1.3$. 

The main effect of increasing $C$ is to generate the aforementioned overshoots in the within-fluid buoyancy profiles; this also steepens the pressure gradient, deepens the central pressure, and increases the magnitude of the vertical velocities in each fluid. 
These effects are small compared to the order-of-magnitude effects associated with varying $\hat{\gamma}_0$: for example, the maximum velocity increases monotonically from $0.3$ to $0.45$ as $C$ increases from $0$ to $1$. 
These effects are qualitatively similar at all $\Ra$, but for $\Ra \gtrsim 10^7$, the individual fluid buoyancy profiles no longer exhibit overshoots, so $C = 0$ provides a better fit with the DNS buoyancy profiles.

\subsection{Scaling of Nusselt number with Rayleigh number}\label{sec:scalingOfNuWithRa}
To investigate the performance of the two-fluid single column model more systematically, the scaling of the Nusselt number for single-column models across the Rayleigh number range $10^2 \leq \Ra \leq 10^{10}$ is compared with the DNS results.
The scaling $\gamma / \nu \propto \Ra^{1/4}$ (section~~\ref{sec:scaling}) is evaluated, along with two choices of the transferred buoyancy, $C = 0$ and $C = 0.5$. 
For each transferred buoyancy, the dimensionless proportionality factor $\hat{\gamma}_0$ was fixed by finding the value which gave the correct Nusselt number at $\Ra = 10^5$.
Fixing this constant at different Rayleigh numbers changes the prefactor of the $\Nu(\Ra)$ scaling, but does not change the scaling itself.

\begin{figure}[t!hb]
\begin{subfigure}{\textwidth}
    \centering
    \includegraphics[width=0.7\textwidth]{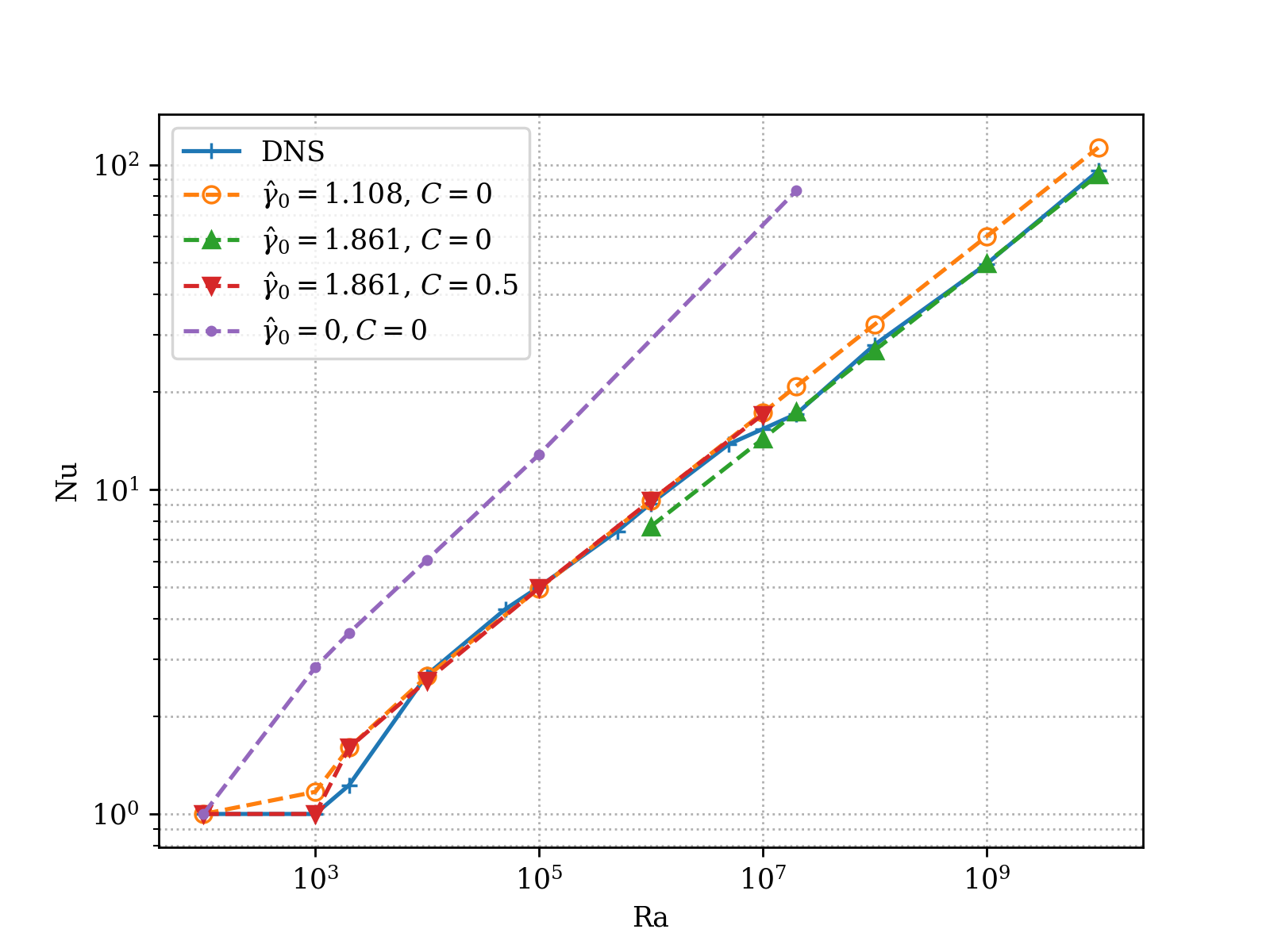}
    \caption{Nusselt number vs. Rayleigh number for two fluid single column models with various values of $\hat{\gamma}_0$ and $C$. 
    The dashed blue curve shows the reference DNS results (as \ref{fig:DNS_scalingVerification}a), while the dashed brown curve shows the results of running the single column model with $C=0$ (mass exchanges transfer the mean buoyancy) and $\gamma = 0$ (no pressure differences between the fluids). 
    The green, purple, and red curves show the results for $\gamma \sim \Ra^{1/4}$, with different values of $\gamma_0$; all give scalings of $\Nu \approx \Ra^{2/7}$.
    Single-column Nusselt numbers are calculated from the buoyancy gradient at the boundaries, and checked against the column-integrated buoyancy flux.
    }
    \label{fig:2F1C_NuVsRa}
\end{subfigure}
\begin{subfigure}{\textwidth}
    \centering
    \includegraphics[width=0.7\textwidth]{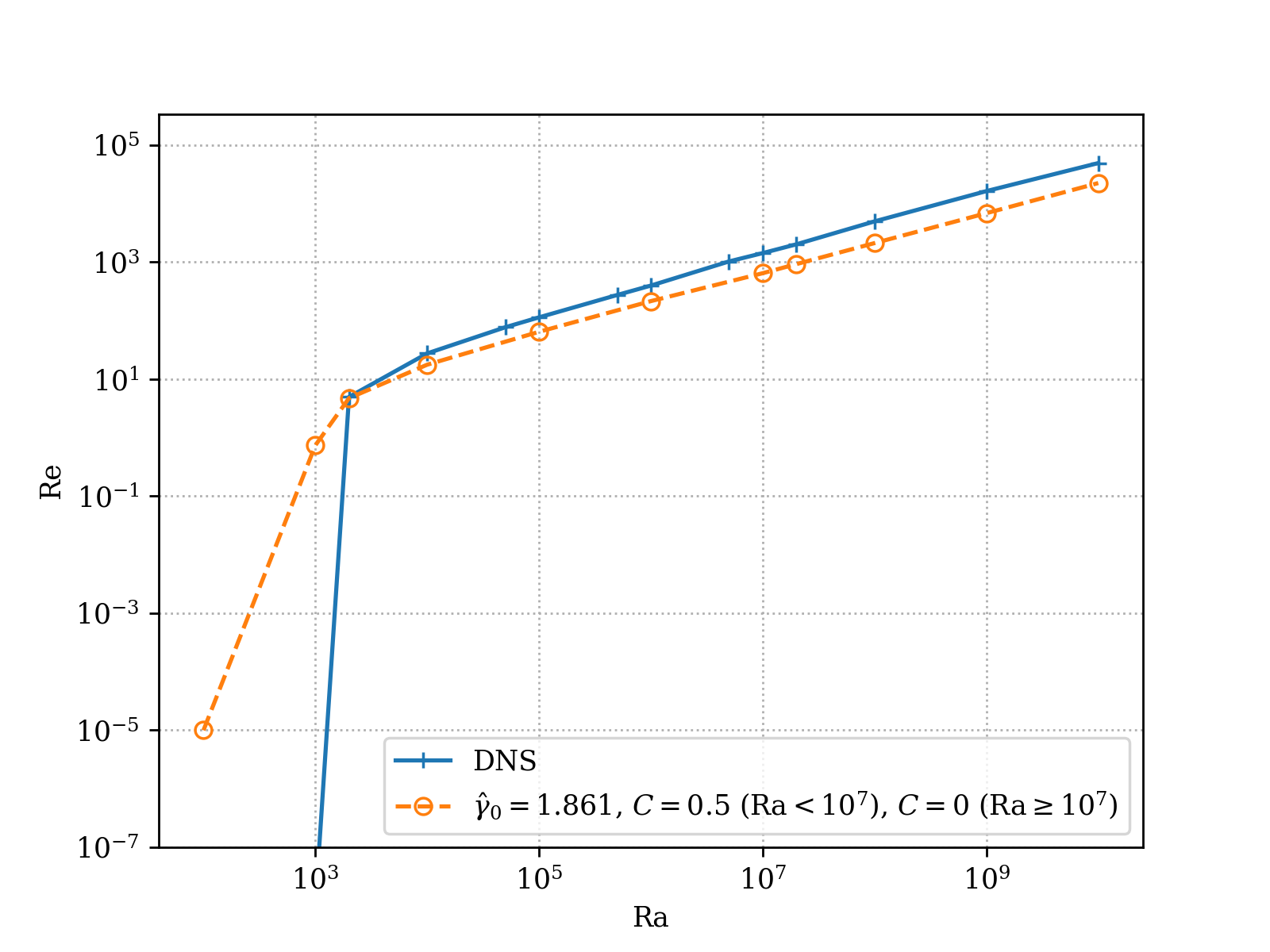}
    \caption{Reynolds number vs. Rayleigh number for $10^2 \leq \Ra \leq 10^{10}$. 
    The dashed blue curve shows the reference DNS results (as \ref{fig:DNS_scalingVerification}b), while the solid orange curve shows results from a two fluid single column model obeying equations \eqref{eq:2FBoussinesq_sigma_closed}-\eqref{eq:2Fclosure_gamma}, with $\gamma/\nu = 1.861 \Ra^{1/4}$ and $C=0.5$ for $\Ra \leq 10^7$, $C=0$ for $\Ra > 10^7$; these constants give the best fit for $\Nu$ as a function of $\Ra$ (figure~\ref{fig:2F1C_NuVsRa}). 
    Both curves exhibit scalings of $\Re \approx \Ra^{1/2}$ for $\Ra \gtrsim 10^4$. 
    Single-column Reynolds numbers are calculated using the maxima of the individual fluid vertical velocity profiles for the velocity scale.}\label{fig:2F1C_ReVsRa}
\end{subfigure}
\caption{}
\end{figure}

Figure~\ref{fig:2F1C_NuVsRa} shows $\Nu$ against $\Ra$ for the different values of $C$ and scalings for $\gamma$. 
The DNS results are shown for comparison, along with results from the single column model run with both tunable parameters set to zero, $C = \hat{\gamma}_0 = 0$. 
All models with $\hat{\gamma}_0 > 0$ perform significantly better than the model with $\hat{\gamma}_0 = 0$, which becomes supercritical for $\Ra < 10^3$ and follows a $\Nu(\Ra)$ scaling with exponent everywhere $>0.33$.

Models with $\gamma / \nu \propto \Ra^{1/4}$ show exceptional agreement with the DNS heat fluxes for $\Ra \geq 10^4$, giving $\Nu \sim \Ra^{2/7}$ with both $C=0$ (green curve) and $C=0.5$ (purple curve). 
This shows that the Nusselt number scaling exponent depends on $\gamma / \nu$ but not on $C$; this makes sense since $C$ is a crude parametrization for \textit{how} the flow produces a given heat flux, and should not affect the scaling of the heat flux itself. 
Below $\Ra = 10^4$, the models with different values of $C$ produce slightly different behaviour: the $C=0$ solutions become supercritical below $\Ra = 10^3$, inconsistent with the known $\Ra_\text{c} \approx 1708$. 
While the $C=0$ simulations are still subcritical at $\Ra = 10^3$, the heat flux at $\Ra = 2\times10^3$ is roughly $30\%$ too high. 
These discrepancies suggest that the scaling used for $\gamma / \nu$ is not quite correct in the low $\Ra$ regime; unsurprising since the scaling argument assumed $\Re \gg 1$. 
For the intended application to highly turbulent atmospheric convection, however, this does not present a severe problem.

The single-column model does not naturally capture the drop in the prefactor of the Nusselt number scaling which occurs as the flow transitions to turbulence around $\Ra \approx 10^7$. 
The drop in the Nusselt number scaling prefactor may not be a robust feature of the convective flow, so it is far more important to get the scaling exponent correct. 
Such drops in the scaling prefactor are found in other RBC experiments \parencite[see][for a 2D numerical and a 3D experimental example, respectively]{ar:JohnstonDoering2009,ar:RocheEtAl2004}, but appear to be dependent directly on the nature of the flow, rather than global in nature like the scaling exponent. 
However, this drop \textit{can} be accurately reproduced by using $C=0.5$ for $\Ra \leq 10^7$ and $C = 0$ for $\Ra > 10^7$, retaining the value of $\hat{\gamma}_0 \approx 1.861$. 
With this parametrization, the Nusselt number is correctly predicted to within $5\%$ across six orders of magnitude of buoyancy forcing, $10^4 \leq \Ra \leq 10^{10}$, and approximately the correct transitional behaviour is found for $\Ra < 10^4$. 
This could be diagnostically incorporated into the parametrization by, for instance, reducing $C$ to $0$ whenever the vertical velocity maximum gives a turbulent $\Re \gtrsim 2\times10^3$.

The Reynolds number in the single-column simulations was estimated from the maximum magnitude of the vertical velocity; this should scale with the large-scale circulation, so makes sense for a bulk Reynolds number.
The scaling behaviour of the Reynolds number is also well-captured (figure~\ref{fig:2F1C_ReVsRa}), in particular giving the same scaling exponent as the DNS. 
Notably, the change in $C$ required to capture the correct behaviour of $\Nu$ does not cause a corresponding kink in the Reynolds number scaling. 
This suggests that $C$ really is just a crude measure of the flow state. 
Future work would hope to capture these flow states dynamically through representing the sub-filter scale variability of the variables within each fluid.

\section{Summary, conclusions, and future work}
In this paper we have shown that the simple two-fluid single column model \eqref{eq:2FBoussinesq_sigma_closed}-\eqref{eq:2Fclosure_gamma} can qualitatively reproduce horizontal-mean DNS buoyancy, vertical velocity, and pressure profiles in all three characteristic regimes of Rayleigh-B\'{e}nard convection. 
A scaling argument for the pressure differences between the fluids allows the model to predict the correct power-law scaling of $\Nu \sim \Ra^{2/7}$, and after measuring a dimensionless constant at one Rayleigh number the magnitude of $\Nu$ can be predicted to within $5\%$ over 6 orders of magnitude of $\Ra$.
The model also captures approximately the correct spin-up behaviour, and approximately the correct critical Rayleigh number. 
The closure set is minimal, requiring only two constants to be set; and not finely-tuned, as both closure constants may be varied significantly from their central values without destroying the solution.

Although we use a similar equation set and identical fluid definitions to \textcite{ar:WellerEtAl2020}, this is the first such study to model a fully turbulent regime with these fluid definitions. 
It is also the first multi-fluid convection study to considerably vary the applied forcing, testing the robustness of the parametrization.

This demonstrates the essential validity of the multi-fluid concept: the model directly captures the dominant overturning circulation of convection, present even in the fully turbulent regime, by allowing for a circulation even in a single column. 
It is important to note that this performance is achieved without even a minimal treatment of fluxes due to variability within each fluid (i.e. conventional `turbulent' or `subfilter' fluxes) apart from the fixed viscosity and Prandtl number of the fluid.

With the current model the mean buoyancy profile (and therefore the Nusselt number), the vertical velocity maxima in each fluid (and therefore the implied Reynolds number), and the pressure profile, cannot all simultaneously have the correct magnitude. 
It is unclear whether this is due to neglected subfilter variability (in the form of exchanged buoyancy or neglected subfilter stresses, for example), or due to inadequate representation of the fluid fraction transfers.
A more accurate and flexible representation of these transfers is essential to progressing beyond single-column modelling.

Future work will test the two-fluid model of this paper in the grey zone of RBC, investigating how the closures scale with resolution, and noting what flow features are missed by the simple closures in a higher dimensional setting. 
Improvements could arise from a partition which better selects the coherent structures, and from representation of within-fluid variability by consideration of higher moments of the flow. 
In particular, DNS data may be used to diagnose $S_{ij}, b^T_{ij}, \vec{u}^T_{ij}$ for various filter scales and fluid definitions. 
Possible closures could be informed by direct analysis of the interactions between coherent structures, boundary layers, and homogeneous, isotropic bulk \parencite{ar:TogniEtAl2015,ar:BerghoutEtAl2021}.

All of the above will develop fundamental understanding of the multi-fluid equations for convection. 
A thorough understanding of the dry convective grey zone, and of possible multi-fluid approaches to its parametrization, will help sharpen the questions for the much thornier problem of moist convection.

\clearpage

\printbibliography

\end{document}